\documentclass[prb,showpacs,preprintnumbers,amsmath,aps]{revtex4}
\usepackage{graphicx,hyperref}% Include figure files
\usepackage{amssymb}
\usepackage{bm}% bold math
\usepackage{subfigure}% bold math
\usepackage{color}

 \let\b=\beta    \let\d=\delta

\newcommand{\beq}{\begin{equation}}
\newcommand{\eeq}{\end{equation}}
\newcommand{\bea}{\begin{eqnarray}}
\newcommand{\eea}{\end{eqnarray}}
\newcommand{\beal}{\begin{aligned}}
\newcommand{\eal}{\end{aligned}}

\begin{document}

\title{Random Field Ising-like effective theory of the glass transition II: 
Finite Dimensional Models}

\author{Giulio Biroli} \email{giulio.biroli@cea.fr}
\affiliation{IPhT, CEA/DSM-CNRS/URA 2306, CEA Saclay, F-91191 Gif-sur-Yvette Cedex, France\\
Laboratoire de Physique Statistique, \'Ecole Normale Sup\'erieure, CNRS, France}

\author{Chiara Cammarota} \email{chiara.cammarota@kcl.ac.uk}
\affiliation{Department of Mathematics, King's College London, Strand, London WC2R 2LS, UK}

\author{Gilles Tarjus} \email{tarjus@lptmc.jussieu.fr}
\affiliation{LPTMC, CNRS-UMR 7600, Sorbonne Universit\'e, 4 Pl. Jussieu, F-75005 Paris, France}

\author{Marco Tarzia} \email{tarzia@lptmc.jussieu.fr}
\affiliation{LPTMC, CNRS-UMR 7600, Sorbonne Universit\'e, 4 Pl. Jussieu, F-75005 Paris, France}

\date{\today}

\begin{abstract}

As in the preceding paper (hereafter referred to as paper~I\cite{paperI}) we aim at identifying the effective theory that describes the fluctuations of the local overlap with an equilibrium reference configuration close to a putative thermodynamic glass transition. We focus here on the case of finite-dimensional glass-forming systems, in particular supercooled liquids. The main difficulty for going beyond the mean-field treatment comes from the presence of diverging point-to-set spatial correlations. We introduce a variational low-temperature approximation scheme that allows us to account, at least in part, for the effect of these correlations. The outcome is an effective theory for the overlap fluctuations in terms of a random-field + random-bond Ising model with additional, power-law decaying, pair and multi-body interactions generated by the point-to-set correlations. This theory is much more tractable than the original problem. We check the robustness of the approximation scheme by applying it to a fully connected model already studied in paper~I. We discuss the physical implications of this mapping for glass-forming liquids and the possibility it offers to determine the presence or not of a finite-temperature thermodynamic glass transition.
\end{abstract}

\maketitle

\tableofcontents

\section{Introduction} 
\label{sec:intro}

Establishing the theory of the glass transition is admittedly a difficult endeavor. A variety of approaches have been put forward, often with quite different views on the origin of the phenomenon and the best way to describe it, but none of them has so far been accepted as providing the definite answer.\cite{GiulioLudo,tarjus_overview,Wolynesbook,KTW,gotze,debenedetti,chandler-garrahan,frustration-review,dyre,wyart-cates} What has been unambiguously established, though, is the mean-field theory of the glass transition and of the glass phase, which has been shown to be exactly realized for liquids in the limit of infinite spatial dimensions.\cite{kurchan-zanponi-etal,kurchan-maimbourg} This conceptual advance has allowed one to connect the phenomena of jamming and of glass formation in a unified framework. However it leaves open the problem of glass-forming liquids in $3$ dimensions because of the anticipated strong effect of spatial fluctuations on the mean-field scenario in finite dimensions, as discussed in the conclusion of paper~I\cite{paperI}. The mean-field description relies on the existence of a complex free-energy landscape with a multitude of metastable states and a two-transition scenario with a key role played at intermediate temperatures  between these transitions by the configurational entropy associated with the metastable states.\cite{KTW,Wolynesbook,RFOT_review,Pedestrians} Yet the very notion of metastability looses its rigorous meaning in the presence of spatial fluctuations because of the ubiquitous occurrence of ``nonperturbative'' nucleation-like events that destroy metastable states via thermal activation and enforce convexity of the free-energy. These events are of course candidates to explain the strong slowdown of relaxation observed in glass-forming liquids approaching the glass transition\cite{KTW} but in spite of interesting attempts\cite{KTW,RFOT_review,BB-PTS,Dzero,Franz_instantons} they have not been incorporated so far in a proper theoretical treatment.

Our goal, described in this paper and the companion one referred to as paper~I,\cite{paperI} is not to provide a full-blown solution of the glass transition problem, as we do not address the, nonetheless central, question of the dynamics of glass-forming liquids. We want instead to identify the effective theory which describes the statistics of the fluctuations of what is thought to be the relevant order parameter for glassy systems---the overlap field with an equilibrium reference configuration---close to the putative thermodynamic glass transition, the random first-order transition (RFOT)\cite{Wolynesbook} predicted by the mean-field treatment. The idea is of course to derive a theory that is simpler than the original problem and that can be studied by powerful tools of statistical mechanics (large-scale numerical simulations, nonperturbative functional renormalization group, etc.). 

In paper~I\cite{paperI} we have shown that in the case of mean-field models of structural glasses---the random energy model (REM)\cite{derrida} and a Kac-like generalization, the fully connected $2^M$-KREM\cite{REM}---an effective theory in terms of an Ising model in an external field in the presence of quenched disorder, including a random field, naturally emerges. In the richer case of the  $2^M$-KREM, the interactions between Ising variables, where a state represents a low or a high overlap with a reference equilibrium configuration, contain multi-body terms in addition to a standard short-range ferromagnetic term.

The aim of the present paper is to extend this effective description to finite-dimensional glass-formers and, above all, to glass-forming liquids (for a pioneering work, see Ref.~[\onlinecite{Stevenson}]). In this case, contrary to mean-field models, the effective theory cannot be worked out exactly, as already discussed in the last part of paper~I.\cite{paperI} The main difficulty is related to the emergence of (possibly long-ranged) point-to-set correlations. We then need to introduce several approximations, guided by the mean-field results and physical intuition.  The outcome is a description of the glass transition in terms of an Ising model in an external field with random-field and random-bond disorder and  long-range, competing, multi-body interactions:
\begin{equation}
\label{eq_effectiveIsing_final}
\begin{aligned}
\beta {\mathcal H}_{\rm eff} = &- \sum_{\langle i,j \rangle} (J_2+\delta J_{ij}) \sigma^i \sigma^j - 
\sum_i \left ( H + \d h_i\right) \sigma^i  + \frac{1}{2} \sum_{i \neq j} \tilde J_2(\vert r_i-r_j\vert) \, \sigma^i \sigma^j
+ \! \sum_{\langle i,j \rangle \neq \langle k,l\rangle } \!\! 
J_4 (\vert r_i-r_k\vert) \, \sigma^i \sigma^j \sigma^k \sigma^l 
+ \ldots \, .
\end{aligned}
\end{equation}
where the Ising variables $\sigma^i=\pm 1$ refer to a low and a high overlap with a reference equilibrium configuration. The uniform source  $H$ plays the role of a ``renormalized'' configurational entropy, $\delta J_{ij}$ and $\d h_i$ are quenched variables with zero mean, and $J_2$, $\tilde J_2$, $J_4>0$ (the first is associated to a ferromagnetic coupling and the two others to antiferromagnetic ones). The ellipses denote multi-body interactions beyond the $4$-body ones and higher-order random terms. The effective parameters of ${\mathcal H}_{\rm eff}$ are related to the original description of the glass-forming system; they may depend somewhat on temperature, but the whole procedure is applied only close to the putative thermodynamic glass transition (RFOT). In this mapping the thermodynamic glass transition of the liquid becomes the conventional {\it first-order} transition of the random-field Ising model. Such a transition may exist above some lower critical dimension, which is equal to $d=2$ in the presence of a random field,\cite{Nattermann} provided that the effective ferromagnetic coupling is sufficiently large compared to the strength of the disorder.

We stress again that the above mapping is valid for the thermal fluctuations of the overlap with a reference equilibrium configuration but does not imply a mapping of the dynamics of glass-forming liquids onto that of the effective random-field Ising theory. It is nonetheless useful to assess the presence or not of a thermodynamic glass transition in finite-dimensional glass-formers.

The rest of the paper is organized as follows. In Sec.~\ref{sec:GL} we introduce the description in terms of overlap (or similarity) between configurations in the case of glass-forming liquids. At this point the overlaps include those between a reference configuration and $n$ so-called constrained replicas of the liquid as well as all the overlaps among these constrained replicas. From the functional of the 1- and 2-particle densities introduced by Morita and Hiroike in the context of the liquid-state theory\cite{morita-hiroike} we derive an effective Hamiltonian that acts only on all the overlaps. Then in Sec.~\ref{sec:approximations} we describe in detail a variational low-temperature approximation scheme that allows us to integrate over the overlaps between constrained replicas in the presence of diverging spatial point-to-set correlations and to obtain an effective theory for the overlaps with the reference equilibrium configuration. In Sec.~\ref{sec:effective_theory},  we explicitly construct this effective theory and show that it has the form of a random-field + random-bond Ising model with multi-body interactions. This generalizes the results of paper~I\cite{paperI} obtained for mean-field glass models. The physical interpretation of this theory and of the various terms entering in the disordered Hamiltonian as well as a crude estimate of the existence of a thermodynamic glass transition in a specific liquid model via the mapping to the effective theory are presented in Sec.~\ref{sec:phys_interpretation}. Next, we come back in Sec.~\ref{sec:REMFC} to a glass model already studied in paper~I, a generalization of the Random Energy Model with a finite number of states: the $2^M$-KREM. We apply the variational low-temperature approximation scheme to this model, first to its fully connected version to check the quality of the approximations with respect to the exact results derived in paper~I and second to its finite-dimensional version to see if the resulting effective theory has the same structure as for glass-forming liquids (which it does). Finally we provide some concluding remarks and some perspectives in Sec.~\ref{sec:conclusions}. Many of the technical details of the calculations are presented in several appendices.

\section{Description of glass-forming liquids in terms of overlap fields} 
\label{sec:GL}

We aim at finding an effective theory for glass-forming liquids in finite dimensions which describes the statistics of the spatial fluctuations of the overlap with a reference equilibrium configuration. Before explaining in detail the (approximate) way to achieve this, it is worth presenting how overlap fields can be introduced in the context of liquids. Consider a liquid described by a pair potential  $v(r)$ between the $N$ constituent atoms (for the simplicity of the presentation all atoms are taken to be equal and the interactions are pairwise and isotropic but the generalization is straightforward). Its Hamiltonian reads
\begin{equation}
\begin{aligned}
 \label{eq_liquid_hamiltonian}
 H[\mathbf{r}^N]=&\frac{1}{2}\sum_{i,j\neq} v(\vert \mathbf{r}_i-\mathbf{r}_j\vert) = \frac 12\int_{\mathbf r} \int_{\mathbf r'}v(\vert \mathbf r-\mathbf{r'}\vert) \hat \rho(\mathbf r)\hat \rho(\mathbf{r'}),
\end{aligned}
\end{equation}
where $i,j=1,\cdots,N$ and $\hat \rho(\mathbf r)\equiv \hat \rho(\mathbf r\vert \mathbf{r}^N) =\sum_{i}\delta^{(d)}(\mathbf r-\mathbf r_i)$, with $d$ the spatial dimension (for practical studies of course, $d=2$ or $3$) .

A microscopic overlap between two liquid configurations $\mathbf{r}_{\alpha}^N$ and $\mathbf{r}_{\beta}^N$ at point $\mathbf r$ can be defined as
\begin{equation}
\hat q(\mathbf{r}\vert \mathbf{r}_\alpha^N,\mathbf{r}_\beta^N)=\int_{\mathbf y} f(y) \hat \rho(\mathbf r +\mathbf y\vert \mathbf{r}_\alpha^N)\hat \rho(\mathbf r-\mathbf y\vert \mathbf{r}_\beta^N)
\end{equation}
where $f(y)$ is a smooth function of range $a$ significantly less than the atomic diameter and accounting for the fact that the similarity between configurations is defined once the fast vibrational motions have been averaged out; $a$ is therefore a tolerance of the order of the typical vibrational amplitude.\cite{Franz-Jacquin} As in paper~I\cite{paperI} we consider a reference equilibrium configuration, and handling the averages over this  configuration leads to the introduction of $n+1$ copies or replicas of the system, labelled by a Greek letter $\alpha=0,\,1,\,\cdots,\, n$ with $n\to 0$; ``$0$'' refers to the reference configuration and Roman letters $a=1,\,\cdots,\, n$ are used for the other configurations, often referred to as ``constrained''. There are thus two types of overlaps: between the reference and the constrained replicas, and among the constrained replicas. The first step is to derive an effective Hamiltonian (a coarse-grained Ginzburg-Landau action) describing the system in terms of collective variables represented by {\it all} the overlaps, which by itself is not an easy task.

 One can formally define a Hamiltonian for local overlap variables $q_{\alpha\beta}(\mathbf r)$, with $\alpha,\beta=0,1,\cdots,n$ and $\alpha \neq \beta$, together with one-particle density fields $\rho_\alpha(\mathbf r)$, as follows: 
\begin{equation}
\label{eq_formal_action_overlaps}
\mathcal H[\{q_{\alpha\beta}\},\{\rho_\alpha\}]=-\frac 1{\beta} \ln \int \prod_{\alpha=0}^n d\mathbf r_\alpha^N \,\prod_{\alpha\beta\neq} \delta[q_{\alpha\beta}-\hat q[ \mathbf{r}_\alpha^N,\mathbf{r}_\beta^N]]\,\prod_{\alpha}\delta[\rho_\alpha-\hat \rho[\mathbf{r}_\alpha^N]]e^{-\beta \sum_\alpha H[\mathbf r_\alpha^N]}\, ,
\end{equation}
%\begin{equation}
%\label{eq_formal_action_overlaps}
%\mathcal H[\{q_{\alpha\beta}\}]=-T \ln \int \prod_{\alpha=0}^n d\mathbf r_\alpha^N \,\prod_{\alpha\beta\neq} \delta[q_{\alpha\beta}-\hat q_{\alpha\beta}(\mathbf{r}\vert \mathbf{r}_\alpha^N,\mathbf{r}_\beta^N)]e^{-(1/T)\sum_\alpha \mathcal H[\mathbf r_\alpha^N]}\, ,
%\end{equation}
where $\beta=1/T$ (we have set the Boltzmann constant $k_B$ to $1$) and $\delta[\cdots]$ is a functional that indicates a delta function in each point $\mathbf r$ of space.
 
After using the integral representation of the delta functions, which leads one to introduce the auxiliary fields $\lambda_{\alpha\beta}(\mathbf r)$ and $\mu_{\alpha}(\mathbf r)$, the above expression can be rewritten (up to an irrelevant constant) as
\begin{equation}
\begin{aligned}
 \label{eq_formal_action_overlaps2}
\mathcal H[\{q_{\alpha\beta}\},\{\rho_\alpha\}]= &
-\frac 1\beta \ln  \int \prod_{\alpha\beta \neq} \mathcal D\lambda_{\alpha\beta} \prod_{\alpha} \mathcal D\mu_{\alpha} \,e^{\frac{\beta}{2} \sum_{\alpha\beta\neq}\int_{\mathbf r}\lambda_{\alpha\beta}(\mathbf r)q_{\alpha\beta}(\mathbf r) -\beta\sum_\alpha \int_{\mathbf r}\mu_{\alpha}(\mathbf r)\rho_{\alpha}(\mathbf r)}  
\\&\times  \int \prod_{\alpha=0}^n d\mathbf r_\alpha^N \,e^{-\frac{\beta}{2} \sum_{\alpha\beta}\int_{\mathbf r} \int_{\mathbf r'}w_{\alpha\beta}(\mathbf r,\mathbf{r'}\vert \bm{\lambda}) \hat \rho(\mathbf r\vert \mathbf{r}_\alpha^N)\hat \rho(\mathbf{r'}\vert \mathbf{r}_\beta^N) +\beta \sum_\alpha \int_{\mathbf r}\mu_{\alpha}(\mathbf r)\hat \rho(\mathbf r\vert \mathbf{r}_\alpha^N) }
\end{aligned}
\end{equation}
where
\begin{equation}
 \label{eq_w}
w_{\alpha\beta}(\mathbf r,\mathbf{r'}\vert\bm{\lambda})=\delta_{\alpha\beta}v(\vert \mathbf r-\mathbf{r'}\vert)+(1-\delta_{\alpha\beta})f(\vert \mathbf r-\mathbf{r'}\vert) \lambda_{\alpha\beta} \Big( \frac{\mathbf r+\mathbf{r'}}{2} \Big)\,.
 \end{equation}
One recognizes in the functional integral in the second line of Eq.~(\ref{eq_formal_action_overlaps2}) the partition function $\mathcal Z[\{w_{\alpha\beta}[\bm{\lambda}]\},\{\mu_\alpha\}]$ of a liquid mixture of $(n+1)$ components with pair interaction potentials $w_{\alpha\beta}$ and chemical potentials $\mu_\alpha$.  Due to the dependence of the $w_{\alpha\beta}$'s  on the auxiliary  field $\lambda_{\alpha\beta}(\mathbf r)$ and the $\mathbf r$ dependence of the $\mu_\alpha$'s, the mixture is however inhomogeneous.
 
Proceeding further now requires approximations.  Keeping in mind that we want to derive a coarse-grained Ginzburg-Landau action for the overlaps, which in the spirit of the Renormalization Group amounts to integrating over short-ranged fluctuations, one evaluates the functional integral over the auxiliary fields in Eq.~(\ref{eq_formal_action_overlaps2}) through a saddle-point procedure. As a result,
\begin{equation}
\begin{aligned}
 \label{eq_formal_action_overlaps_saddle-point}
\mathcal H[\{q_{\alpha\beta}\},\{\rho_\alpha\}] \approx - \frac 1\beta\ln \mathcal Z[\{w_{\alpha\beta}[\bm{\lambda}^*]\},\{\mu_\alpha^*\}] -\frac{1}{2} \sum_{\alpha\beta\neq}\int_{\mathbf r}\lambda^*_{\alpha\beta}(\mathbf r)q_{\alpha\beta}(\mathbf r)+\sum_\alpha \int_{\mathbf r}\mu_{\alpha}^*(\mathbf r) \rho_{\alpha}(\mathbf r) \, ,
\end{aligned}
\end{equation}
where $\lambda_{\alpha\beta}^*(\mathbf r\vert \{q_{\gamma\delta}\})$ is a functional of the overlap fields which is obtained from the solution of the saddle-point equation 
\begin{equation}
\begin{aligned}
 \label{eq_saddle-point_lambda}
q_{\alpha\beta}(\mathbf{r})=\int_{\mathbf y} f(y)  \rho_{\alpha\beta}^{(2)} \Big( \mathbf r +\frac{\mathbf y}{2},\mathbf r-\frac{\mathbf y}{2} \Big \vert \{w_{\gamma\delta}(\bm{\lambda}^*[\mathbf q])\} \Big)
\end{aligned}
\end{equation}
with $\rho_{\alpha\beta}^{(2)}(\mathbf r,\mathbf r'\vert \bm{\lambda})$ the 2-point density-density correlation functions of the mixture with one-point density fields $\rho_\alpha(\mathbf r)$, and $\mu_\alpha^*(\mathbf r)$ is the chemical potential of the liquid mixture leading to these one-point densities. In this saddle-point approximation the Hamiltonian $\mathcal H[\{q_{\alpha\beta}\},\{\rho_\alpha\}]$ is thus related to the Morita-Hiroike functional $ \Gamma_{MH}$ of the 1- and 2-particle densities\cite{morita-hiroike} for the replicated $(n+1)$-component liquid mixture. More precisely it is given by 
\begin{equation}
\begin{aligned}
 \label{eq_formal_action_overlaps_saddle-point2}
\mathcal H[\{q_{\alpha\beta}\},\{\rho_\alpha\}] = \Gamma_{MH}[\{\rho_{\alpha\beta}^{(2)}(\mathbf r,\mathbf r'\vert \bm{\lambda}^*[\mathbf q])\},\{\rho_\alpha\}]-\frac{1}{2} \sum_{\alpha}\int_{\mathbf r} \int_{\mathbf r'}v(\vert \mathbf r-\mathbf r'\vert)\rho_{\alpha\alpha}^{(2)}(\mathbf r,\mathbf r'\vert \bm{\lambda}^*[\mathbf q]) \, ,
\end{aligned}
\end{equation}
After introducing the total pair correlation functions $h_{\alpha\beta}$ that are defined through $\rho_{\alpha\beta}^{(2)}(\mathbf r,\mathbf r'\vert \bm{\lambda})=\rho_{\alpha}(\mathbf r)\rho_{\beta}(\mathbf r')[1+h_{\alpha\beta}(\mathbf r,\mathbf r'\vert \bm{\lambda})]$, the Morita-Hiroike functional $ \Gamma_{MH}$ can be cast in the  form
\begin{equation}
\begin{aligned}
 \label{eq_2PI}
\Gamma_{MH}[\{\rho_{\alpha\beta}^{(2)}(\mathbf r,\mathbf r'\vert \bm{\lambda}^*)\},\{\rho_\alpha\}]= &\sum_\alpha \int_{\mathbf r}  \rho_{\alpha}(\mathbf r)[\ln( \rho_{\alpha}(\mathbf r))-1]
+\frac 1{2}\sum_{\alpha\beta} \int_{\mathbf r}\int_{\mathbf r'}\rho_{\alpha}(\mathbf r)\rho_{\beta}(\mathbf r')(1+h_{\alpha\beta}(\mathbf r,\mathbf r'\vert \bm{\lambda}^*))w_{\alpha\beta}(\mathbf r,\mathbf r'\vert \bm{\lambda}^*) 
\\&+ \frac 1{2}\sum_{\alpha\beta} \int_{\mathbf r}\int_{\mathbf r'}\rho_{\alpha}(\mathbf r)\rho_{\beta}(\mathbf r')\Big [(1+h_{\alpha\beta}(\mathbf r,\mathbf r'\vert \bm{\lambda}^*))\ln(1+h_{\alpha\beta}(\mathbf r,\mathbf r'\vert \bm{\lambda}^*))-h_{\alpha\beta}(\mathbf r,\mathbf r'\vert \bm{\lambda}^*)\Big ] \\&+
\frac 12\sum_{p\geq 3} \frac{(-1)^p}{p}\sum_{\alpha_1\cdots \alpha_p}\int_{\mathbf r_1}\rho_{\alpha_1}(\mathbf r_1)\cdots \int_{\mathbf r_p}\rho_{\alpha_p}(\mathbf r_p)h_{\alpha_1\alpha_2}(\mathbf r_1,\mathbf r_2\vert \bm{\lambda}^*)\cdots h_{\alpha_p\alpha_1}(\mathbf r_p,\mathbf r_1\vert \bm{\lambda}^*)
\\& + 2\rm{-PI}
\end{aligned}
\end{equation}
where $2\rm{-PI}$ denotes the sum of all 2-particle irreducible diagrams.\cite{morita-hiroike} Neglecting the latter leads to the  HNC approximation, well known in liquid-state theory.\cite{hansen-mcdo} Within the HNC approximation, the saddle-point equations read
\begin{equation}
\begin{aligned}
 \label{eq_SP_HNC}
&h_{\alpha\beta}(\mathbf r,\mathbf r'\vert \bm{\lambda}^*)=c_{\alpha\beta}(\mathbf r,\mathbf r'\vert \bm{\lambda}^*)+\sum_\gamma \int_{\mathbf r''}\rho_\gamma(\mathbf r'')c_{\alpha\gamma}(\mathbf r,\mathbf r''\vert \bm{\lambda}^*)h_{\gamma\beta}(\mathbf r'',\mathbf r'\vert \bm{\lambda}^*)
\\&c_{\alpha\beta}(\mathbf r,\mathbf r'\vert \bm{\lambda}^*)=-\ln[1+ h_{\alpha\beta}(\mathbf r,\mathbf r'\vert \bm{\lambda}^*)]+h_{\alpha\beta}(\mathbf r,\mathbf r'\vert \bm{\lambda}^*)-\frac 1T w_{\alpha\beta}(\mathbf r,\mathbf r'\vert \bm{\lambda}^*)\,,
\end{aligned}
\end{equation}
and are also valid for $\alpha=\beta$. (The chemical potentials $\mu_a^*$ are obtained from the minimization of $\Gamma_{MH}$ with respect to $\rho_a$.) The combination of Eq.~(\ref{eq_saddle-point_lambda}) and Eq.~(\ref{eq_SP_HNC}) allows one to obtain $\bm{\lambda}^*[\mathbf q]$ and finally $\mathcal H[\{q_{\alpha\beta}\},\{\rho_\alpha\}]$.

It should be stressed that $\mathcal H[\{q_{\alpha\beta}\},\{\rho_\alpha\}]$ is {\it not} the full free-energy (or effective action in the language of quantum field theory) of the system. The fluctuations of the variables $q_{\alpha\beta}(\mathbf r)$ still have to be integrated over in order to describe the thermodynamical properties of the system. The saddle-point approximation used above corresponds to a coarse-graining of the overlap fields, and the $q_{\alpha\beta}(\mathbf r)$'s are then taken as locally averaged over a small region of space around point $\mathbf r$.

One can further simplify the Hamiltonian for the $q_{\alpha\beta}$'s in the spirit of a Ginzburg-Landau theory by performing a gradient expansion around a purely local contribution, 
\begin{equation}
\beta \mathcal H[\{q_{\alpha\beta}\}]= \int_{\mathbf r}  \bigg \{U(\{q_{\gamma\delta}(\mathbf r)\}) + (1/2)\sum_{\alpha\beta \neq}Z_{\alpha\beta}(\{q_{\gamma\delta}(\mathbf r)\})  [\partial q_{\alpha\beta}(\mathbf r)]^2 +  {\rm O}(\partial^4)\bigg \}\,, 
\end{equation}
where the $\rho_\alpha$'s are taken as uniform and not explicitly displayed. The potential $U(\{q_{\gamma\delta}\})$ and the functions $Z_{\alpha\beta}(\{q_{\gamma\delta}\})$ are obtained from  the HNC approximation and therefore retain much of the physics of the liquid. An additional and more drastic approximation is to expand these functions in powers of the overlaps, keeping only the first terms.\cite{Dzero,Franz-Jacquin} This leads to the form proposed in Ref.~[\onlinecite{Dzero}]:
\begin{equation}
\begin{aligned}
 \label{eq_replica_action_wolynes}
\beta\mathcal H[\{q_{\alpha\beta}\}]=& \int_{\mathbf r} \bigg \{\frac{c}{2} \sum_{\alpha\beta \neq} [\partial q_{\alpha\beta}(\mathbf r)]^2
 + \sum_{\alpha\beta \neq} V(q_{\alpha\beta}(\mathbf r))  
 -\frac{u}{3} \sum_{\alpha\beta\gamma \neq}
q_{\alpha\beta}(\mathbf r)q_{\beta\gamma}(\mathbf r)q_{\gamma\alpha}(\mathbf r)\bigg \}\, ,
\end{aligned}
\end{equation}
where $V(q)=(t/2)q^2-[(u+w)/3]q^3+(y/4)q^4$ with $u,w,y >0$ and the primary dependence on temperature is given by $t\approx (T-T_0)/T_0$, with $k_BT_0$ a typical energy scale of the liquid. The mean-field analysis of the above Hamiltonian leads to the well-established mean-field scenario of glass formation with two distinct glass-transition temperatures, a dynamical one at $t_d=w^2/(4y)$ [hence, $T_d=T_0(1+t_d)$] and a thermodynamic one (the RFOT) at $t_K=2w^2/(9y)$ [hence, $T_K=T_0(1+t_K)$].\cite{Dzero} The above simplified expression of the effective Hamiltonian for all the overlap fields will therefore be sufficient to derive the form of the sought-after effective theory and we will consider it as our starting point. However, one should keep in mind that Eq.~(\ref{eq_replica_action_wolynes}) is somehow too crude to provide reliable estimates for the parameters entering the final effective theory, especially at low temperature near the putative RFOT. An alternative way to extract these parameters would be from the full HNC computation or from computer simulations of finite-size glass-forming liquid models (see below).

Even in the form described by Eq.~(\ref{eq_replica_action_wolynes}), the problem remains challenging and, despite some progress already mentioned in paper~I ({\it e.g.}, instanton calculations\cite{Dzero,Franz_instantons}, Kac model analysis\cite{Franz_Kac}, and real-space RG approaches\cite{mkrg,Castellana,Angelini}), no satisfactory way of handling the associated large-scale physics has been provided so far. This comes in part from the complicated  replica matrix structure of the overlap fields $q_{\alpha\beta}$ when $n\to 0$. Our present strategy is to focus on the overlaps $p_a\equiv q_{0a}$ between the reference and the constrained replicas and integrate out the other overlap fields, $q_{ab}$, to obtain an effective theory for the former. Formally, this gives
\begin{equation}
\begin{aligned}
 \label{eq_action_formal}
\mathcal S[\{p_{a}\}]=-\ln \int \prod_{ab \neq} \mathcal D q_{ab} \,e^{\beta \mathcal H[\{p_a\},\{q_{ab}\}]}\, ,
\end{aligned}
\end{equation}
with $\mathcal H$ given by Eq. (\ref{eq_formal_action_overlaps_saddle-point2}) or  (\ref {eq_replica_action_wolynes}). This elimination of the $q_{ab}$'s is what we have previously achieved when studying the situation where an external source $\epsilon$ is linearly coupled to the $p_a$'s. In the extended temperature/source phase diagram there may be a line of first-order transition terminating in a critical point. Near the latter, only the overlaps $p_a$ are critical and the $q_{ab}$'s, which stay ``massive'', can be  integrated out in a cavalier way without altering the critical physics.\cite{noi_Tc} Here, instead, we are interested in the problem of describing the ideal (RFOT) glass transition ($\epsilon=0$, $T \to T_K$). A more careful treatment must be applied.  One indeed expects, beyond the mean-field treatment, the divergence of point-to-set correlation length(s) near the putative $T_K$.\cite{BB-PTS,Franz_instantons}  This implies that fixing the overlap with the reference configuration in localized regions of space may generate very strong correlations among the constrained replicas, making the integration over the $q_{ab}$'s much more delicate.

\section{Low-temperature approximations} 
\label{sec:approximations}

As recalled above,  the construction of an effective theory for the statistics of the fluctuations of the overlap field with an equilibrium configuration in the vicinity of the mean-field thermodynamic glass transition (RFOT) is a difficult process in finite-dimensional systems, contrary to the case of the mean-field glasses studied in paper~I\cite{paperI} where an essentially exact derivation could be performed. The key point is to account for the possibly infinite point-to-set correlation length(s). As discussed in Sec. IV of paper~I this is already contained in a saddle-point integration of the $q_{ab}$'s, provided one could solve the problem for all fixed configurations of the $p_a$'s. Here we choose a slightly different route to approximately determine $\mathcal S[\{p_{a}\}]$ from $\mathcal H[\{p_a\},\{q_{ab}\}]$. This is what we detail below.

\subsection{From continuous to two-state variables}
\label{sec:2-state_variables}

As fluctuations are expected to lower the RFOT temperature below its mean-field value $T_K^{MF}$ (or even destroy it), we focus on a low-temperature version of the problem for $T \lesssim T_K^{MF}$. We therefore consider an action of the form of Eq.~(\ref{eq_replica_action_wolynes}) with a strong first-order character, {\it i.e.}, such that the low-overlap minimum (at $q=0$) and the high-overlap minimum (at $q=q_\star$) of the potential $U(q)=V(q)+(u/3)q^3=(t/2)q^2-(w/3)q^3+(y/4)q^4$ are well-defined with $q_\star$ and the peak difference $s_c=U(q_\star)-U(0)$ being of $O(1)$. (Here, the notation $s_c$, although the same as that used for the configurational entropy, actually just denotes a source/field favoring the low-overlap state and, as such, it can become negative.)
In the following we shall call it ``bare configurational entropy'' to distinguish it from the physical configurational 
entropy that includes fluctuations of the overlap field.  
Furthermore, to simplify the problem and get rid of some of the short-range fluctuations, we consider the limit where the overlaps become ``hard'' $2$-state variables, in the sense that the $q_{\alpha\beta}$'s can only take two values, $0$ or $q_\star$, the Boltzmann weight for any other value being negligible.\cite{footnote_2-state}

After redefining the variables as $p_a=q_\star \tau_a$ and $q_{ab}=q_\star \tau_{ab}$ where the $\tau_a$'s and the $\tau_{ab}$'s take the values $0$ and $1$, incorporating the factors of $q_\star$ in a redefinition of the various parameters ($c \to q_\star^2 c$, $u \to q_\star^3 u$), and replacing the continuum space description by a lattice one which is appropriate for hard variables, we obtain a new replicated action,
\begin{equation}
\begin{aligned}
 \label{eq:hamiltonian_hard}
\mathcal S_{\rm rep} [\{\tau_a\},\{\tau_{ab}\}] &= \sum_{a=1}^n 
{\cal S}_0 [\tau_a] + 
{\cal S}_{\rm int} [\{\tau_a\},\{\tau_{ab}\}] \, ,
\end{aligned}
\end{equation}
with
\begin{equation}
\label{eq:S0Sint}
\begin{aligned}
{\cal S}_0 [\tau_a] & = 2  
\Big [\frac{c}{2}\sum_{ \langle i ,j \rangle}(\tau_a^i-\tau_a^j)^2 + \sum_i \Big( s_c-\frac{u}{3} \Big) \tau_a^i \Big ] \, , 
\\
{\cal S}_{\rm int}
[\{\tau_a\},\{\tau_{ab}\}] & = \sum_{ab\neq} 
\Big [\frac{c}{2}\sum_{\langle i, j \rangle}(\tau_{ab}^i-\tau_{ab}^j)^2 + 
\sum_i \Big( s_c-\frac{u }{3} -u\, \tau_a^i\tau_b^i \Big)\tau_{ab}^i\Big ]  -\frac{u}{3} \sum_{abc\neq}\sum_i
\tau_{ab}^i\tau_{bc}^i\tau_{ca}^i \, ,
\end{aligned}
\end{equation}
where $i,j=1, \cdots,N$ denote the lattice sites and the energy scale $T_0$ has been set to $1$. Note that the restriction to only hard variables is exact for some specific models for structural glasses, and in particular for the REM\cite{derrida} and the $2^M$-KREM\cite{REM}  studied in paper~I\cite{paperI}.

To derive an effective theory for the overlaps with a reference configuration, now the $\tau_a^i$'s, one needs to perform the integration over the $\tau_{ab}^i$'s while keeping the former fixed and compute the ``partition function''
\begin{equation} \label{eq:Ztaua}
\mathcal Z[\{\tau_a\}]= \sum_{\{\tau_{ab}^i=0,1\}} \exp(-\mathcal S_{\rm rep}[\{\tau_{ab}\}\vert \{\tau_a\}])\,. 
\end{equation}
The action $\mathcal S_{\rm rep}[\{\tau_a\}]= \ln \mathcal Z[\{\tau_a\}]$ [see also Eq. (\ref{eq_action_formal})], when expanded in an increasing number of unrestricted sums over replicas\cite{LeDoussal,Tarjus-Tissier1,Tarjus-Tissier2} (see paper~I),
\begin{equation}
\begin{aligned}
\label{eq:expansion_freereplica-sums}
\mathcal S_{\rm rep}[\{\tau_a\}]=\sum_{a=1}^n \mathcal S_1 [\tau_a]-
\frac 12 \sum_{a,b=1}^n \mathcal S_{2}[\tau_a,\tau_b]+\cdots  \,,
\end{aligned}
\end{equation}
generates, under some general conditions, the cumulants of an effective Hamiltonian for the overlap with the reference configuration, where this configuration  plays the role of a quenched disorder:\cite{noi_Tc} $\mathcal S_{1}$, $\mathcal S_{2}$, etc., are then the first, second, etc., cumulants of the effective  Hamiltonian averaged over this quenched disorder.

\subsection{Physical constraints on the $\tau_{ab}$'s}
\label{sec:constraints}

 Computing the above constrained partition function is still a very hard task in general. On physical ground, one expects that if on a given site $i$ two replicas $a$ and $b$ have a high overlap with the reference one, they should also have a high overlap between them ($\tau_a^i=\tau_b^i=1 \Rightarrow \tau_{ab}^i=1$) whereas if one replica has a high overlap with the reference configuration and the other one has a low one, then the overlap between them should be low ($\tau_a^i=1\,,\; \tau_b^i=0 \Rightarrow \tau_{ab}^i=0$, and similarly for $\tau_a^i=0$, $\tau_b^i=1$). One may also naively expect that if $a$ and $b$ have a low overlap with the reference one, since at a coarse-grained level described by a saddle point of Eq.~(\ref{eq_replica_action_wolynes}) there are exponentially many (in the coarse-graining volume) possible distinct states, the chances that $a$ and $b$ are in the same state is negligible and their mutual overlap should also be low: $\tau_a^i=\tau_b^i=0 \Rightarrow \tau_{ab}^i=0$. However, enforcing such a strict constraint misses the effect of spatial correlations, more specifically point-to-set-like correlations associated with ``amorphous order'', which become important close to $T_K$.
  
 All of this suggests that we should use a parametrization of the $\tau_{ab}$'s that takes into account all previous  constraints but the last one.  We, thus,  introduce the following parametrization: 
\begin{equation} 
\label{eq:parametrization}
\tau_{ab}^i=\tau_a^i \tau_b^i +\eta_{ab}^i(1-\tau_a^i)(1- \tau_b^i) \qquad \qquad \textrm{for~} a \neq b \, ,
\end{equation} 
with the new variables $\eta_{ab}^i$ taking also the values $0$ and $1$. [In the following we will refer to the approximation consisting in setting $\tau_{ab}^i=\tau_a^i \tau_b^i$ ({\it i.e.}, $\eta_{ab}^i = 0$) as the ``zeroth order'' approximation.] The replicated action can now be obtained from
\begin{equation}
\begin{aligned}
 \label{eq:Z_approx}
e^{- {\cal S}_{\rm rep} [ \{ \tau_ a\} ]}
\simeq &  \prod_i \prod_{ab\neq}  
\Big [\frac12(\delta_{\tau_a^i ,1}+\delta_{\tau_b^i ,1}-\delta_{\tau_a^i ,1}\delta_{\tau_b^i ,1}) +
\delta_{\tau_a^i ,0}\delta_{\tau_b^i ,0}\Big ] 
\sum_{\{\eta_{ab}^i=0,1\}} e^{-\mathcal S_{\rm rep} [\{\eta_{ab}\}\vert \{\tau_a\}]} \,.
\end{aligned}
\end{equation}
where $\delta_{s,t}$ is a Kronecker symbol. The term inside the square brackets arises when one introduces the reparametrization of the $\tau_{ab}$'s variables in terms of the $\eta_{ab}$'s (it equals one if $\tau_a = \tau_b = 0$ and $1/2$ otherwise, which then compensates the factor of $2$ coming from the sum over the $\eta_{ab}$'s), and $\mathcal S[\{\eta_{ab}\}\vert \{\tau_a\}]$ is obtained from Eq.~(\ref{eq:hamiltonian_hard}) by the change of variables in Eq.~(\ref{eq:parametrization}). Note again that this approximation is exact for some specific models for structural glasses, and in particular for the REM and the $2^M$-KREM studied in paper~I.

The problem of computing ${\cal S}_{\rm rep} [ \{ \tau_ a\} ]$ still remains hardly tractable. There are two reasons for that:
The first is that one should be able to %perform the trace over the $\eta_{ab}^i$'s scanning 
scan over all the possible configurations of the constrained overlaps $\tau_a^i$'s, which is of course an impossible task.
The second is that, even for a given specific configuration of the $\tau_a^i$'s, performing the trace over the $\eta_{ab}^i$'s would require to be able to solve a generic interacting (and strongly inhomogeneous) finite-dimensional model. This leads us to the last
two approximations.

\subsection {Variational ($1$-RSB) approximation}
\label{sec:variational}

We replace the trace over the $\eta_{ab}^i$'s by a variational approximation where we keep only those configurations $[\eta_{ab}^{i}]_\star$ that minimize ${\cal S}_{\rm rep}[\{\eta_{ab}\}\vert \{\tau_a\}]$. Since the system is considered below the mean-field RFOT $T_K^{MF}$ we search for a variational approximation in the form of a one-step replica-symmetry breaking ($1$-RSB) solution.\cite{1RSB} (Note that we are not interested in this work by the possible appearance at a still lower temperature of a so-called ``Gardner transition'' from a stable glass to a marginally stable glass with full replica-symmetry breaking.\cite{kurchan-zanponi-etal})
In other words, for any given inhomogeneous configurations $\{\tau_a^i\}$ of the overlap between the reference and the constrained replicas, we evaluate the trace over the overlap among the constrained replicas $\{\eta_{ab}^i\}$, Eq.~(\ref{eq:Z_approx}), at the level of a saddle-point approximation 
(of a 1-RSB form), as discussed in Sec.~IV of paper~I for the Kac limit.

For clarity, let us first focus on the $1$-replica action $\mathcal S_1 [\tau]$ [see Eq. (\ref{eq:expansion_freereplica-sums})]. It is then sufficient to set all replica fields equal, $\tau_a^i =\tau^i$ $\forall \, a=1, \cdots,n$, keep only the term of order $n$ in the logarithm of the partition function, and take the limit $n\rightarrow 0$ in the end, as in the standard replica trick. After inserting Eq.~(\ref{eq:parametrization}) into Eqs.~(\ref{eq:hamiltonian_hard}) and (\ref{eq:S0Sint}), we obtain at leading order in $n$
\begin{equation} 
\label{eq:Seta}
\begin{aligned}
{\mathcal S}_{\rm rep} [\{ \eta_{ab} \} \vert \tau] &= n \Big [\frac{c}{2}\sum_{\langle i, j \rangle}(\tau^i-\tau^j)^2 + s_c \sum_i  \tau^i \Big ]
+ \sum_{ab\neq} \Big [ \frac{c}{2}\sum_{\langle i, j \rangle}[(1-\tau^i)\eta_{ab}^i-(1-\tau^j) 
\eta_{ab}^j]^2 \\& +  
\sum_i (1-\tau^i)\Big(-c \sum_{j/i} \tau^j +s_c-\frac{u}{3} \Big)\eta_{ab}^i \Big ]
 -\frac{u}{3} \sum_{abc\neq}\sum_i (1-\tau^i) \eta_{ab}^i\eta_{bc}^i\eta_{ca}^i \, .
\end{aligned}
\end{equation}
where $\sum_{j/i}$ denotes the sum over the nearest neighbors of site $i$ on the lattice and $(1-\tau^i)$ plays the role of a dilution variable.  Note that on all the sites where $\tau^i = 1$, $\eta_{ab}^i$ is not constrained [see Eq.~(\ref{eq:parametrization})] and can be $0$ or $1$ with equal probability.  The  sum over all configurations of the $\eta_{ab}^i$ on those sites can be straightforwardly performed. This yields a term of the form $n(n-1) \ln 2 \sum_i \tau^i$ in the exponential which exactly cancels the entropic factor of Eq.~(\ref{eq:Z_approx}). We have yet to account for the sum over the $\eta_{ab}^i$'s for all the sites where $\tau^i=0$.

We will then approximate the $1$-replica action by
\begin{equation}
\label{eq:first_cum}
\mathcal S_{1} [\tau] \simeq  \mathcal S_{\rm rep} [\{[\eta_{ab}]_\star\}\vert \tau] \,.
\end{equation}
where $[\eta_{ab}^{i}]_\star$ minimizes Eq. (\ref{eq:Seta}) for all the sites $i$ where $\tau^i=0$. (The sum over the configurations $\eta_{ab}^i$ on sites where $\tau^i=1$ has already been accounted for in Eq.~(\ref{eq:first_cum}) as explained above.)

The higher-order terms of the expansion in number of free replica sums can be treated along the same lines. For instance, for the $2$-replica term, one needs to introduce two groups of $n_1$ and $n_2$ replicas (with $n_1 + n_2 = n$), such that $\tau_a^i = \tau_1^i$ for $a=1,\cdots,n_1$ and $\tau_a^i = \tau_2^i$ for $a=n_1+1,\ldots,n$, keep only the terms of order $n_1 n_2$ in the expression of the effective action, and take the limit $n_1,n_2\rightarrow 0$ in the end. By restricting the sum over the $\tau_{ab}^i$'s according to Eq.~(\ref{eq:parametrization}) we find that the matrix $\eta_{ab}^i$ now has 4 different blocks which are defined as
\begin{equation} \label{eq:2replicamatrix}
\eta_{ab}^i = 
\begin{bmatrix}
\eta_{a_1 b_1}^i & \eta_{a_1 b_2}^i \\
\eta_{a_2 b_1}^i & \eta_{a_2 b_2}^i
\end{bmatrix}
\qquad \qquad \textrm{with~}
\left \{
\begin{array}{ll}
\tau_{a_1 b_1}^i & = \tau_1^i + \eta_{a_1 b_1}^i (1 - \tau_1^i)\\
\tau_{a_2 b_2}^i & = \tau_2^i + \eta_{a_2 b_2}^i (1 - \tau_2^i)\\
\tau_{a_1 b_2}^i & = \tau_1^i \tau_2^i + \eta_{a_1 b_2}^i (1 - \tau_1^i) (1 - \tau_2^i)
\end{array}
\right .
\end{equation}
where the indices $a_1,b_1,\ldots$ vary from $1$ to $n_1$ and the indices $a_2,b_2,\ldots$ vary from $1$ to $n_2$. This form is inserted in ${\mathcal S}_{\rm rep} [\{ \eta_{ab} \} \vert \tau_1,\tau_2]$ and we again search for a variational 1-RSB variational approximation for the matrix $\eta_{ab}^i$ that  minimizes the appropriate component of ${\mathcal S}_{\rm rep} [\{ \eta_{ab} \} \vert \tau_1,\tau_2]$.

The problem however still remains too difficult to handle as the solution $[\eta_{ab}^{i}]_\star$ of the minimization of Eq. (\ref{eq:Seta}) and its generalizations for higher-order cumulants are strongly nonuniform in the presence of arbitrary profiles of the $\tau_a^i$'s. This leads us to the last approximation.

\begin{figure}
\includegraphics[scale=0.36]{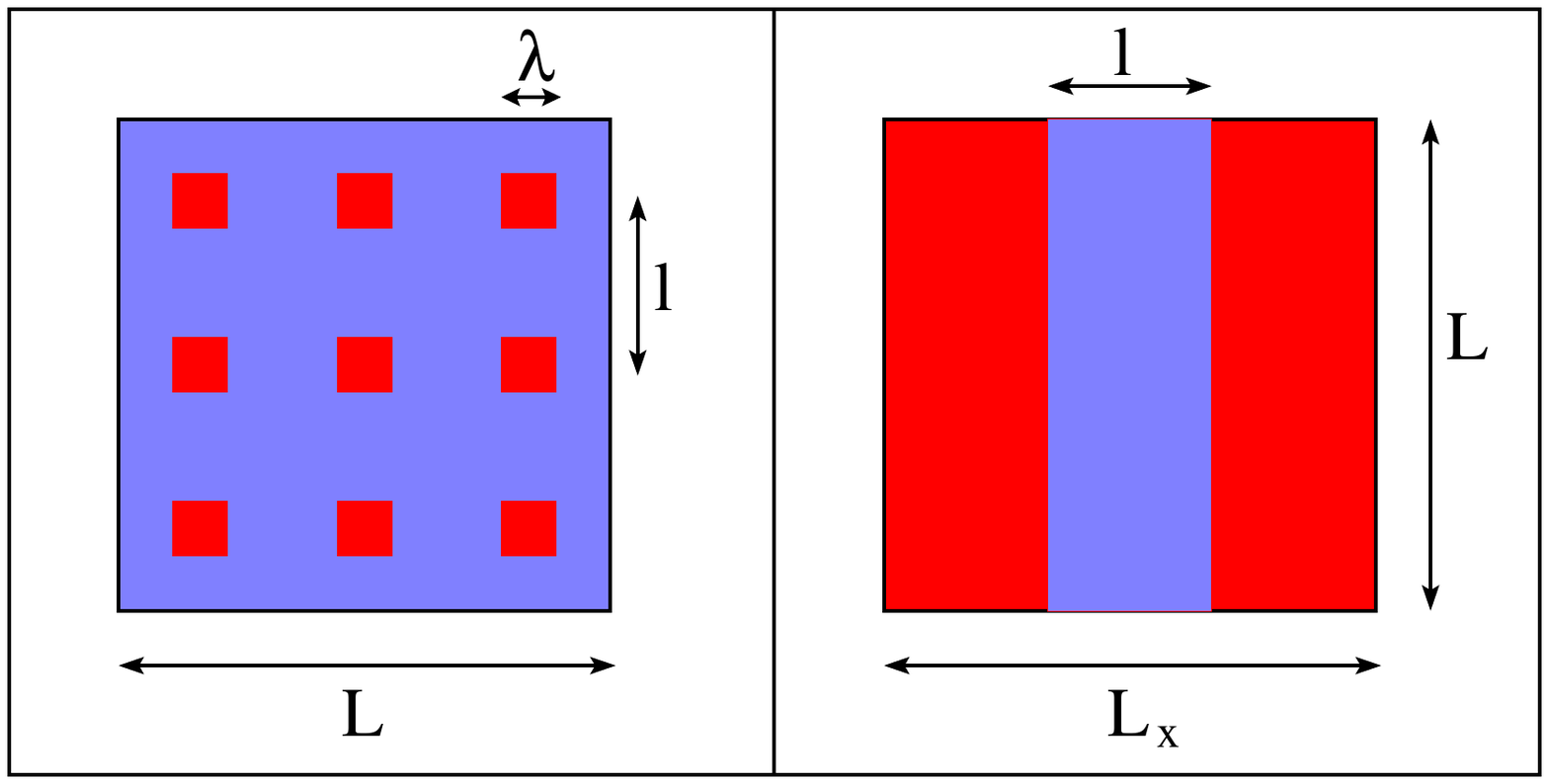}
\caption{$2$-$d$ sketches of the specific pattern of the $\tau^i$'s chosen for the implementation of the variational approximation:
Periodic cluster arrangement described in Sec.~\ref{sec:approximations} (left panel) and slab geometry treated in Appendix~\ref{app:slab} 
(right panel). $\tau^i=1$ in the red regions and $0$ in the light blue regions.}
\label{fig:sketch}
\end{figure}

\subsection {$1$-replica action: Periodic cluster ansatz}
\label{sec:periodic_cluster}

Let us start again with the $1$-replica action $\mathcal S_{1}$. Since finding the variational approximation for an arbitrary overlap profile $\{\tau^i\}$ with the reference configuration is an impossible task, we restrict the analysis to specific patterns of the $\tau^i$'s with enough symmetry for allowing a determination of the corresponding $[\eta_{ab}^{i}]_\star$'s, and we reproduce the result so obtained by means of an effective external field and effective $2$- and multi-body interactions in a {\it translationally invariant theory}.  Doing this we nonetheless account for the effect of the correlations induced by the underlying ``amorphous order'' characterized by diverging point-to-set correlation lengths.

The choice of the specific overlap patterns is of course arbitrary. Our idea is to focus on those patterns which mimic the typical configurations of the 
overlap field with a reference equilibrium configuration of a glass-forming liquid close to its putative thermodynamic glass transition. On general grounds one intuitively expects that these configuration are highly inhomogeneous and are characterized by well defined regions of high- and low-overlap separated by rather sharp interfaces (see for instance Ref.~[\onlinecite{phase_separation}]). We then consider a periodic arrangement of the $\tau^i$'s in which cubes of side $\lambda$ are regularly placed on the lattice with a distance $\ell$ between the centers of two neighboring cubes, with $\ell \gg \lambda$. On the sites of the cubes, $\tau^i=1$, whereas on all sites outside the cubes, $\tau^i=0$ (see the left panel of fig.~\ref{fig:sketch} for a sketch). On a $d$-dimensional hyper-cube of linear size $L$, the total number of sites is ${\cal N} = L^d$, the total number of cubes is ${\cal N}_c = (L/\ell)^d$, the total number of sites with $\tau^i=1$ is $\lambda^d {\cal N}_c = (L \lambda/\ell)^d$, and the total number of links between sites where $\tau^i =1$ and sites where $\tau^i = 0$ is ${\cal N}_L = 2 d \lambda^{d-1} {\cal N}_c$. To check the robustness of our procedure we  have also considered a slab geometry, as shown in the right panel of Fig.~\ref{fig:sketch}, where $\tau^i=1$ except in a slab of width $\ell$: details are given in Appendix~\ref{app:slab}.

To proceed further we compute ${\mathcal S}_{\rm rep} [\{ \eta_{ab} \} \vert \tau]$ for the above ``periodic cluster'' geometry and find the configurations $[\eta_{ab}^{i}(\lambda,\ell)]_\star$ that minimize the expression. On the sites where $\tau^i = 0$ we set $\eta_{ab}^i$ to be of a $1$-RSB form, {\it i.e.}, we divide the $n$ replicas in $n/m$ blocks of size $m \times m$ such that if two replicas $a$ and $b$ are in the same block, $\eta_{ab}^i = 1$, whereas if $a$ and $b$ do not belong to the same block, $\eta_{ab}^i = 0$. Note that $m$ is a variational parameter which must fixed by minimizing the action for a given choice of the pattern of the overlap with the reference configuration ($m=1$ corresponds to the ``zeroth-order'' approximation, $\tau_{ab}^i = \tau_a^i \tau_b^i$). One could in principle let $m$ be site dependent. However, we have checked that this yields a negligible correction, 
and from now on we will only consider the case of a uniform $m$.

The gradient term in Eq.~(\ref{eq:Seta}) is different from zero only on the ${\cal N}_L$ links connecting a site where $\tau^i = 1$ and a site where $\tau^i = 0$ ({\it i.e.}, only on the sites placed at the surface of the cubes):
\begin{displaymath}
\sum_{ab\neq} \frac{c}{2}\sum_{\langle i, j \rangle}[(1-\tau^i)\eta_{ab}^i-(1-\tau^j) 
\eta_{ab}^j]^2 = \frac{c}{2} n (m-1) {\cal N}_L 
\, .
\end{displaymath}
For the other terms of Eq.~(\ref{eq:Seta}) one finds
\begin{displaymath}
\begin{aligned}
\sum_{ab\neq} \sum_i (1-\tau^i)\Big(
-c \sum_{j/i} \tau^j + 
s_c-\frac{u}{3} \Big)\eta_{ab}^i &= n(m-1) \Big[ -c \, {\cal N}_L + \Big(s_c - \frac{u}{3} \Big) \big (
{\cal N} - \lambda^d {\cal N}_c \big) \Big]  \, , \\
-\frac{u}{3} \sum_{abc\neq}\sum_i (1-\tau^i) \eta_{ab}^i\eta_{bc}^i\eta_{ca}^i
& = - n (m-1)(m-2) \frac{u}{3}  \big( {\cal N} - \lambda^d {\cal N}_c \big)  \, .
\end{aligned}
\end{displaymath}
After putting all these terms together, minimizing with respect to the variational parameter $m$ yields
\begin{displaymath}
m_\star = \left \{
\begin{array}{ll}
1 & {\rm ~~for~}
\ell \ge \ell_{\rm pin} \, ,\\
1 - \frac{3}{2u} \left( \frac{c d \lambda^{d-1}}{\ell^d - \lambda^d} - s_c \right) & {\rm ~~for~}
\ell < \ell_{\rm pin} \, ,
\end{array}
\right .
\qquad {\rm with~~~} 
\ell_{\rm pin} = \left\{ 
\begin{array}{ll}
\lambda \left( 1 + \frac{cd}{\lambda s_c}\right)^{\frac{1}{d}} & {\rm ~~for~} s_c \ge 0 \, ,\\
\infty & {\rm ~~for~} s_c < 0 \, .
\end{array}
\right .
\end{displaymath}
Here $\ell_{\rm pin}$ plays the role of the ``pinning'' correlation length which is a point-to-set correlation length. It diverges at the mean-field RFOT, $T_K^{MF}$, and stays infinite in the (mean-field) ideal glass phase below. [Since $m$ cannot become negative, one obtains an upper bound on $c$ ensuring that $m_\star (\ell = \lambda+1) > 0$, above which our approximations are ill-defined. The most stringent constraint is realized for $\lambda=1$, which gives $c \le (2 u/3 + s_c)(2^d-1)/d$.]

By inserting the above results into Eq.~(\ref{eq:Seta}) we find that the $1$-replica component of the effective Hamiltonian reads
\begin{equation} 
\label{eq:DS1}
{\cal S}_1 (\lambda,\ell) 
= \frac{c}{2}\sum_{\langle i, j \rangle}(\tau^i-\tau^j)^2 + s_c  \sum_i \tau^i 
+ \Delta {\cal S}_1 \, , {\rm ~~~~~with~~~~}
\Delta {\cal S}_1 = \left \{ 
\begin{array}{ll} 
0 & {\rm ~~for~} \ell \ge \ell_{\rm pin} \, ,\\
\frac{3 {\cal N}_c}{4 u} \, \frac{[s_c (\ell^d - \lambda^d) - c d \lambda^{d-1}]^2}{(\ell^d - \lambda^d)}
& {\rm ~~for~} 0 < \ell < \ell_{\rm pin} \, .
\end{array}
\right .
\end{equation}
The first part of the expression coincides with the ``zeroth-order'' result, {\it i.e.}, what one would obtain by assuming that $\tau_{ab}^i = \tau_a^i \tau_b^i$ on all sites, whereas $\Delta {\cal S}_1$ can be interpreted as the excess contribution due to the fluctuations of the $\tau_{ab}$'s beyond the mean-field level. Since we are interested by the liquid significantly below $T_K^{MF}$, we will mainly focus on the case where $s_c <0$ and $\ell_{\rm pin} = \infty$.

Guided by the mean-field results of paper~I, we now search for an ansatz for the $1$-replica action that allows us to reproduce the above result by means of a translationally invariant theory with an effective external field and effective $2$- and multi-body interactions. Below, we show that the appropriate form of the $1$-replica component of the effective Hamiltonian is
\begin{equation} 
\label{eq:DS1eff}
{\cal S}_{1}[\tau] 
= \frac{c}{2}\sum_{\langle i, j \rangle}(\tau^i-\tau^j)^2 + s_c  \sum_i \tau^i 
+ \Delta {\cal S}_{1,\rm eff}[\tau] 
\end{equation}
with
\begin{equation} 
\label{eq:Heff-1}
\Delta{\cal S}_{1,\rm eff}[\tau] \approx \mu \sum_i (1 - \tau^i) + w_2 \sum_{\langle i, j \rangle}
\tau^i (1 - \tau^j) + \frac{1}{2} \sum_{i \neq j} W(\vert  r_i-r_j\vert ) \tau^i 
\tau^j + \frac{w_4}{L^d} \!\!\! \sum_{\langle i, i^\prime \rangle \neq \langle j j^\prime \rangle}
\!\!\! [\tau^i (1 - \tau^{i^\prime}) \tau^j (1 - \tau^{j^\prime})]_{\rm sym}  
\, ,
\end{equation}
where we have used the following notation:
\begin{displaymath}
[\tau^i (1 - \tau^{i^\prime}) \tau^j (1 - \tau^{j^\prime})]_{\rm sym} \equiv \frac{1}{4}
\big[ \tau^i (1 - \tau^{i^\prime}) \tau^j (1 - \tau^{j^\prime}) + \tau^{i^\prime} (1 - \tau^i) \tau^j (1 - \tau^{j^\prime})
+ \tau^i (1 - \tau^{i^\prime}) \tau^{j^\prime} (1 - \tau^j) + \tau^{i^\prime} (1 - \tau^i) \tau^{j^\prime} (1 - \tau^j) \big] \, .
\end{displaymath}
We determine the expressions of $\mu$, $w_2$, $w_4$, and the functional form of $W(x)$ by requiring that, for the periodic-cluster pattern chosen above, $\Delta {\cal S}_{1,\rm eff}[\tau]$ reproduces the dependence of $\Delta {\cal S}_1$ on $\ell$ and $\lambda$.

The $4$-body interaction term in Eq.~(\ref{eq:Heff-1}) corresponds to an infinite-range coupling between two {\it links} $\langle i, i^\prime \rangle$ and $\langle j, j^\prime \rangle$ between sites with $\tau^i = \tau^j = 1$ and sites $\tau^{i^\prime} = \tau^{j^\prime} = 0$. Note also that if the pairwise interaction potential $W(x)$ were finite at short distance, the interaction between sites belonging to the same cube would yield a term proportional to $\lambda^{2d}/\ell^d$. Since there is no such term in the expression of $\Delta {\cal S}_1$, we posit that $W(x)=0$ for $x<a$, with $a$ of the order of few lattice spacings and such that $\lambda < a \ll \ell$. Hence, for $a>\lambda$ only sites belonging to different cubes are coupled through the potential $W(x)$.
 
Taking the continuum limit (for $\ell/\lambda \gg 1$) and comparing the asymptotic behavior of  $\Delta {\cal S}_1$, Eq.~(\ref{eq:DS1}), with the one of the effective $1$-replica action in Eq~(\ref{eq:Heff-1}), one obtains
\begin{displaymath} 
\begin{split}
&\frac{\Delta {\cal S}_{1,{\rm eff}}}{L^d}  \approx \mu - \frac{ \lambda^d \mu- 
d \lambda^{d-1} w_2}{\ell^d} + 
\frac{\lambda^{2d} \Omega_d}{2 \ell^d} \int_1^{L/\ell} x^{d-1} W( \ell x) \, {\rm d}x
+ \frac{ d^2 \lambda^{2 d - 2} w_4}{\ell^{2d}} \equiv \\&
\frac{\Delta {\cal S}_1}{L^d} = \frac{3}{4u} \bigg[ s_c^2 - \frac{s_c^2 \lambda^d + 2 c d s_c \lambda^{d-1}}{\ell^d} + 
\frac{c^2 d^2 \lambda^{2d - 2}}{\ell^{2d}} \bigg(1 
+ \Big(\frac{\lambda}{\ell} \Big)^d - \Big( \frac{\lambda}{\ell} \Big)^{2d} + \ldots
\bigg) \bigg]  \, ,
\end{split}
\end{displaymath}
where $\Omega_d= 2 \pi^{d/2} / \Gamma(d/2)$ is the surface area of the sphere of unit radius in $d$ dimensions. This equation yields $\mu = 3 s_c^2 / (4 u)$, $w_2 = - 3 c s_c/(2 u)$, and $w_4 = 3 c^2/(4 u)$. The terms of the right-hand side decaying as $1/\ell^{rd}$ with $r \ge 3$ can be approximately 
reproduced (although the $\lambda$-dependence cannot be captured exactly) by setting
\begin{equation} 
W(x) = \theta(x-a) \sum_{r=2}^\infty \tilde{w}_r/x^r
\end{equation} 
with $\tilde{w}_r \approx (-1)^r 3 c^2 d^3 (r-1)/(2 u \Omega_d)$. In the following we will drop all the terms with $r > 2$ and will keep only the dominant term which behaves as $\tilde w_2/r^{2d}$ with $\tilde w_2\approx 3c^2d^3/(2 u \Omega_d)$.

\begin{figure}
\includegraphics[scale=0.36]{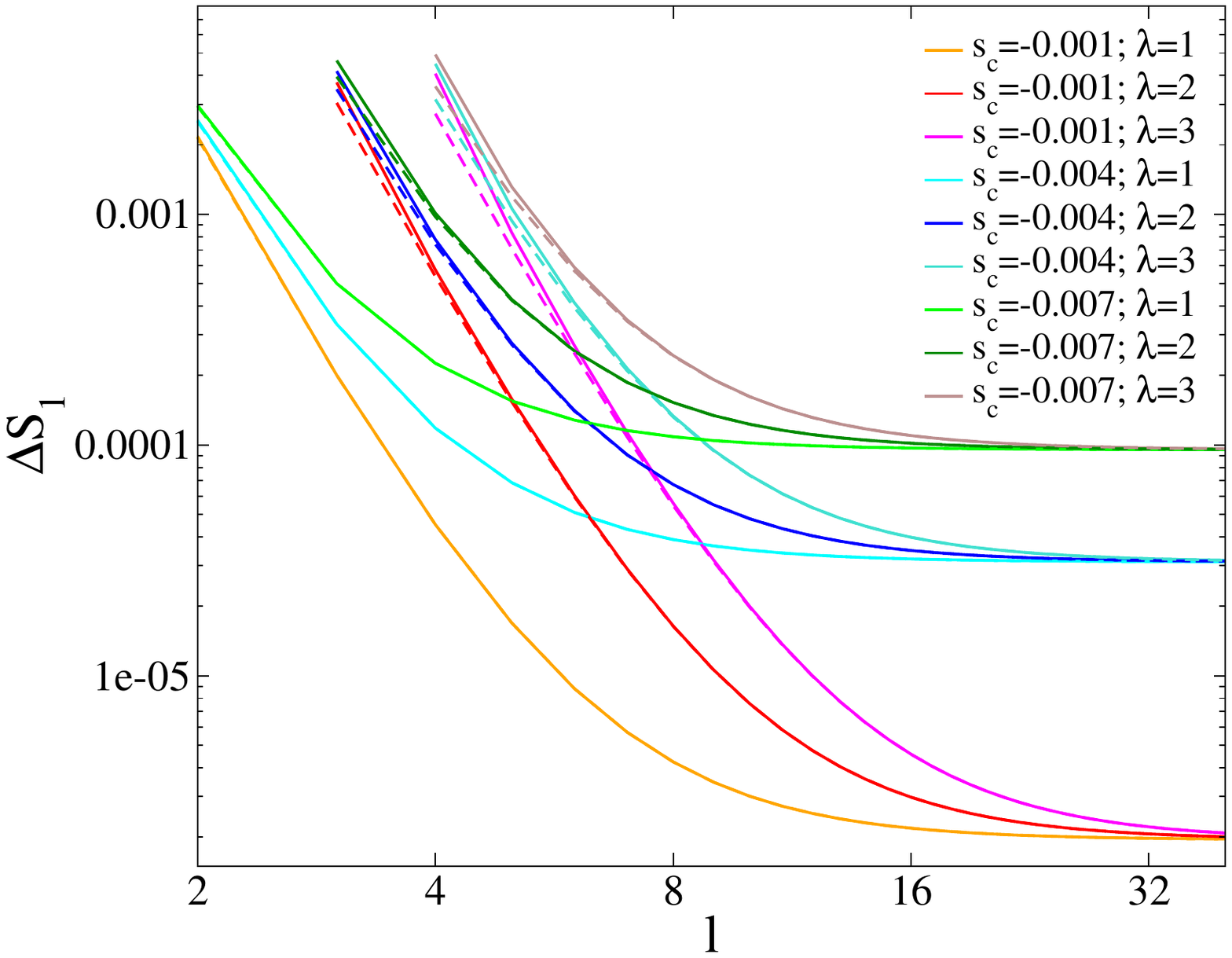}
\caption{Periodic cluster ansatz: Correction to the $1$-replica action beyond the ``zeroth-order'' approximation, 
$\Delta \mathcal S_1$, as a function of $\ell$ for several values of $\lambda$ and $s_c$ (close to, but below, the mean-field thermodynamic glass transition). 
%$-0.004$ and $-0.007$ (close to the ``renormalized'' Kauzmann transition, $s_c \approx -0.00786$)  and for 
The full lines correspond to Eq.~(\ref{eq:DS1}), and the dashed ones are obtained from the translationally invariant Hamiltonian with effective 
long-range $2$- and multi-body interactions and an effective external field, Eq.~(\ref{eq:Heff-1}), in three dimensions. 
The values of the other parameters are $u = 0.385$, $c=0.081$, $a=4$.}
\label{fig_amorphous_effective}
\end{figure}

Note that the calculation above can be carried out numerically on the $d$-dimensional hyper-cubic lattice, without taking the continuum limit ($\ell/\lambda \gg 1$).\cite{footnote_continuum} We have done it in three dimensions and it yields, both qualitatively and quantitatively, very similar results to the asymptotic analysis. As an illustration, $\Delta {\cal S}_1/ L^d$ is plotted in Fig.~\ref{fig_amorphous_effective} as a function of $\ell$ for different values of $s_c$ and $\lambda$, showing a reasonably good agreement with the result obtained for the translationally invariant effective Hamiltonian. [The behavior of $\Delta {\cal S}_1$ above $T_K^{MF}$ ({\it i.e.}, $s_c > 0$) can also be reproduced by the effective first cumulant in Eq.~(\ref{eq:Heff-1}) with the same coefficients $w_2$, $w_4$, and $\tilde{w}_r$ as given above, but one has to introduce a finite interaction range $\ell_{\rm pin}$ for the $2$- and $4$-body interaction terms.]

In order to check the robustness of the procedure, we have repeated the same calculations for a different choice of the overlap pattern, {\it i.e.}, a slab geometry where $\tau^i = 1$ except in a stripe of width $\ell$ where $\tau^i = 0$. In Appendix~\ref{app:slab} we show that the resulting effective Hamiltonian found for the slab pattern has exactly the same structure as Eq.~(\ref{eq:Heff-1}), albeit with some small differences due to the strongly anisotropic character of the slab geometry. The analysis nonetheless confirms that our procedure is quite robust with respect to the specific choice of the overlap pattern.

\subsection{$2$-replica action and higher orders}
\label{sec:2replica_approx}

We illustrate our computation of higher orders of the expansion in number of free replica sums by considering the $2$-replica action (see above and paper~I). Within the present variational 1-RSB approximation, the diagonal $n_1 \times n_1$ square  blocks $[\eta_{a_1 b_1}^i]_\star$ of the matrices~(\ref{eq:2replicamatrix}) (respectively, the diagonal $n_2 \times n_2$ square  blocks $[\eta_{a_2 b_2}^i]_\star$) can be determined at the level of the $1$-replica action given the pattern of the $\tau_1^i$'s alone, independently of the configuration of the $\tau_2^i$'s (resp., of the $\tau_2^i$'s alone, independently of the $\tau_1^i$'s). In consequence, the only remaining variables to be found are those contained in the rectangular $n_1 \times n_2$ blocks $\eta_{a_1 b_2}^i$ of the matrices~(\ref{eq:2replicamatrix}) on the sites $i$ where both groups of replicas have a vanishing overlap with the reference configuration~[see Eq.~(\ref{eq:parametrization})]. Although one could envisage to generalize the variational procedure based on the periodic-cluster ansatz for these off-diagonal blocks, this is now quite involved due to the presence of several groups of replicas, and we choose for simplicity to calculate the $2$-replica action $\mathcal S_2$ only within the ``zeroth-order'' approximation: We assume that on the sites where $\tau_1^i =  0$ and $\tau_2^i = 0$ the elements of the rectangular $n_1 \times n_2$ matrices $\eta_{a_1 b_2}^i$  are also equal to zero. % one also has that $\eta_{a_1 b_2}^i = 0$. 
This choice is motivated by the mean-field results obtained in paper~I, and we will explicitly check in Sec.~\ref{sec:REMFC-check} that for the fully connected $2^M$-KREM in the Kac limit, going beyond this  zeroth-order assumption only yields sub-leading corrections.

After some simple algebra one therefore finds
\begin{equation} 
\label{eq_S2}
\begin{aligned}
\mathcal S_{2} [\tau_1,\tau_2] \approx 2 c \sum_{ \langle i, j \rangle} \tau_1^i \tau_1^j \tau_2^i \tau_2^j + 2 \sum_i \Big( \frac{u}{3} 
- s_c - c d \Big) \tau_1^i \tau_2^i \, . 
\end{aligned}
\end{equation}
Within the present approximation the higher-order terms of the expansion in number of free replica sums are also given by their ``zeroth-order'' expression and can be easily obtained from Eqs.~(\ref{eq:hamiltonian_hard}) and~(\ref{eq:S0Sint}).
In particular, for the simplified effective Hamiltonian in Eq.~(\ref{eq_replica_action_wolynes}) which is truncated at the cubic order, one simply obtains that
\begin{equation}
\label{eq_S3}
\begin{aligned}
\mathcal S_{3} [\tau_1,\tau_2,\tau_3] \approx - 2 u \sum_i \tau_1^i \tau_2^i \tau_3^i 
\end{aligned}
\end{equation}
and that all higher-order cumulants are equal to zero.

\section{Construction of the Effective Theory} 
\label{sec:effective_theory}

We are now in a position to derive the sought-after effective theory. Below, we proceed to the explicit construction of the effective Hamiltonian
describing the statistics of the local fluctuations of the overlap with an equilibrium configuration by starting from the $1$- and $2$-replica components of the action that have been derived in the preceding Section. 

\subsection{$1$- and $2$-replica parts of the action and effective disordered theory} 
\label{sec:effective_theory-S1}

Collecting the results obtained above, we can express the $1$-replica action as
\begin{equation}
\begin{aligned}
\label{eq_cum1_effect}
\mathcal S_{1} [\tau] \approx & \frac{c}{2}\sum_{\langle i,j \rangle}(\tau^i-\tau^j)^2 - \sum_i  \tilde{\mu} \tau^i 
- w_2 \sum_{\langle i, j \rangle} \tau^i \tau^j %\\ & 
	+ \frac{\tilde{w}_2}{2} \! \sum_{\vert  r_i-r_j\vert  >a}  \! \frac{\tau^i \tau^j}{\vert  r_i-r_j\vert ^{2d}}
+ \frac{w_4}{L^d} \!\!\! \sum_{\langle i, i^\prime \rangle \neq \langle j ,j^\prime \rangle}
\!\!\! [\tau^i (1 - \tau^{i^\prime}) \tau^j (1 - \tau^{j^\prime})]_{\rm sym}
\, ,
\end{aligned}
\end{equation}
with
\begin{equation} \label{eq:S1eff_couplings}
\begin{aligned}
&\tilde \mu =  -s_c+\frac{3(2 c d  + s_c)} {4 u}s_c  \, ,\\&
w_2 = - \frac{3 c}{2 u}s_c \, ,\\&
\tilde{w}_2 = \frac{3 c^2 d^3}{2 u \Omega_d}   \, ,\\&
w_4 = \frac{3 c^2}{4 u} \, .
\end{aligned}
\end{equation}
whereas the $2$-replica action is given by Eq.~(\ref{eq_S2}).

As we have already discussed, an expansion in increasing number of free replica sums, $\mathcal S_{\rm rep}[\{\tau_a\}]= \sum_{a=1}^n \mathcal S_1 [\tau_a]-\frac 12 \sum_{a,b=1}^n \mathcal S_{2}[\tau_a,\tau_b]+\cdots$ as in Eq.~(\ref{eq:expansion_freereplica-sums}), is, modulo some constant, equivalent to the expansion in cumulants of a disordered Hamiltonian (see also paper~I). More specifically, for a disordered Hamiltonian ${\mathcal H}_{\rm eff}[\tau]$, the associated replicated theory,
\begin{displaymath} 
e^{-\mathcal S_{\rm rep}[\{\tau_a\}]}=\overline{e^{-\beta \sum_{a=1}^n\mathcal H_{\rm eff}[\tau_a]}}\,,
\end{displaymath}
where the overline denotes an average over the quenched disorder, leads to an expansion in number of free replica sums with the identification $\mathcal S_1 [\tau_a]=\overline{\beta \mathcal H_{\rm eff}[\tau_a]}$, $\mathcal S_2 [\tau_a,\tau_2]=\overline{\beta \mathcal H_{\rm eff}[\tau_a]\beta \mathcal H_{\rm eff}[\tau_b]}-\overline{\beta \mathcal H_{\rm eff}[\tau_a]}\,\overline{\beta \mathcal H_{\rm eff}[\tau_b]}$, etc. One immediately sees that an identification of this kind is only possible if, {\it e.g.}, $\mathcal S_{2}$ is nonnegative.

From Eqs.~(\ref{eq_cum1_effect}) and (\ref{eq_S2}), one therefore infers the following effective disordered Hamiltonian: 
\begin{equation} 
\label{eq:Heff-final-tau}
\begin{split}
\beta {\mathcal H}_{\rm eff}[\tau] &= \frac{c}{2}\sum_{\langle i,j \rangle}(\tau^i-\tau^j)^2 - \sum_i ( \tilde{\mu}+ \delta \tilde{\mu}_i) \tau^i 
- \sum_{\langle i, j \rangle}(w_2+\delta w_{2,ij}) \tau^i \tau^j \\ 
& \qquad + \frac{\tilde{w}_2}{2} \! \sum_{\vert  r_i-r_j\vert >a}  \! \frac{\tau^i \tau^j}{\vert  r_i-r_j\vert ^{2d}}
+ \frac{w_4}{L^d} \!\!\! \sum_{\langle i, i^\prime \rangle \neq \langle j ,j^\prime \rangle}
\!\!\! [\tau^i (1 - \tau^{i^\prime}) \tau^j (1 - \tau^{j^\prime})]_{\rm sym} + \ldots \, ,
\end{split}
\end{equation}
where the random variables, $\delta \tilde{\mu}_i$ and $\delta w_{2,ij}$, have a zero mean ($\overline{\delta \tilde{\mu}_i} = \overline{\delta w_{2,ij}} = 0$) and variances given by
\begin{equation} 
\label{eq_variances-tau}
\begin{split}
\overline{\delta \tilde{\mu}_i  \delta \tilde{\mu}_j} & = 2\Big( \frac{u}{3} - s_c - c d \Big) \delta_{ij} \, , \\
\overline{\delta w_{2,ij}\delta w_{2,kl}} & = 2c (\delta_{ik}\delta_{jl}+\delta_{il}\delta_{jk}) \, \\
\overline{\delta \tilde{\mu}_i \delta w_{2,jk}} & = 0 \, .
\end{split}
\end{equation}
Note that the requirement of positive variances imposes the condition $u/3 >s_c+ cd$. 

Taking into account the higher-order terms of the expansion in number of free replica sums, $\mathcal S_3 [\tau_1,\tau_2,\tau_3]$, etc., would lead to nonzero higher-order cumulants of the random variables (which otherwise are simply Gaussian distributed).
For the specific case of the simplified model considered here, one has from Eq.~(\ref{eq_S3})
\begin{equation}
	\overline{\delta \tilde{\mu}_i  \delta \tilde{\mu}_j \delta \tilde{\mu}_k}^c = - 2 u \delta_{ij} \delta_{ik} \, ,
\end{equation}
with the other third-order cumulants and all the higher-order ones equal to zero.

\subsection{Effective disordered Ising Hamiltonian}

After changing variables to Ising ones, $\tau^i=(1+\sigma^i)/2$ with $\sigma^i=\pm1$, Eq.~(\ref{eq:Heff-final-tau}) provides the following effective disordered Hamiltonian for Ising spins (up to an additive random energy term independent of the $\tau^i$'s): 
\begin{equation} 
\label{eq:Heff-final}
\begin{split}
\beta {\mathcal H}_{\rm eff}[\sigma] &= \mathcal S_0 - \sum_i \left ( H + \d h_i\right) \sigma^i  - \sum_{\langle i,j \rangle} (J_2 +\delta \! J_{ij}) \sigma^i \sigma^j + \frac{\tilde J_2}{2} \sum_{\vert  r_i-r_j\vert >a} \frac{\sigma^i \sigma^j}{\vert  r_i-r_j\vert ^{2d}} \\
& \qquad
+ \frac{J_4}{L^d} 
\!\!\! \sum_{\langle i, i^\prime \rangle \neq \langle j , j^\prime \rangle}
\!\!\! [(1+\sigma^i) (1-\sigma^{i^\prime}) (1+ \sigma^j) (1-\sigma^{j^\prime})]_{\rm sym}  + \ldots \,,
\end{split}
\end{equation}
where $a$ is of the order of a few lattice spacings and 
\begin{equation} 
\label{eq:coefficients}
\begin{aligned}
H & 
= -\frac{1}{2} \Big( s_c + 3\frac{c^2 d^2 - s_c^2a^d}{4 u a^d} \Big) \, , 
\\
J_2  & 
= \frac{c}{4} \Big( 1 - \frac{3 s_c}{2 u} \Big) \, , 
\\
\tilde J_2  & 
= \frac{3 c^2 d^3}{8 u \Omega_d} \, , 
\\
J_4  & 
= \frac{3 c^2}{64 u} \, .
\end{aligned}
\end{equation}
The random-field and random-bond variables, $\delta h_i$ and $\delta \! J_{ij}$, have a zero mean ($\overline{\d h_i} = \overline{\d J_{ij}} = 0$), and their variances are given by
\begin{equation} 
\label{eq_variances}
\begin{split}
\overline{\d h_i  \d h_j} & = \Delta_h \delta_{ij} +\frac c8 {\mathbb C}_{ij}\,, \;\; \;{\rm with} \;\; \; \Delta_h=\frac 12 \Big( \frac{u}{3} - s_c - \frac{c d}{2} \Big), \\
\overline{\delta \! J_{ij} \delta \! J_{kl}} & = \frac{c}{8} (\delta_{ik}\delta_{jl}+\delta_{il}\delta_{jk}) \, \\
\overline{\d h_i \delta J_{jk}} & = \frac{c}{8} ( \delta_{ij} {\mathbb C}_{ik}+ \delta_{ik}{\mathbb C}_{ij} ) \, ,
\end{split}
\end{equation}
where ${\mathbb C}_{ij}$ is the connectivity matrix of the lattice, equal to $1$ for nearest neighbors and $0$ otherwise. The third cumulant
of the random field is $\overline{\d h_i  \d h_j \d h_k} = -(u/4) \delta_{ij} \delta_{ik}$, and $\mathcal S_0$ is a random term that does not depend on the Ising variables.\cite{S0}

The above Hamiltonian is that of a random-field + random-bond Ising model defined on a $d$-dimensional lattice, in which the ferromagnetic nearest-neighbor interactions compete with longer-range and multi-body terms, as announced in the Introduction [see Eq.~(\ref{eq_effectiveIsing_final})]. We note that its structure is more complex than that of the rough magnetic analogy derived in Ref.~[\onlinecite{Stevenson}]. We now discuss in more detail the physical interpretation and the expected influence of the various terms present in the effective disordered Hamiltonian.

\section{Physical interpretation of the effective theory}
\label{sec:phys_interpretation}

\subsection{Meaning and effect of the different contributions to the effective Hamiltonian}

The above treatment maps the description of the fluctuations of the overlaps in a glass-forming liquids near the putative thermodynamic glass transition (RFOT) onto an effective random-field + random-bond Ising model where the Ising variables describe a low or high local overlap with a reference equilibrium configuration. For a theory described by an effective Hamiltonian (action) of the generic form given in Eq.~(\ref{eq:Heff-final}), one can for instance take as control parameters the ferromagnetic tendency of the interactions, described by $J_2$, the applied uniform source, $H$, and the on-site strength of the random field, described by $\Delta_h$, keeping all the other parameters fixed. For a large enough $J_2$ (compared to the disorder strength and to the amplitude of the competing antiferromagnetic terms) the model has a paramagnetic-to-ferromagnetic critical transition point, provided one is above the lower critical dimension which is $d=2$ in the presence of random fields.\cite{Nattermann} This critical point occurs at a specific value  $J_{2,c}(\Delta_h)$ [or $\Delta_{h,c}(J_2)$] and a critical magnetic field $H_c$. The latter is equal to zero when the model has a statistical $Z_2$ inversion symmetry, {\it e.g.}, if one neglects the third-order cumulant of the random field and the covariance between the random bond and the random field; otherwise, $H_c$ may be (slightly) different from zero. In the following we will often neglect for simplicity the effect of the third cumulant and of the cross-correlations between random bond and random field. Then, for $H$ going through $0$ and either $J_2>J_{2,c}$ at constant $\Delta_h$ or $\Delta_h<\Delta_{h,c}$ at constant $J_2$, there is a first-order transition associated with the presence of a ferromagnetic phase and the coexistence between two oppositely magnetized states (if $H_c\neq 0$, coexistence occurs for a value of the magnetic field that depends on $J_2$ or on $\Delta_h$). 

\begin{figure}
\includegraphics[scale=0.36]{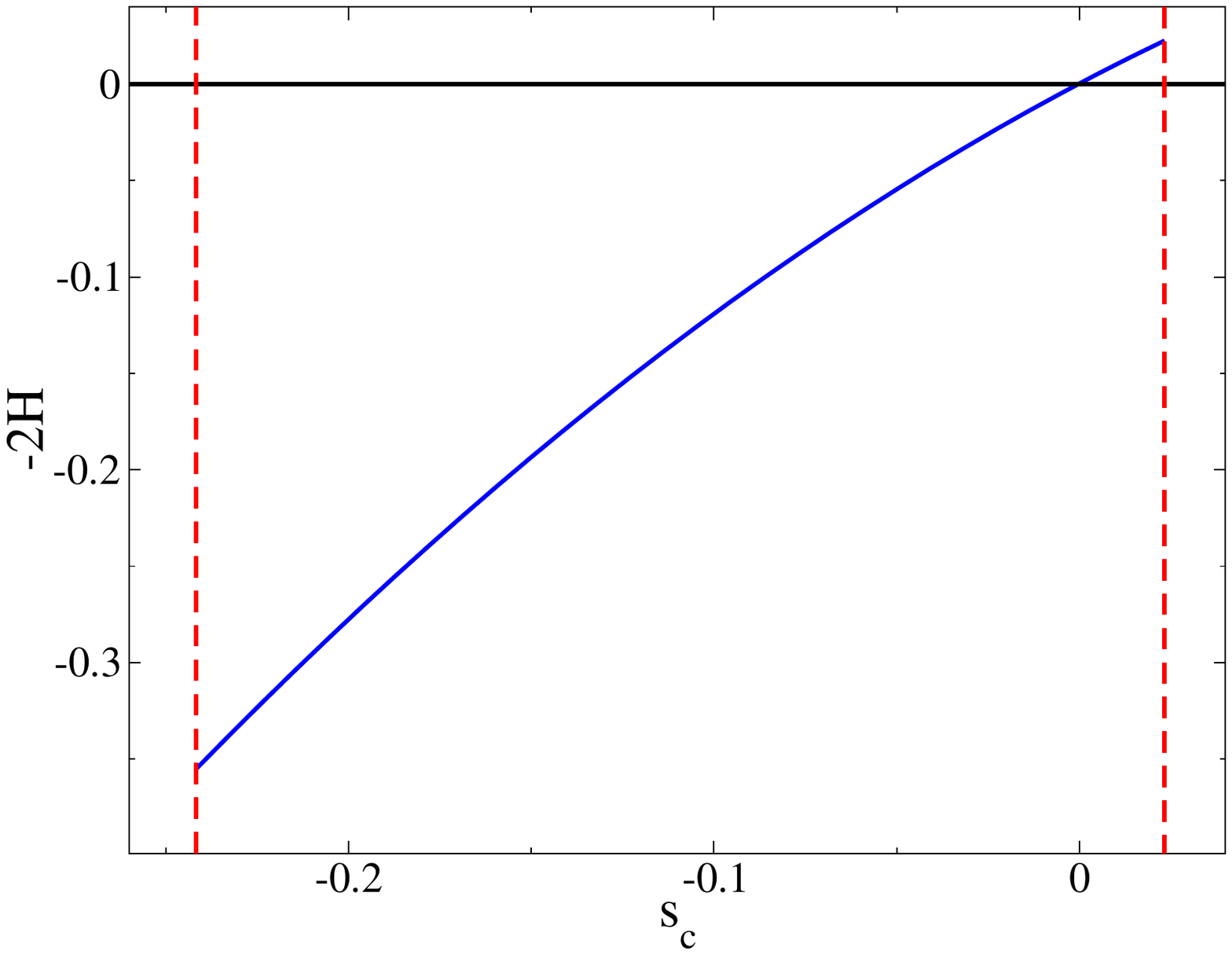}
\caption{Effective external field $-2H$ as a function of the bare configurational entropy $s_c$. The position of the mean-field critical point is at $s_c=0$ whereas the ``renormalized'' value  for which $H=0$ is at $s_c \approx -0.00786$ (given by Eq.~(\ref{eq_critical_condition}) in three dimensions for $u = 0.385$, $c=0.081$, and $a=4$. The dashed red vertical lines represents the limits of validity of our approach ($s_c <  u/3-cd$ to have positive variances and $s_c > c d /(2^d-1) -2u/3$ to ensure that the variational parameter $m_\star$ stays positive).
For this choice of the parameters, the ratio $\Delta_h/\tilde{J} \approx 11.7$, which is incompatible with the existence of $T_c$.}
\label{fig:estimate}
\end{figure}

However, from the above mapping, $J_2$, $\Delta_h$, and $H$ cannot be taken as independent parameters, nor can all the other parameters be taken as fixed. In the original description of glass-forming liquids in terms of overlaps (see Sec.~\ref{sec:GL}), the only (indirectly) controlable parameter is the bare configurational entropy $s_c$ which strongly decreases as the temperature decreases, all the other (liquid-specific) parameters being only weakly dependent on temperature in the domain under study. One can then see from Eqs.~(\ref{eq:coefficients}) and (\ref{eq_variances}) that  $J_2$, $H$, and $\Delta_h$ vary with $s_c$. However, $H$ varies more strongly than the other two: The variation of $H$ with $s_c$ is illustrated in Fig.~\ref{fig:estimate}. Decreasing the temperature in the original glass-forming liquid therefore corresponds to a trajectory in the ($J_2$, $H$, $\Delta_h$) diagram. If it exists, the thermodynamic glass transition (RFOT) of the liquid becomes a first-order transition of the Ising model, which takes place in the simplest case where the statistical $Z_2$ symmetry holds when $H$ goes through zero. Then, when $H<0$, the system is negatively magnetized, i.e., it is in a low-overlap state (a liquid), and when $H>0$ the system is positively magnetized, which corresponds to a high-overlap state (a glass). The transition, which corresponds to the coexistence between low- and high-overlap states, exists provided the trajectory of the liquid in the ($J_2$, $H$, $\Delta_h$) diagram crosses $H=0$ for $J_2>J_{2,c}$ or $\Delta_h<\Delta_{h,c}$. This will be further discussed below. Note also that the description of the ideal glass may require some refinement but we are mostly interested in the existence of the transition itself.

The zeroth-order (mean-field-like) approximation that does not take into account the point-to-set correlations amounts to having a simpler random-field + random-bond Ising Hamiltonian with $H=-s_c/2, J_2=c/4, \tilde J_2=0, J_4=0$ and random variables having the same variances as in Eqs.~(\ref{eq_variances}). As anticipated, $H$ is then simply proportional to the mean-field configurational entropy, $s_c$, and when accounting for the presence of point-to-set correlations, $H$, or rather $-2H$, plays the role of a renormalized configurational entropy. The bare $s_c$ is generically corrected by a positive contribution [$3 (c^2 d^2 a^{-d}- s_c^2)/(4 u)$ is positive at least when $\vert s_c \vert$ is small enough] which then tends to lower the mean-field transition temperature $T_K^{MF}$. In the zeroth-order description, provided $c$ is not too small, the random-bond disorder does not destroy the ferromagnetic trend. (It is interesting to notice that the fluctuations of the ferromagnetic coupling can become much bigger than the average value in the $c \to 0$ limit. In this limit, in contrast with what we have just described, it seems that a spin-glass-like physics may emerge, as advocated in Ref. [\onlinecite{Moore}].)

The effect of the point-to-set correlations, which are associated with some form of incipient  ``amorphous order'' induced by the specific patterns of local overlaps with a reference configuration, is to generate additional interactions between the Ising variables, in particular, an antiferromagnetic one with coupling $\tilde J_2>0$. This interaction has the effect of decreasing the ferromagnetic tendency and, as a consequence,  of lowering the transition temperature. (Note however that,  when $s_c<0$, there is also an additional ferromagnetic coupling proportional to $s_c$.) Before discussing the $4$-body link-link infinite-range interaction with amplitude $J_4>0$ that is also generated by the point-to-set correlations, we first stress that the scale-free character of these additional interactions stems from the fact that we have considered the liquid below the putative mean-field RFOT $T_K^{MF}$ and therefore characterized by diverging (mean-field-like) point-to-set correlation lengths. Above $T_K^{MF}$ the range of the interactions is cut off by the finite (mean-field-like) point-to-set lengths, {\it e.g.}, the length $\ell_{\rm pin}$ introduced in Sec. \ref{sec:periodic_cluster}. If, when working below $T_K^{MF}$, we find that the transition is destroyed in the effective disordered theory, this in turn implies that the mean-field description we took as our starting point is not a valid proxy for describing the low-energy configurations of the original glass-forming system in finite dimensions and that the approximate mapping to the effective Hamiltonian itself cannot be used beyond showing that the RFOT scenario is not self-consistent. Conversely, if the transition is not destroyed, but just modified and shifted to lower temperature, then our variational approximation scheme is better justified.

\subsection{The $4$-body link-link infinite-range interaction} 
\label{sec:4body}

The infinite-range $4$-body link-link interaction corresponds to a repulsive coupling ($J_4 > 0$) between pairs of links connecting sites with low and high overlap with the reference configuration, thereby disfavoring the formation of interfaces between the high- and the low-overlap phases (or equivalently,  between the negatively and positively magnetized phases). {\it A priori}, such a term could have a dramatic impact on the existence of the transition and on its properties. For instance, one may naively think that it could turn the effective theory into a mean-field model due to the infinite-range nature of the interactions. It appears however that the effect is not as severe, as indicated by the following observations:
\\

\noindent
\textbf{1.}
The $4$-body interaction is strictly non-negative and is identically zero ({\it i.e.}, minimal) only in both the perfectly ferromagnetically ordered states (either $+1$/high-overlap  with $\sigma^i=+1$ $\forall i$ or $-1$/low-overlap configurations with $\sigma^i=-1$ $\forall i$), since $(1 + \sigma^i)(1 - \sigma^{i^\prime})$ is then zero on each link $\langle i, i^\prime \rangle$. Conversely, its associated energy cost is maximum for antiferromagnetically ordered configurations where sites with $\sigma^i = +1$ are surrounded by sites with $\sigma^{i'}=-1$ (however these configurations have a very high-energy and their thermodynamic weight is very small). In the paramagnetic (liquid) phase, the energy cost generated by the $4$-body term is inversely proportional to the correlation length (that of the effective theory, not a mean-field-like point-to-set correlation length). In fact, if the correlation length is large, the probability that two neighboring spins have the same sign is large as well, which reduces the interfaces and the associated energy cost. The $4$-body term clearly disfavors the formation of interfaces and does not destroy the establishment of long-range ferromagnetic order.
\\

\begin{figure}
\includegraphics[scale=0.4]{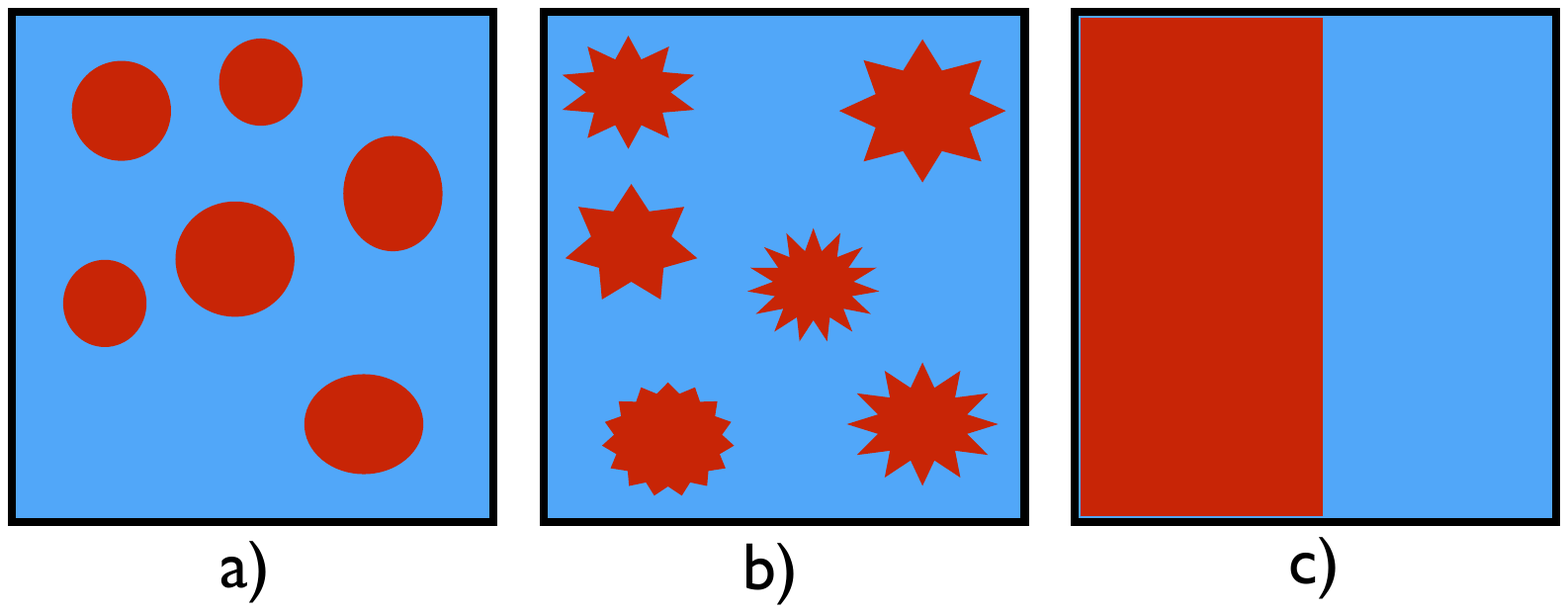}
\vspace{-3.7cm}
\caption{Sketch of potentially relevant spin configurations of the effective model in Eq.~(\ref{eq:Heff-final}) close to the putative thermodynamic transition ({\it i.e.}, $H \approx 0$, $J_2$ strong enough and $\overline{\d h_i^2}$  not too big), showing clusters of $+1$ spins in red (high overlap with the reference configuration) in the background of $-1$ spins in light blue (low overlap with the reference configuration): a) smooth interfaces, b) rough interfaces, and c) one flat interface.}
\label{fig:clusters}
\end{figure}

\noindent
\textbf{2.}
Consider a set of parameters for which the system happens to be close to the putative thermodynamic transition ($H \approx 0$). If the effective ferromagnetic coupling $J_2$ is strong enough and the effective disorder $\overline{\d h_i^2}$  is not too big, typical configurations will be formed by  
clusters of spins with (mostly) $\sigma^i=+1$ in the background of spins with (mostly) $\sigma^i=-1$, as sketched in Fig.~\ref{fig:clusters}. The $4$-body coupling is proportional to the square of the number of links between sites with $\sigma^i=+1$  and sites with $\sigma^i=-1$. Thus, the (intensive) energy cost of such configurations is roughly $J_2 \rho S + J_4 \rho^2 S^2$ ($\rho$ being the density of the cluster and $S$  their typical surface area), resulting in an effective increase of the surface tension and in an increase of the energy barriers associated with nucleation. Since such an increase goes as $S^2$, the energetic cost associated to the creation of interfaces is lower for smooth ones (Fig.~\ref{fig:clusters}a) than for rough ones (Fig.~\ref{fig:clusters}b). For instance, in the case of the flat interface of Fig.~\ref{fig:clusters}c the $4$-body interaction only gives a sub-leading contribution which can be neglected in the thermodynamic limit (see also Appendix~\ref{app:slab}). In conclusion, it seems that the $4$-body interaction could have a strong impact on the properties of the interfaces, thereby possibly modifying the critical behavior of the effective model but not on the very existence of a transition. (Of course, this term also disfavors and slows down the nucleation processes of the effective model. However, since there is no obvious mapping between the dynamics of the glass-forming liquid and that of the effective model, this does not provide any direct information on the dynamical processes of the original glassy model.)
\\

\noindent
\textbf{3.}
The $4$-body link-link interaction can be rewritten as a sum of two competing contributions, 
\begin{equation} \label{eq:4body}
\begin{split}
%\frac{w_4}{L^d} \!\!\! \sum_{\langle i, i^\prime \rangle \neq \langle j ,j^\prime \rangle}
%!\!\! [\tau^i (1 - \tau^{i^\prime}) \tau^j (1 - \tau^{j^\prime})]_{\rm symm}
%& = 
\frac{J_4}{L^d} \!\!\! \sum_{\langle i, i^\prime \rangle \neq \langle j ,j^\prime \rangle}
\!\!\! [(1 + \sigma^i) (1 - \sigma^{i^\prime}) (1 + \sigma^j) (1 - \sigma^{j^\prime})]_{\rm sym} %\\
%& 
= - 2 d J_4 \sum_{\langle i,j \rangle} \sigma^i \sigma^j + 
\frac{J_4}{L^d} \!\!\! \sum_{\langle i, i^\prime \rangle \neq \langle j ,j^\prime \rangle}
\sigma^i \sigma^{i^\prime} \sigma^j \sigma^{j^\prime} \, .
\end{split}
\end{equation}
The first term is a standard ferromagnetic nearest-neighbor interaction which strengthens the ferromagnetic coupling (thus favoring the establishment of ferromagnetic long-range order and, as a result, the existence of a thermodynamic glass transition), while the second term is an infinite-range antiferromagnetic coupling between {\it pairs of nearest-neighbor spins} (which, on the contrary, frustrates the formation of magnetically ordered phases and suppresses the glass transition).

From Eq.~(\ref{eq:4body}), after performing a Hubbard-Stratonovich and a saddle-point calculation, the effective disordered Ising Hamiltonian can be expressed as
\begin{equation} \label{eq:Heff_psi} 
\b {\mathcal H}_{\rm eff} (\{\sigma^i\},\psi)  = \mathcal S_0 - \sum_{\langle i,j \rangle} [J_2+2dJ_4 (1 + \psi) + \delta \! J_{ij}] \sigma^i \sigma^j - 
\sum_i \left ( H + \d h_i\right) \sigma^i  + \frac{\tilde J_2}{2} \sum_{\vert  r_i-r_j\vert >a} \frac{\sigma^i \sigma^j}{|i-j|^{2d}} 
 + \ldots \, ,
\end{equation}
where $\psi$ is the average value of $\sigma^i \sigma^j$ for nearest-neighbor sites, which is computed by using the Boltzmann-Gibbs measure provided by the effective Hamiltonian~(\ref{eq:Heff_psi}). It must therefore be determined self-consistently,
\begin{displaymath}
\psi = \langle \sigma^i \sigma^j \rangle_{\rm eff} \equiv \frac{1}{Nd} \,
\frac{ {\rm Tr}_{\{ \sigma \}} \Big ( \sum_{\langle i, j \rangle} \sigma^i \sigma^j \Big)  
e^{- \b {\cal H}_{\rm eff}(\psi)} }
{ {\rm Tr}_{\{ \sigma \}}  
e^{- \b {\cal H}_{\rm eff}(\psi)}} \, .
\end{displaymath}
The effective ferromagnetic coupling is thus  ``renormalized'' by a factor which depends on the spin configuration itself [$J_2 \to J_2 +2dJ_4 (1 + \langle \sigma^i \sigma^j \rangle_{\rm eff})$] and takes its strongest possible value in the ferromagnetically ordered phases ($\langle \sigma^i \sigma^j \rangle_{\rm eff} \approx +1$) and its lowest possible value in the antiferromagnetic phase (however, as noticed above, antiferromagnetic states are very rare configurations). In the paramagnetic phase $\langle \sigma^i \sigma^j \rangle_{\rm eff} \gtrsim 0$ and, as  already discussed, it is larger, the larger the correlation length. This observation leads us to the same conclusions as before, namely that the $4$-body coupling  cannot {\it a priori} destroy the transition, but could modify some of its properties (especially the nucleation process and the properties of the interfaces). Note that the infinite-range character of the interaction {\it does not} lead to a mean-field model because of the fact that it couples different pairs of nearest neighbors. 
\\

All in all, although the above arguments are only qualitative, they strongly suggest that the $4$-body link-link interaction may have an impact on the properties of the interfaces but not on the existence of a thermodynamic glass transition. This point could of course be fully settled by studying directly the effective theory in Eq.~(\ref{eq:Heff-final}), {\it e.g.}, by large-scale numerical simulations, but this goes beyond the scope of the present paper.

\subsection{When does a thermodynamic glass transition exists? A rough estimate} 
\label{sec:effective_theory-estimation}

Although the effective random-field + random-bond Ising theory can now be investigated by powerful methods such as large-scale computer simulations and nonperturbative renormalization group techniques, it is instructive to provide a rough estimate of when a thermodynamic glass transition (RFOT) persists in the presence of spatial fluctuations in $3$-dimensional systems. To do this we somehow project the effective theory onto the simpler short-range random-field Ising model (RFIM) on a cubic lattice that has been thoroughly investigated.\cite{middleton-fisher,martinmayor-fytas,Nattermann}

In order to take into account the $4$-body link-link interaction, we use Eq.~(\ref{eq:Heff_psi}), and set $\psi = 0$, {\it i.e.}, $J_2 \to J_2 + 2 d J_4$, which provides a lower bound for the value of the effective ferromagnetic coupling (see Sec.~\ref{sec:4body} for a more detailed discussion). The $2$-body power-law decaying antiferromagnetic interaction has the tendency to suppress the ferromagnetic order. Since its spatial decay is nonetheless relatively fast ($1/r^6$ in $d=3$), we take its effect  into account by renormalizing the strength of the short-range ferromagnetic coupling, as  
$J_2^{{\rm eff}} d =(J_2 + 2 d J_4) d - \tilde J_2 \Omega_d/2 \int_{a}^\infty r^{-d-1} {\rm d}r$,
which gives
\begin{equation} 
\label{eq:J2eff}
J_2^{{\rm eff}} = \frac{c}{4} \Big( 1 - \frac{3 s_c}{2 u} + \frac{3 c d}{8 u} - \frac{3 c d}{4 u a^d} \Big) \, .
\end{equation}
We will also neglect the randomness of the bonds (which seems reasonable provided that $c$ is not too small) and neglect also higher-order disorder terms, to only retain the local part of the variance of the random source, $\Delta_h=u/6 -s_c/2- cd/4$.

For the standard short-range RFIM with independent and identically distributed (i.i.d.) random fields sampled from a centered Gaussian distribution, a (first-order) transition can only take place in zero external field, $H=0$, which, by using the above expressions, imposes that
\begin{equation} 
\label{eq_critical_condition}
\frac{3 s_c^2}{4 u} - s_c + \frac{3 c^2 d^2}{4 u a^d} = 0 \, ,
\end{equation}
leading to $s_{c,\rm crit}=-(2u/3)\big (\sqrt{1+[9c^2d^2/(4u^2a^d)]}-1\big )$ (only the negative solution is physically meaningful): See also Fig.~\ref{fig:estimate} for $d=3$. (Note that for $d \to \infty$ the critical value of $s_c$ approaches continuously the mean-field one, $s_c^{\rm crit} = 0$,  since the second term in the square root is exponentially small for any $a>1$.)

A phase transition in the short-range RFIM is present (for $d>2$) if the variance of the random field $\Delta$ is not too large compared to the ferromagnetic coupling $J$. From numerical simulations on the RFIM with a Gaussian distribution of the random fields,\cite{middleton-fisher,martinmayor-fytas,Nattermann} one knows that in $d=3$ this requires that $\sqrt{\Delta}/J\lesssim 1.2$. It then allows us to provide a bound for the existence of a thermodynamic glass transition by requiring that 
\begin{equation}
\label{eq_ratio_RFIM}
\frac{\sqrt{\Delta_h}}{J_2^{\rm eff}}\lesssim 1.2\;\;\;\; {\rm for}\;\;\;\; s_c=s_{c,\rm crit} \,,
\end{equation}
where the factors of $\beta$ that are present in the effective theory and not in the parameters $\Delta$ and $J$ of the RFIM cancel out in the ratio appearing in Eq.~(\ref{eq_ratio_RFIM}). Combined with the expression of $J_2^{\rm eff}$ in Eq.~(\ref{eq:J2eff}) and that of $\Delta_h$ in Eq.~(\ref{eq_variances}), this leads to
\begin{equation}
\label{eq_ratio_RFOT}
\frac{128 u^2 \Big (2u - 6 s_{c,\rm crit} - 9 c\Big )}{3 c^2 \Big( 8 - 12 s_{c,\rm crit} + 9 c- 18 c a^{-3}\Big)^2} \lesssim 1.44 \,,
\end{equation}
with $s_{c,\rm crit}$ given above.

For instance, if one considers for illustration the set of parameters given in Ref.~[\onlinecite{Dzero}] to empirically reproduce some key features of the phenomenology of the fragile glass-forming liquid OTP from the field-theory in Eq.~(\ref{eq_replica_action_wolynes}), {\it i.e.}, $w=2.73 q_\star^3$, $y=1.82q_\star^4$, and $u = 0.385 q_\star^3$, with $q_\star \simeq 1$, and if one sets $a=4$, one finds from the bound in Eq.~(\ref{eq_ratio_RFOT}) that there is no thermodynamic glass transition for any (allowed) value of $c$ when large-scale fluctuations are taken into account via the effective theory. On the other hand, in a previous study of the critical point that terminates the transition line in the extended ($T$,$\epsilon$) phase diagram where one introduces a source $\epsilon$ for the overlaps with the reference configuration, critical point which we showed to be in the RFIM universality class,~\cite{noi_Tc} we found with the same OTP parameters of Ref.~[\onlinecite{Dzero}] (but not quite the same effective theory) that  $\sqrt{\Delta_h}/J_2^{\rm eff} \approx 0.47$. Contrary to what we have obtained here for the thermodynamic glass transition in $\epsilon=0$, this is compatible with the existence of a terminal critical point. Although the two results are not in principle mutually incompatible,\cite{footnote_incompatible} we want to stress that the estimate based on the empirical parameters for OTP is a very crude one,\cite{footnote_bare-parameters} and one should not give too much weight to the output. This rather serves as an illustration of what could be done with the effective theory if one had a better estimate of the parameters: In principle, the effect of the large-scale and/or nonperturbative spatial fluctuations can now be accounted for in a much easier way.

To go beyond the crude estimate of the parameters of the effective theory used above, two routes seem promising. The first one builds on the derivation of the overlap field description based on liquid-state theory that was explained in Sec.~\ref{sec:GL}. Starting from the HNC approximation to the Morita-Hiroike functional in Eq.~(\ref{eq_2PI}) one could use the saddle-point procedure discussed in Sec. IV of paper~I in conjunction with the periodic cluster ansatz introduced above. This would provide a way to relate, at least approximately, the microscopic Hamiltonian of the liquid to the parameters needed for the effective theory. A second procedure would be to use the effective theory in direct conjunction with simulation data obtained on glass-forming liquid models in systems of finite (in practice, small) size. Indeed, the limitation then imposed on the spatial extent of the fluctuations in principle allows a determination of the key parameters entering the effective theory from the numerical study of the overlap fluctuations in relatively small systems of interacting particles (typically $100$ or less).\cite{extract}

\section{ Back to the Kac-like random energy model ($2^M$-KREM)} 
\label{sec:REMFC}

\subsection{A check of the low-temperature approximations on the fully connected model}
\label{sec:REMFC-check}

In order to check the validity of the low-temperature approximations put forward in Sec.~\ref{sec:approximations} and used to derive the effective Hamiltonian describing the fluctuations of the overlap with an equilibrium reference configuration near the putative thermodynamic glass transition of finite-dimensional supercooled liquids, we apply the very same procedure to the fully connected $2^M$-KREM,\cite{REM} for which we have already determined the (quasi) exact effective theory in paper~I. (Recall that the model is exactly solvable without having recourse to the effective theory and therefore represents a useful benchmark.)

For completeness, we repeat the definition of the $2^M$-KREM. On each of the $N$ sites $i$ of the lattice (with $N \to \infty$), there are $2^M$ configurations, $\mathcal C_i=\{1,\cdots,2^M\}$, and on each link $(i,j)$ we define i.i.d. Gaussian random energies $E_{ij}=E(\mathcal C_i,\mathcal C_j)$ with $\overline {E_{i j} (\mathcal{C}_i, \mathcal{C}_j)} = 0$ and $\overline {E_{i j} (\mathcal{C}_i, \mathcal{C}_j) E_{i j} (\mathcal{C}_i^\prime, \mathcal{C}_j^\prime)} = M \delta_{\mathcal{C}_i,\mathcal{C}_i^\prime} \delta_{\mathcal{C}_j,\mathcal{C}_j^\prime}$. The Hamiltonian of the model is simply given by
\begin{equation}
\mathcal{H} = \frac{1}{2 \sqrt{N}} \sum_{i \neq j} E_{i j} (\mathcal{C}_i, \mathcal{C}_j) \,.
\end{equation}
The standard mean-field limit corresponds to $M\to \infty$, but the model is solvable for finite $M$ when it is considered on a fully connected lattice.

As already noticed in paper~I, the overlap with an equilibrium reference configuration, $\mathcal{C}_i^0$, is a $2$-state binary variable, $p_a^i = 1$ if $\mathcal{C}_i^a = \mathcal{C}_i^0$  and zero otherwise. In consequence, the first low-temperature approximation of Sec.~\ref{sec:2-state_variables} is {\it exactly} satisfied by the model:
\begin{equation}
p_a^i=\tau_a^i\;\;, \;\;  q_{ab}^i=\tau_{ab}^i\,.
\end{equation}
We also remark again that if $\mathcal{C}_i^a = \mathcal{C}_i^0$ and $\mathcal{C}_i^b = \mathcal{C}_i^0$ ({\it i.e.}, $p_a^i = p_b^i = 1$), then 
$\mathcal{C}_i^b = \mathcal{C}_i^a$ ({\it i.e.}, $q_{ab}^i = 1$). Similarly, if $\mathcal{C}_i^a = \mathcal{C}_i^0$ and $\mathcal{C}_i^b \neq \mathcal C_i^0$ ({\it i.e.}, $p_a^i = 1$ and $p_b^i = 0$), then $\mathcal{C}_i^b \neq \mathcal{C}_i^a$ ({\it i.e.}, $q_{ab}^i = 0$). The same is true, of course, if $\mathcal C_i^a \neq \mathcal{C}_i^0$ and $\mathcal{C}_i^b = \mathcal{C}_i^0$. The only undetermined case corresponds to $\mathcal{C}_i^a \neq \mathcal{C}_i^0$ and $\mathcal{C}_i^b \neq \mathcal{C}_i^0$. As a result, the parametrization in Eq.~(\ref{eq:parametrization}),
\begin{equation}
q_{ab}^i=p_a^ip_a^i+\eta_{ab}^i (1-p_a^i) (1-p_b^i)\,,
\end{equation}
with $\eta_{ab}^i=0,1$ is also {\it exactly} satisfied by the model. For simplifying the calculations, we have used the so-called annealed approximation to handle the averages over the random energies. This approximation is exact above the thermodynamic glass transition temperature $T_K$ and at $T_K$ but deviates from the exact result below $T_K$. As our interest is mainly in the location of the transition and not on the properties of the ideal glass phase, this has negligible  consequences on our conclusions. We stress on the other hand that we do {\it not} use the annealed approximation when computing the averages over the disorder represented by the reference configuration. It is crucial in this case to properly perform the quenched calculation (see also paper~I).

The approximations that we want to test more specifically are therefore the variational determination of the $\eta_{ab}^i$'s and the choice of specific periodic patterns of the $p_a^i$'s to handle the effect of long-ranged point-to-set correlations (see Secs. \ref{sec:variational}, \ref{sec:periodic_cluster}, and \ref{sec:2replica_approx}). For the clarity of the presentation we relegate most details of the derivation of the effective theory to Appendix~\ref{app:REMFC}. 

The ``zeroth-order'' part of the replicated action ${\mathcal S}_{\rm rep} [p_a]$ can be easily obtained by using the naive guess $q_{ab}^i = p_a^i p_b^i$ ({\it i.e.}, $\eta_{ab}^i=0$) on all sites and performing the trace over the configurations ${\cal C}_i^\alpha$, as shown in Appendix~\ref{app:REMFC-ann}:
\begin{equation} 
\label{eq:S1S2-REM}
\begin{aligned}
{\mathcal S}_{\rm rep}^{{(0)}} [ \{ p_a \}] & =  \sum_{a=1}^n \bigg (\!\! - \! \frac{M \beta^2}{8 N}\sum_{ i \neq j } p_a^i p_a^j + 
\big[ \psi^{(0)} (2^M) + \gamma \big]  \sum_i p_a^i  \bigg )  
- \frac 12 \sum_{a,b=1}^n \bigg (\frac{M \beta^2}{4 N} \sum_{i  \neq j } p_a^i p_a^j p_b^i p_b^j \\& 
- \Big[  \psi^{(1)} (2^M) + \frac{\pi^2}{6} \Big ] \sum_{i} p_a^i p_b^i \bigg )+\cdots \, ,
 \end{aligned}
\end{equation}
where the ellipses denote higher-order terms in the expansion in increasing number of free replica sums and the limit $n \to 0$ has been taken; the functions $\psi^{(0)}$ and $\psi^{(1)}$ are the polygamma functions, defined as the logarithmic derivatives of the $\Gamma$-function, $\psi^{(m)} (z) = \textrm{d}^{m+1} \ln \Gamma (z) / \textrm{d} z^{m+1}$, and $\gamma = - \Gamma^\prime (1)$ is the Euler constant. By using Stirling's formula, the polygamma functions can be expanded at large $M$ as $\psi^{(0)} (2^M) \approx \ln (2^M - 1)$ and $\psi^{(1)} (2^M) \approx 1/(2^M-1)$. (In the standard mean-field limit, $M \to \infty$, the ``zeroth-order'' approximation gives the exact result to the leading order.\cite{paperI} The number of local states accessible to the system, $2^M$, becomes so large that the probability that two replicas both having zero overlap with the reference configuration fall in the same state is extremely low, so that the naive guess $q_{ab} = p_a p_b$ is essentially correct.)

We go beyond the zeroth order by evaluating the $\eta_{ab}^i$'s through a $1$-RSB variational ansatz and a restriction to specific patterns of the $p_a^i$'s. Due to the absence of geometry in a fully connected lattice, we use for specific patterns random configurations of the $p_a^i$'s. Specifically, we set the overlap with the reference equilibrium configuration for all constrained replicas to be $1$ on the first $cN$ sites ({\it i.e.}, $\forall a$, $p_a^i = 1$ for $i=1,\ldots,cN$) and $0$ on all the other $(1-c)N$ sites ({\it i.e.}, $\forall a$, $p_a^i = 0$ for $i=cN+1,\ldots,N$). The difference with the computation made in paper~I is that now, instead of computing the partition function of the constrained system as a function of $c$ (essentially) exactly, we perform a variational calculation by considering a 1-RSB ansatz for the matrices $q_{ab}^i$ on the sites where $p^i=0$: More precisely, we divide the replicas $a=1, \ldots, n$ in $n/m$ blocks of $m$ replicas and set $q_{ab}^i = 1$ if $a$ and $b$ belong to the same block and zero otherwise (note that $m=1$ gives back the ``zeroth-order'' approximation).

As in Sec.~\ref{sec:periodic_cluster}, the minimization of the $1$-replica action with respect to the variational parameter $m$ yields $\Delta {\cal S}_1 (c)/N$, the correction to the ``zeroth-order'' part of the $1$-replica action due to the fluctuations of the $q_{ab}^i$ (see Appendix~\ref{app:REMFC-S1} for details):
\begin{equation} 
\label{eq:DS1_REMFC}
\frac{\Delta \mathcal{S}_1 (c)}{N} 
=  \frac{M (1 - c)}{8} \left( \beta \sqrt{1+c} - \beta_K^{(0)} \right)^2 \qquad \textrm{for}~~\beta > 
\beta_K^{(0)} \, ,
\end{equation}
where $\beta_K^{(0)} = \sqrt{8 [ \psi^{(0)} (2^M) + \gamma]/M}=1/T_K^{(0)}$ and $T_K^{(0)}$ is the thermodynamic glass transition (RFOT) temperature obtained at the ``zeroth-order'' [see Appendix~\ref{app:REMFC-ann}]. (Note that $\beta_K^{(0)} \to \sqrt{8 \ln 2}$ for $M \to \infty$, which is the exact value in the standard mean-field limit.) In the following we will focus on the case $\beta \ge \beta_K^{(0)}$, {\it i.e.}, $T\le T_K^{(0)}$. A discussion of the situation $\beta < \beta_K^{(0)}$ is presented in Appendix~\ref{app:REMFC-S1}.

The strategy already followed in Sec.~\ref{sec:periodic_cluster} consist in finding a translationally invariant theory for the $1$-replica action with an effective Hamiltonian that  allows one to reproduce the result in Eq.~(\ref{eq:DS1_REMFC}) by means of an effective external source (chemical potential) and effective $2$- and multi-body interactions:
\begin{equation} 
\label{eq:Heff_REMFC-expansion} 
\Delta \mathcal{S}_{1,\rm eff} [p]= \textrm{cst} -
\mu \sum_i p_i  - \sum_{q=2}^\infty \frac{w_q}{q! N^{q-1}} \sum_{i_1, \ldots, i_q \neq} p_{i_1} \cdots p_{i_q}   \, ,
\end{equation}
which for the specific (random) pattern of the $p^i$'s chosen above simply gives $\Delta \mathcal{S}_{1,\rm eff} (c)/N = {\rm cst} -\mu c - \sum_{q=2}^{\infty} (w_q/q!) c^q$. After expanding Eq.~(\ref{eq:DS1_REMFC}) in powers of $c$ around $c=0$, one immediately finds the values of the effective coefficients, $\mu$ and $w_q$, by a term by term identification with Eq.~(\ref{eq:Heff_REMFC-expansion}) in the limit $N\to \infty$. The calculation is detailed in Appendix~\ref{app:REMFC-S1}. 

In the present case where we start from a ``low-temperature'' approximation scheme and study a mean-field (fully connected) model, we are interested in a region near but below the thermodynamic glass transition (RFOT). As a result, the arguments used in paper~I to justify a truncation of the expansion in $c$, {\it i.e.}, that the configurations of the $p_a^i$'s dominating the thermodynamics have a small concentration $c$ of sites where $p_a^i=1$, and that one can reproduce the shape of $\mathcal{S}_1 (c)/N$ with a few monomials
are no longer valid. One therefore needs in principle all the coupling constants $w_q$ to reconstruct the exact behavior of $\Delta \mathcal{S}_1 (c)/N$.  We nonetheless find that taking into account a finite number of terms, say, $5$ to $10$, provides a reasonably good approximation: see Fig.~\ref{fig:cfr_REMFC}. In practice, we will only keep interactions up to the $10$-body term in the $p^i$'s variables, {\it i.e.}, with the explicit expressions of the effective coefficients (given here only up to the $4$-body term), 
\begin{equation} 
\label{eq:S1_fin_REMFC}
\mathcal{S}_1 [p] \approx 
\frac{M \beta \beta_K^{(0)}}{8} \sum_i p^i
- \frac{M \beta}{4N} \bigg ( \beta - \frac{5 \beta_K^{(0)}}{8} \bigg)  
\sum_{i \neq j} p^i p^j
- \frac{3 M \beta \beta_K^{(0)}}{64 N^2} \sum_{i,j,k \neq} p^i p^j p^k 
 + \frac{13 M \beta \beta_K^{(0)}}{512 N^3} \sum_{i,j,k,l \neq} p^i p^j p^k p^l + \ldots \, .
\end{equation}
The $4$-body interaction obtained above for the fully connected model is the counterpart of the link-link infinite-range interaction found for the effective theory of glass-forming liquids in finite dimensions [see Eq.~(\ref{eq_cum1_effect})], but it is now a bulk term, due to the fully-connected nature of the microscopic model under study.

\begin{figure}
\includegraphics[scale=0.36]{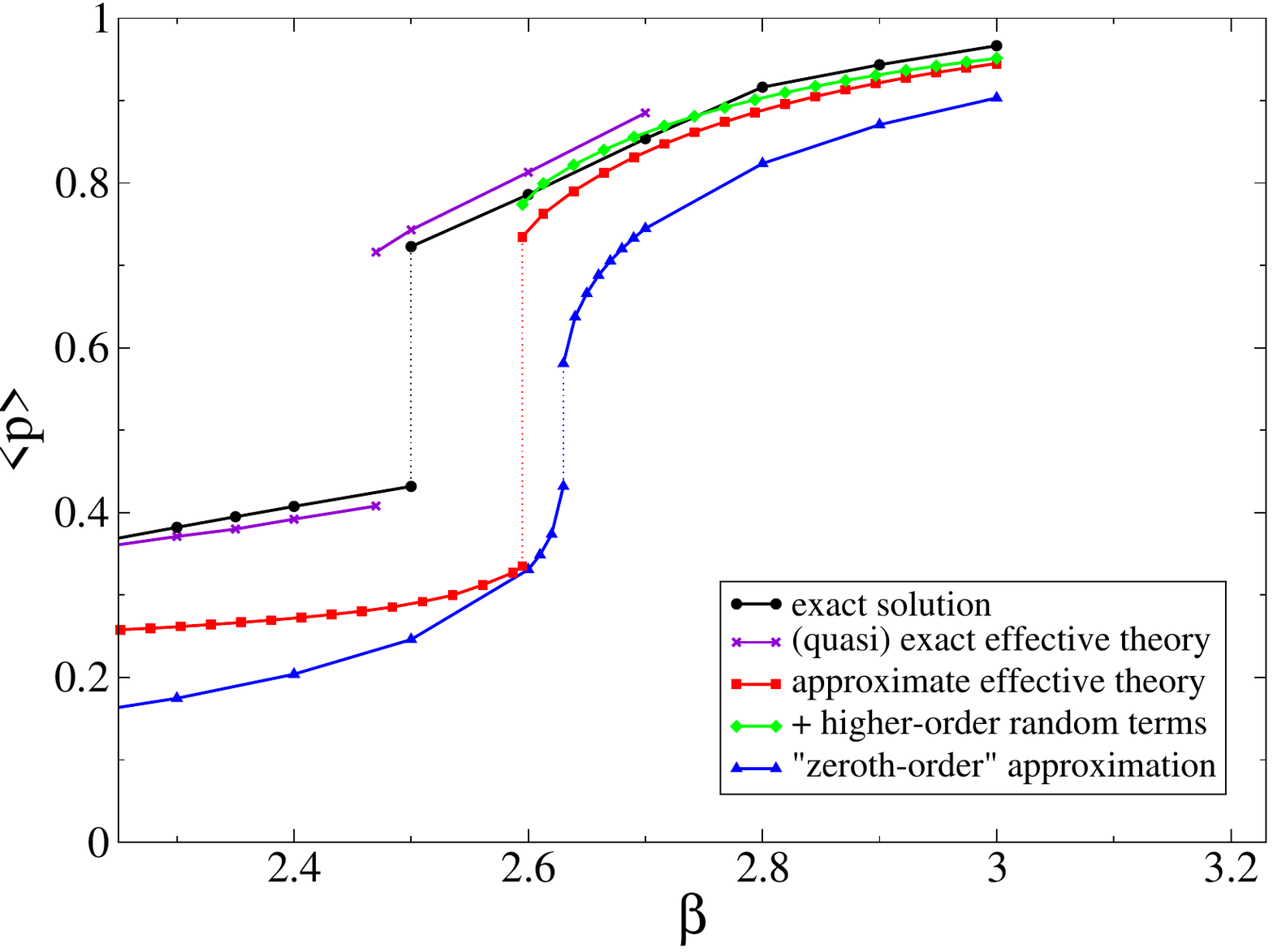}
\caption{Comparison between the approximate effective theory and the exact result for the fully connected $2^M$-KREM with $M=3$: Mean overlap $\langle p \rangle$ with a reference configuration versus $\beta=1/T$. The black curve (circles) represents the values of $p_0$ and $p_1$ given by the exact solution of the model (called $q_0$ and $q_1$ in App.~A1 of paper~I). The violet curve (crosses) corresponds to the results given by the (quasi) exact effective theory of the model obtained in paper~I: The effective Hamiltonian is exact at and above $T_K$ but is somewhat approximate below $T_K$ because of our use of the annealed approximation for handling the random energies (see paper~I). The red curve (squares) corresponds to the prediction of the approximate effective Hamiltonian, where we neglect the third cumulant of the random field and of the covariance between random field and random bond: The solution is derived in App.~B of paper~I.  The discrepancy between the two curves on the low-$\beta$ side of the transition is due to high-temperature terms that are not included in the low-$T$ approximation scheme. The roughly constant shift actually corresponds to a term that can be calculated in the limit $T\to \infty$: The exact value of the overlap approaches $\langle p \rangle \to 1/2^M$ for $T \to \infty$ ({\it i.e.}, the probability that two randomly chosen configuration on a given site are in the same state), while the overlap goes to zero when computed by means of the effective random-field + random-bond Ising model. This explains the discrepancy of about $1/2^M \approx 0.11$ on the low-$\beta$ side of the transition. On the high-$\beta$ side, the green curve (diamonds) is the prediction of the approximate effective disordered theory when one accounts for higher-order correlations in the disorder (see Appendix~B of paper~I). Finally, the blue curve (triangles) corresponds to the ``zeroth-order'' approximation with no account of the glassy  correlations, Eq.~(\ref{eq:Heff-REMFC-ann}).} 
\label{fig_REM}
\end{figure}

We now turn to the computation of the second cumulant of the effective action and divide the $n$ replicas into two groups of $n_1$ and $n_2$ replicas respectively. We take the same overlap pattern as the one described in paper~I, with $c_1 N$ sites where $p_1^i=1$ and $p_2^i=0$, $c_2 N$ sites where $p_1^i=0$ and $p_2^i=1$, $c_{12} N$ sites where $p_1^i = p_2^i = 1$, and $c_0 N$  sites where $p_1^i = p_2^i = 0$ (with $c_0 = 1-c_1-c_2-c_{12}$). The second cumulant can be computed by keeping only the terms of order $n_1 n_2$ in the expression of the effective action and by taking the limit $n_1 , n_2 \to 0$ in the end. The calculation, which is shown in full detail in Appendix~\ref{app:REMFC-S2}, gives
\begin{equation} \label{eq:S1S2_REMFC}
\begin{split}
{\cal S}_2 [p_1,p_2] &\approx  \frac{M \beta^2}{4 N} \sum_{i \neq j} p_1^i p_1^j p_2^i p_2^j 
- \bigg ( \frac{\beta}{\beta_K^{(0)}} \bigg)^{\!2} \Big[ \psi^{(1)} (2^M) + \frac{\pi^2}{6} \Big]
\bigg[  \sum_{i} p_1^i  p_2^i 
+ \frac{1}{2N} \sum_{i\neq j} p_1^i p_2^i ( p_1^j  + p_2^j) \\
& \qquad \qquad \qquad  \qquad \qquad \qquad
+  \frac{1}{4N^2} \sum_{i ,j, k\neq} p_1^i p_1^j p_2^i p_2^k 
-  \frac{1}{8N^2} \sum_{i ,j, k\neq} p_1^i p_2^i ( p_1^j  p_1^k + p_2^j p_2^k ) + \ldots
 \bigg] \, ,
 \end{split}
\end{equation}
where the ellipses stand for higher-order terms. Note that the terms in the last square bracket that appear in addition to the first one, $\sum_{i} p_1^i  p_2^i$, are absent in the ``zeroth-order'' replicated action in Eq.~(\ref{eq:S1S2-REM}) and therefore contribute to the corrections to the latter. In the large-$M$ limit their contribution is negligible, and they also turn out to also be unimportant for $M=3$ (see Fig.~\ref{fig_REM}), giving credit to the approximation used in Sec.~\ref{sec:approximations} for obtaining the second cumulant. Finally, higher-order cumulants can be calculated along the same lines.

As in Sec.~\ref{sec:effective_theory}, we can infer from the above $1$-replica and $2$-replica (and higher-order if needed) actions the approximate effective disordered theory for the fully connected $2^M$-KREM. We straightforwardly obtain
\begin{equation} 
\label{eq:Heff-final-tauKREM}
\begin{split}
\beta {\mathcal H}_{\rm eff}[p] &= -  \sum_i (\tilde{\mu} + \delta \tilde{\mu}_i) p^i - \frac 12\sum_{i, j \neq} \Big ( 
	\frac{w_2}{N}+\frac{\delta w_{2,ij}}{\sqrt N} \Big ) p^i p^j - \frac{w_3}{3!N^2} \sum_{i, j,k \neq} p^ip^jp^k %\\ & \qquad 
	- \frac{w_4}{4!N^3} \!\!\! \sum_{i,j,k,l\neq} p^ip^jp^kp^l  + \ldots\, ,
\end{split}
\end{equation}
where the nonrandom coefficients are given by 
\begin{equation} 
\label{eq:coefficientsKREM}
\begin{aligned}
\tilde \mu & 
= - \frac{M \beta \beta_K^{(0)}}{8}  \, , 
\\
w_2  & 
=  \frac{M \beta}{2} \bigg ( \beta - \frac{5 \beta_K^{(0)}}{8} \bigg) \, , 
\\
w_3  & 
= \frac{9 M \beta \beta_K^{(0)}}{32} \, , 
\\
w_4  & 
= -\frac{39M \beta \beta_K^{(0)}}{64} \, ,
\end{aligned}
\end{equation}
and where the random variables, $\delta \tilde{\mu}_i$ and $\delta w_{2,ij}$, have a zero mean ($\overline{\delta \tilde{\mu}_i} = \overline{\delta w_{2,ij}} = 0$) and variances given by
\begin{equation} 
\label{eq_variances-tauKREM}
\begin{split}
\overline{\delta \tilde{\mu}_i  \delta \tilde{\mu}_j} & = - \bigg ( \frac{\beta}{\beta_K^{(0)}} \bigg)^{\!2} \Big[ \psi^{(1)} (2^M) + \frac{\pi^2}{6} \Big]\delta_{ij} \, , \\
\overline{\delta w_{2,ij}\delta w_{2,kl}} & = \frac{M \beta^2}{2} (\delta_{ik}\delta_{jl}+\delta_{il}\delta_{jk}) \, \\
\overline{\delta \tilde{\mu}_i \delta w_{2,jk}} & = - \bigg ( \frac{\beta}{\beta_K^{(0)}} \bigg)^{\!2} \Big[ \psi^{(1)} (2^M) + \frac{\pi^2}{6} \Big]\frac{(\delta_{ij}+\delta_{ik})}{\sqrt N} \, .
\end{split}
\end{equation}
There are also random $3$, $4$, and higher-order interaction terms that we do not take into account in the simplified approximation. There is an unpleasant feature in the above expressions: The variance of the random ``chemical potential'' $\delta \tilde \mu_i$ is negative, which is unphysical. This variance however does not scales with $M$ as the variance of the two-body interactions and is already small even for $M=3$. (The same is already true for the zeroth-order result: see Appendix~\ref{app:REMFC-ann}.) This hitch is not an intrinsic problem of the mapping itself but is an artifact of the low-temperature approximations used here to derive the effective theory. It disappears when the the partition function of the ($2$-replica) constrained systems is computed exactly, as done in paper~I: In fact, in this case we find that the variance of the random chemical potential is also exponentially small in $M$ and subleading with respect to the random bond term, but it is positive, $\overline{\delta \tilde{\mu}_i  \delta \tilde{\mu}_j}  = M \beta^2/2^{M+1}$.

We now go from the overlap variables, $p^i = 0,1$, to the Ising ones, $\sigma^i = \pm 1$, via the relation $p^i = (\sigma^i + 1)/2$. We obtain from Eqs. (\ref{eq:Heff-final-tauKREM}-\ref{eq_variances-tauKREM}) the following effective disordered Hamiltonian (truncated up to
the $4$-body interaction term in the $\sigma^i$'s variables):
\begin{equation} 
\label{eq:Heff-finalKREM}
\begin{split}
	\beta {\mathcal H}_{\rm eff}[\sigma] &=\mathcal S_0 - \sum_i \left ( H + \d h_i\right) \sigma^i  - \frac 12\sum_{i, j \neq} \Big( \frac{J_2}{N}+\frac{\delta J_{2,ij}}{\sqrt N} \Big) \sigma^i \sigma^j - \frac{J_3}{3!N^2} \sum_{i, j,k \neq} \sigma^i\sigma^j\sigma^k\\ 
& \qquad - \frac{J_4}{4!N^3} \!\!\! \sum_{i,j,k,l\neq} \sigma^i\sigma^j\sigma^k\sigma^l  + \ldots\, ,
\end{split}
\end{equation}
where the coupling constants are given by $H= \tilde{\mu}/2 + \sum_{m\geq 2} w_m/[(m-1)! \, 2^m]$ and $J_n = \sum_{m\geq n}  w_m/[(m-n)! \, 2^m]$,  %$J_n = n! \sum_{m=n}^{m_{\rm max}} \binom{m}{n} w_m/(m! 2^m)$ 
which after truncating the sums to $m_{\rm max} = 10$ gives
\begin{equation} 
\label{eq:coefficients_REMFC}
\begin{split}
%H & =  \frac{M \beta}{8} \left( \beta - \frac{265}{256} \beta_K^{(0)} \right) \, , \\
	H & \approx  \frac{M \beta}{8} \left( \beta - 0.986 \beta_K^{(0)} \right) \, , \\
%J_2 &= \frac{M \beta}{8} \left( \beta - \frac{127}{256} \beta_K^{(0)} \right) \, ,\\
J_2 & \approx \frac{M \beta}{8} \left( \beta - 0.4423 \beta_K^{(0)} \right) \, ,\\
%J_3 & =  - \frac{3 M \beta \beta_K^{(0)}}{1024} \, , \\
J_3 &  \approx  0.0144 M \beta \beta_K^{(0)} \, , \\
%J_4 & =  - \frac{39 M \beta \beta_K^{(0)}}{1024} \, , 
J_4 & \approx  - 0.0145 M \beta \beta_K^{(0)} \, .
\end{split}
\end{equation}
The random field $\delta h_i$ and the random coupling $\delta J_{2,ij}$ have zero mean and variances given by
\begin{equation} 
\label{eq_variances-spinKREM}
\begin{split}
\overline{\delta h_i  \delta h_j} & = \bigg [ \frac{M \beta^2}{32} - \frac{1}{4} \bigg( \frac{\beta}{\beta_K^{(0)}} \bigg)^{\!2} 
	\Big [ \psi^{(1)} (2^M) + \frac{\pi^2}{6} \Big ] \bigg ]\delta_{ij} + \frac{M \beta^2}{32 N} - \frac{1}{4N} \bigg( \frac{\beta}{\beta_K^{(0)}} \bigg)^{\!2} 
        \Big [ \psi^{(1)} (2^M) + \frac{\pi^2}{6} \Big ]\, , \\
\overline{\delta J_{2,ij}\delta J_{2,kl}} & = \frac{M \beta^2}{8} (\delta_{ik}\delta_{jl}+\delta_{il}\delta_{jk}) \, \\
\overline{\delta h_i \delta J_{2,jk}} & =  \frac 18\bigg [\frac {M\beta^2}{4}- \bigg ( \frac{\beta}{\beta_K^{(0)}} \bigg)^{\!2} \big[ \psi^{(1)} (2^M) + \frac{\pi^2}{6} \big] \bigg ]\frac{\delta_{ij}+\delta_{ik}}{\sqrt N} \, .
\end{split}
\end{equation}
As before, $\mathcal S_0$ is a random term that does not depend on the Ising variables. The requirement of a positive definite variance matrix imposes that $\overline{\d h_i^2}>0$, which in turn implies that $M\gtrsim2$.

When applied to the fully connected $2^M$-KREM, the low-temperature variational approximation scheme of Sec.~\ref{sec:approximations} again yields an effective random-field + random-bond Ising Hamiltonian with multi-body interactions that has the same structure as the exact effective theory derived in paper~I, with similar expressions of the effective parameters. Because it is defined on a fully connected lattice the approximate  disordered effective theory (as well as the exact one) can be solved analytically. This is done in Appendix~B of paper~I for the simpler case where we neglect the cross-correlation term between the random-field and the random-bond variables and keep only the diagonal terms of the variances.

As already discussed, the effective disordered theory has a transition for a uniform source $H_c$ when the ferromagnetic tendency is strong enough relative to the strength of the disorder. There is then a line of first-order transition terminating in a critical point that is in the RFIM universality class. Changing the temperature in the $2^M$-KREM amounts to following a given trajectory in the disordered Ising model, and the thermodynamic glass transition of the former corresponds to crossing the first-order transition line in the latter (the jump in the mean overlap $\langle p \rangle$ with a reference configuration is equivalent to a jump in the magnetization $m$). The thermodynamic glass transition therefore exists if the trajectory in the disordered Ising model crosses a coexistence line below the critical point. Because the $3$-body coupling $J_3$ is very small, as are the effects of the third cumulant of the random field and of the covariance between random field and random bond, the critical point and the coexistence line take place for $H\approx 0$.

In Fig.~\ref{fig_REM} we compare for the fully connected $2^M$-KREM with $M=3$ the prediction for the mean overlap $\langle p \rangle=(1+m)/2$ from the approximate effective disordered theory derived by means of our variational approximation scheme (where we neglect the third cumulant of the random field and the covariance between random field and random bond) to the exact temperature dependence obtained in paper~I.  As can be seen, the agreement is good. The approximate treatment predicts a thermodynamic glass transition with a jump of $\langle p \rangle$ that reproduces quite well the exact behavior. The transition temperature for the approximate effective model is at $\beta \approx 2.595$ ($H_c \approx 0.0022$), {\it i.e.}, $T_K\approx 0.385$, slightly below but quite close to the exact value, $T_K\approx 0.40$. The discrepancy found on the low-$\beta$/high-$T$ side of the transition stems from the low-$T$ nature of the variational approximation scheme and essentially corresponds to a constant term in the mean overlap that can be calculated in the limit where $T \to \infty$. The residual discrepancy between the exact and the approximate descriptions is mostly due to the fact that the effective theory has been truncated at the level of the $4$-body interaction term. In addition, we have plotted the prediction obtained by including higher-order correlations in the effective disorder, but one can see that the effect is extremely small. We also find that the zeroth-order description gives a rather poor account and that introducing the effect of the glassy correlations induced by pinning the overlaps (the mean-field analog of the point-to-set correlations) is crucial to improve the prediction of the effective theory.

All in all this comparison shows that the variational treatment used to derive the effective theory for the fluctuations of the overlap with an equilibrium reference configuration at low temperature (below the mean-field RFOT, $T< T_K^{MF}$) provides a good description of glassy systems in the fully connected limit. This gives credit to its use as a general approximation scheme in the context of finite-dimensional systems including glass-forming liquids.

\subsection{Approximate effective theory for the finite-dimensional lattice version of the $2^M$-KREM}
\label{sec:finiteKREM}

In this Section we consider the finite-dimensional lattice version of the $2^M$-KREM. We apply our low-temperature variational approximations to construct the approximate effective theory which describes the fluctuations of the overlap with an equilibrium reference configuration in this model. By doing this we want to check whether the resulting effective theory  for this finite-dimensional glass model has the same structure as the one found above for supercooled liquids. In addition, the model has been recently investigated through a real-space renormalization group method,\cite{Angelini} and it is thus interesting to compare the predictions of the two approaches.

Below we will present the main results only. The calculations are carried out in full detail in Appendix~\ref{app:REM-Kac}. The model is defined on a $d$-dimensional hyper-cubic lattice, very similarly to its fully connected counterpart. On each site $i$ we define a state variable $\mathcal{C}_i$, which can take $2^M$ possible value, $\mathcal{C}_i=1, \ldots, 2^M$. Two neighboring sites $i$ and $j$ interact via a coupling $E_{\langle i, j \rangle} (\mathcal{C}_i, \mathcal{C}_j)$ that is an i.i.d.~Gaussian random variable, such that $\overline {E_{\langle i, j \rangle} (\mathcal{C}_i, \mathcal{C}_j)} = 0$ and $\overline {E_{\langle i, j \rangle} (\mathcal{C}_i, \mathcal{C}_j) E_{\langle i, j \rangle} (\mathcal{C}_i^\prime, \mathcal{C}_j^\prime)} = M \delta_{\mathcal{C}_i,\mathcal{C}_i^\prime} \delta_{\mathcal{C}_j,\mathcal{C}_j^\prime}$. The Hamiltonian of the system simply reads 
\begin{equation}
\mathcal{H} = \frac 1{\sqrt{d}} \sum_{\langle i,j\rangle} E_{\langle i, j \rangle} (\mathcal{C}_i, \mathcal{C}_j)\,,
\end{equation}
where the factor $1/\sqrt{d}$ is introduced to have a well-defined limit when $d \to \infty$.

As for the fully connected case, the first two approximations of Secs.~\ref{sec:2-state_variables} and \ref{sec:constraints} are exactly satisfied by the model. Applying then our variational approximation scheme described in Secs. ~\ref{sec:variational}, \ref{sec:periodic_cluster}, and \ref{sec:2replica_approx} we obtain an effective theory having exactly  the same form as that derived for glass-forming liquids near the putative $T_K$ [see Eq.~(\ref{eq:Heff-final})]. The resulting Hamiltonian corresponds to a random-field + random-bond Ising model with antiferromagnetic power-law decreasing pairwise interactions and an infinite-range $4$-body link-link coupling. The expressions of the parameters of the effective theory, uniform field $H$, couplings $J_2$, $\tilde J_2$, and $J_4$, as well as the variances of the random variables are given in Eq.~(\ref{eq:coefficients_REMKac}) of Appendix~\ref{app:REM-Kac}. As for the effective theory of glass-forming liquids, we find that the mean-field configurational entropy, which is essentially proportional to $H$, is renormalized by a positive factor, implying that the thermodynamic glass transition temperature is lowered with respect to its mean-field value. (However, it is easy to check that the for $d \to \infty$ the expressions in Eq.~(\ref{eq:coefficients_REMKac}) give back the known mean-field result.)

\begin{figure}
\includegraphics[scale=0.36]{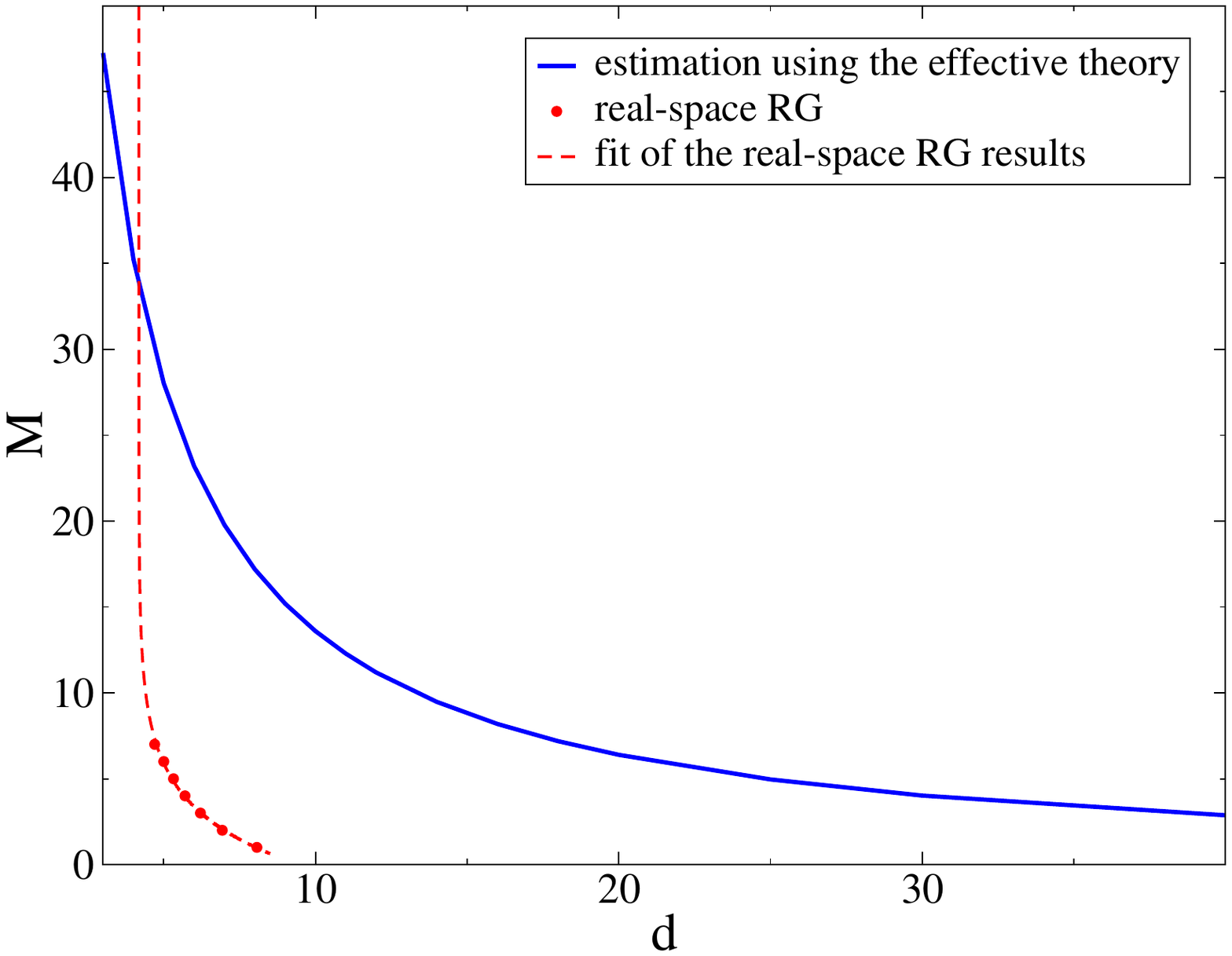}
\caption{Estimate for the minimum value of $M$  for the thermodynamic glass transition (RFOT) in the finite-dimensional lattice version of the  $2^M$-KREM. The blue curve shows the result of the effective random-field Ising theory: A RFOT exists only for $M$ above this curve. The red circles correspond to the values of $d$ and $M$ for which an ideal-glass fixed point is found within the real-space RG analysis of Ref.~[\onlinecite{Angelini}], and the red dashed line is an exponential fit of these results which yields a lower critical dimension $d_L \approx 4.18$.}
\label{fig:estimate_rem}
\end{figure}

The effective disordered Hamiltonian can be used to study the existence of a thermodynamic glass transition in the $2^M$-KREM and can be conveniently studied by computer simulation. Here we rather provide a rough estimate of the minimum value of $M$ for which the transition persists by simplifying further the effective disordered Hamiltonian to that of the short-range RFIM by following the same procedure as in Sec.~\ref{sec:effective_theory-estimation}. The results are listed in the table of Appendix~\ref{app:REM-Kac-estimate} and are plotted in Fig.~\ref{fig:estimate_rem}. For comparison, we also plot the results previously obtained through a real-space renormalization group (RG) analysis.\cite{Angelini} In both cases one finds a monotonous decrease of the minimum value of $M$ with $d$, but there are significant differences between the two sets of results.  One is in the value of the lower critical dimension for the existence of the thermodynamic glass transition ($d_L = 2$ from the RFIM-like effective theory {\it vs} $d_L \approx 4.18$ from the real-space RG analysis) and the other in the behavior at large dimensions: The real-space RG approach predicts that a much smaller value of $M$ is needed to have a thermodynamic glass transition compared to the estimate obtained from our effective theory. Although both methods are approximate and one cannot {\it a priori} tell whether one or the other is better, we note that the real-space RG procedure is not expected to provide accurate results in large $d$.

\section{Discussion and Conclusion}
\label{sec:conclusions}

In this work we have presented the derivation of a $2$-state ({\it i.e.}, Ising-like) disordered effective theory which describes the fluctuations of what is thought to be the relevant order parameter for glassy systems, {\it i.e.}, the overlap field with a random equilibrium configuration, close to the putative thermodynamic glass transition temperature. In the companion paper\cite{paperI} we have focused on archetypal mean-field models for the glass transition, in particular the
Random energy Model\cite{derrida} and its extension to a finite number of states, \cite{REM} the $2^M$-KREM, on a fully connected lattice. The effective Hamiltonian for these mean-field models can be worked out  (essentially) without any approximation. We have shown that in both cases the effective theory for the fluctuations of the overlap with a reference configuration is a random-field + random-bond Ising model. We have argued that this result is very general and should apply (possibly with some minor model-dependent adjustments) to any mean-field model in the ``universality class'' of structural glasses, {\it i.e.}, with a complex free-energy landscape characterized by a multitude of metastable states and two distinct glass transitions, a dynamical and a thermodynamical one.

In the present paper, we have shown that such an effective description in terms of a random-field + random-bond Ising model also applies to supercooled liquids and simpler models in finite dimensions close to, and below, the mean-field thermodynamic glass transition. Of course, the effective theory cannot be worked exactly in this case, and our derivation is based on a low-temperature variational approach. The new physics appearing in finite dimensions is the presence of point-to-set spatial correlations. This is what makes the derivation of the effective theory more involved than in the mean-field limit. A physical consequence of these correlations is the appearance of effective pair and multi-body interactions that decay as power laws with distance below the mean-field glass transition (RFOT). The generic effect of these additional interactions is to depress the thermodynamic glass transition temperature and, possibly, to change its behavior. Finite-dimensional fluctuations tend to increase the strength of the effective disorder, to reduce the ferromagnetic tendency of the interactions, and to move the external field playing the role of a renormalized configurational entropy further away from zero. 

The great merit of such a mapping to an effective disordered theory for Ising variables is that it is much easier to investigate than the original glass problem.  By having coarse-grained over some of the fluctuations we derive a theory in which understanding the role of the long-range and/or nonperturbative fluctuations is now feasible. The presence of a random-field type of disorder immediately tells us for instance that such fluctuations wash out the existence of a  thermodynamic glass transition in dimensions $d\leq 2$.\cite{footnote_2D} It also emphasizes the role of the disorder strength, which is associated with fluctuations of the local configurational entropy, and that of the diverging point-to-set correlations, which generate new effective interactions. 

As we have already pointed out, the mapping is only approximate for finite-dimensional glass-formers and one may then wonder what could go wrong in our derivation of the effective disordered theory? The two main assumptions that we have made are that the variational ``low-$T$'' approximations provide a good zeroth-order description and that one can truncate to a small number of terms both the effective description in terms of many-body interactions and the expansion in  cumulants of the effective disorder. We have checked these assumptions in the case of the exactly solvable fully connected $2^M$-KREM and found that they provide results in good agreement with the exact ones. Yet, this model is a mean-field one and, as a consequence,  the comparison does not address the validity of all aspects of the variational procedure. We have also noted that when the amplitude $c$ of the spatial gradient terms in the Ginzburg-Landau-like description of supercooled liquids in terms of overlaps [Eq. (\ref{eq_replica_action_wolynes})] is small, the standard deviation of the random couplings may dominate the nonrandom ferromagnetic value, and the physics of the disordered model leaves the realm of the random-field Ising model for that of an Ising spin glass in an external field, with a quite different phenomenology. This nonetheless appears to be only present in the limit of a very small amplitude ($c\ll 1$) and to be furthermore always absent in the other model studied, the $2^M$-KREM. It seems therefore fair to conclude that the present mapping does not provide any mechanism that would {\it generically} destroy the thermodynamic glass transition predicted at the mean-field level (at least for $d>2$). The existence or not of such a transition rather appears to be system-dependent, and its investigation thus requires a {\it quantitative} analysis.

Even taking now for established the generic form of the effective disordered theory for the thermodynamic glass transition, a quantitative derivation of the parameters entering in the theory is not an easy task for supercooled liquids. We have indicated some ways to estimate them by combining liquid-state theory and a low-temperature variational approximation scheme. A potentially powerful approach would be to use data obtained from numerical simulations of {\it finite-size} systems made of realistic $3$-dimensional glass-forming liquids  as input. Finite sizes have indeed the effect of suppressing spatial fluctuations and one expects that the effective disordered theory at the ``bare level'', {\it i.e.}, without having turned on the long-range and nonperturbative fluctuations, precisely describes such situations.\cite{extract} This then provides a means to determine the effective parameters of the theory from simulation data on the fluctuations of the overlap field in finite-size glass-forming systems. Rather small systems, as already studied,\cite{Stevenson,Berthier2,Berthier3,Parisi_Bea} would be sufficient.

One should finally reiterate that our approach in terms of an effective disordered Ising model is not meant to provide a full solution to the glass transition problem, as there is no obvious mapping from the dynamics of the glass-forming liquid to that of the effective theory. Yet it allows one to study the existence and properties of a thermodynamic glass transition, and it sheds light on the conditions under which such a transition can be destroyed. One could envisage a further, and rather ambitious, step to map also the dynamics by upgrading the formalism to a supersymmetric formulation of the dynamics, as for instance done by Rizzo in the vicinity of the (mean-field) dynamical transition,\cite{Rizzo} and trying to derive an effective dynamical disordered theory in this framework. However, we leave this for future work.

\begin{acknowledgments}
We acknowledge support from the ERC grant NPRGGLASS and the Simons Foundation grant on ``Cracking the Glass Problem'' (No. 454935, GB).
\end{acknowledgments}

\appendix

\section{Variational approximation for the effective Hamiltonian using a slab geometry}
\label{app:slab}

In order to assess the consistency of our procedure, it is important to check that the effective Hamiltonian found in Sec.~\ref{sec:effective_theory} by using the low-temperature approximations is robust enough when different geometries of the pattern of the $\tau_a^i$'s are considered. To this aim, we repeat in this appendix the steps of the approximation scheme proposed in Sec.~\ref{sec:approximations} for a slab geometry (instead of a periodic cluster pattern), taking the Ginzburg-Landau functional for the overlap fields in glass-forming liquids [Eq.~\ref{eq_replica_action_wolynes}] as a starting-point description. 

Here, we will only compute the $1$-replica component of the effective action. In the following we thus set $\tau_a^i = \tau^i$~$\forall a$ and keep only the terms of order $n$. We consider a $d$-dimensional hypercubic lattice of linear size $L_x$ along the $x$-direction and of linear size $L$ along the other directions. We set $\tau^i=0$ in a slab of width $\ell$ in the $x$-direction ({\it e.g.}, between $x=1$ and $x=\ell$) and $\tau^i=1$ outside the slab (see the right panel of Fig.~\ref{fig:sketch}). On the sites where $\tau^i = 0$ we set $\eta_{ab}^i$ to be of a $1$-RSB form, {\it i.e.}, we divide the replicas in
$n/m$ blocks of size $m \times m$ such that if $a,b$ are in the same block, then $\eta_{ab}^i = 1$, whereas if $a,b$ do not belong to the same block then $\eta_{ab}^i = 0$. The variational parameter $m$ must be fixed by minimizing the action. (Recall that $m=1$ gives back the ``zeroth-order'' approximation, $\tau_{ab}^i = \tau_a^i \tau_b^i$).

The gradient term involving the $\eta_{ab}^i$'s in Eq.~(\ref{eq:Seta}) is different from zero only on the links connecting a site with $\tau^i = 1$ and a site with $\tau^i = 0$. As a result,
\begin{displaymath}
\frac{c}{2}  \sum_{ab\neq} \sum_{\langle i,j \rangle}[(1-\tau^i)\eta_{ab}^i-(1-\tau^j) 
\eta_{ab}^j]^2 = n (m - 1) L^{d-1} c \, .
\end{displaymath}
Similarly, we have that
\begin{displaymath}
\begin{split}
\sum_{ab\neq} \sum_i (1-\tau^i)  \Big(s_c-\frac{u}{3} -c \sum_{j/i} \tau^j  \Big)\eta_{ab}^i 
& = n (m - 1) L^{d-1} \left[ \left(s_c-\frac{u}{3} \right) \ell - 2 c \right] \, , \\
- \frac{u}{3} \sum_{abc\neq}\sum_i (1-\tau^i)
\eta_{ab}^i\eta_{bc}^i\eta_{ca}^i & = - n (m - 1) (m - 2) \frac{ u L^{d-1} \ell}{3} \, .
\end{split}
\end{displaymath}
Putting all these terms together and taking the derivative with respect to $m$ leads to
\begin{displaymath}
m_\star = \left \{
\begin{array}{ll}
1 & {\rm ~~for~}
\ell \ge \ell_{PS} \, , \\
1 - \frac{3(c - \ell s_c)}{2 u \ell} & {\rm ~~for~}
\ell < \ell_{PS} \, ,
\end{array}
\right .
\qquad {\rm with~~~} 
\ell_{PS} = \left\{ 
\begin{array}{ll}
c/s_c & {\rm ~~for~} s_c \ge 0 \, , \\
\infty & {\rm ~~for~} s_c < 0 \, ,
\end{array}
\right .
\end{displaymath}
where $\ell_{PS}$ plays the role of the mean-field point-to-set correlation length. [Similarly to the case of the periodic pattern described in Sec.~\ref{sec:periodic_cluster}, we also find in the present case an upper bound on $c$ by ensuring that $m_\star$ is always positive: $c < (2 u)/3 + s_c$.]

After inserting the above results into Eq.~(\ref{eq:Seta}) we immediately find that the $1$-replica part of the effective action reads
\begin{equation} \label{eq:DS1-slab}
{\cal S}_1 %[\tau^i] 
= \frac{c}{2}\sum_{\langle i, j \rangle}(\tau^i-\tau^j)^2 + \sum_i s_c \tau^i 
+ \Delta {\cal S}_1 \, , {\rm ~~~~~with~~~~}
\Delta {\cal S}_1 = \left \{ 
\begin{array}{ll} 
0 & {\rm ~~for~} \ell \ge \ell_{PS} \, ,\\
\frac{3 L^{d-1}}{4u} \left ( \ell s_c^2 - 2 c s_c + \frac{c^2}{\ell} \right)
& {\rm ~~for~} 0 < \ell < \ell_{PS}  \, ,
\end{array}
\right .
\end{equation}
Since we consider the system only below the mean-field RFOT $T_K^{MF}$, we will focus in the following on the case where $s_c  \lesssim 0$, which corresponds to $\ell_{PS} = \infty$.

Our strategy now consists in checking that the approximate ansatz of a translationally invariant theory with an effective external field and effective $2$- and multi-body interactions as introduced in the main text allows us to reproduce the above result. More specifically, we consider a contribution to the $1$-replica effective action in the same form as in Eq.~(\ref{eq:Heff-1}), {\it i.e.},
\begin{displaymath} 
\Delta \mathcal S_{1,\rm eff} = -\mu_+ \sum_i \tau_i + \mu_- \sum_i (1 - \tau^i) + w_2 \sum_{\langle i, j \rangle}
\tau^i (1 - \tau^j) + \frac{1}{2} \sum_{i \neq j} W(|i-j|) \tau^i 
\tau^j + \frac{w_4}{L^d} \!\!\! \sum_{\langle i, i^\prime \rangle \neq \langle j j^\prime \rangle}
\!\!\! [\tau^i (1 - \tau^{i^\prime}) \tau^j (1 - \tau^{j^\prime})]_{\rm sym}
\, .
\end{displaymath}
For the slab geometry, the contributions from the external uniform field yield respectively $-L^{d-1}(L_x-\ell) \mu_+$ and $L^{d-1} \ell \mu_-$, while the nearest-neighbor coupling gives $L^{d-1} w_2$. The $4$-body link-link infinite-range interaction only gives a sub-leading contribution, $w_4 L^{d-2}$, for the geometry considered here and can be neglected in the thermodynamic limit (see also Sec.~\ref{sec:4body}). The shape of the pairwise interaction needed to reproduce  Eq.~(\ref{eq:DS1-slab}) turns out to be very similar but slightly different  than that found for the periodic pattern considered in Sec.~\ref{sec:periodic_cluster}: $W(x)$ vanishes at short distances ({\it e.g.}, for $x<a$) and $W(x) \approx \tilde{w}_2/x^{2d-1}$ for large $x$.

In the continuum limit, the interaction between two vertical planes with $\tau=1$ at distance $z$ is given by (in the following we specify to the $3$-$d$ case)
\begin{displaymath}
L^{d-1} \tilde{w}_2 \int_1^\infty \frac{2 \pi r \, {\rm d} r}{(r^2 + z^2)^{5/2}} = \frac{2 \pi \tilde{w}_2 L^{d-1}}{3(1+z^2)^{3/2}} \, .
\end{displaymath}
In consequence, the interaction between all the vertical planes with $\tau=1$ located to the left of the slab and all the vertical planes with $\tau=1$
located to the right of the slab is
\begin{displaymath}
\frac{2 \pi \tilde{w}_2 L^{d-1}}{3} \int_0^\infty {\rm d} z^\prime \int_{\ell + 1 + z^\prime}^{\infty} \frac{{\rm d} z }{(1+z^2)^{3/2}} \approx 
\frac{\pi \tilde{w}_2 L^{d-1}}{3 \ell} + \ldots \, .
\end{displaymath}
Finally, the self-interaction between all the sites with $\tau=1$ on the left of the slab (and, equivalently, on the right) gives a bulk term proportional to $L^{d-1}(L_x - \ell)$, which can be roughly estimated as
\begin{displaymath}
\frac{L^{d-1} (L_x - 1) \tilde{w}_2}{2} \int_a^\infty \frac{4 \pi r^2 {\rm d}r}{r^5} = \frac{\pi \tilde{w}_2 L^{d-1}(L_x-1)}{a^2} \, .
\end{displaymath}
Putting all these terms together, we obtain that
\begin{displaymath}
\Delta \mathcal S_{1,\rm eff} = L^{d-1} \Big[ (L_x - \ell) \Big( -\mu_+ + \frac{\pi \tilde{w}_2}{a^2} \Big) + \mu_- \ell + w_2
+ \frac{\pi \tilde{w}_2}{3 \ell} + \ldots \Big] \, .
\end{displaymath}
Hence, in order to reproduce the functional dependence of $\Delta {\cal S}_1$ given in Eq.~(\ref{eq:DS1-slab}), we
need to set $\mu_+ = 9 c^2/(4 u a^2)$, $\mu_- = 3 s_c^2 / (4 u)$, $w_2 = -3 c s_c/(2u)$, and $\tilde{w}_2 = 9 c^2/(4 \pi u)$.
The $1$-replica action thus reads
\begin{equation}
\label{eq_cum1_effect-slab}
\mathcal S_1[\tau^i] \approx \frac{c}{2}\sum_{\langle i,j \rangle}(\tau^i-\tau^j)^2 + \sum_i (s_c - \tilde{\mu}) \tau^i 
- w_2 \sum_{\langle i, j \rangle}
\tau^i \tau^j   
+ \frac{\tilde{w}_2}{2} \sum_{|i- j|>a}  \frac{\tau^i \tau^j}{\vert  r_i-r_j\vert ^{2d-1}}
+ \ldots
\, ,
\end{equation}
with 
\begin{equation} 
\label{eq:coefficients-slab}
\begin{aligned}
&\tilde{\mu}= -s_c+\frac{3(2 c d s_c + s_c^2 + 3 c^2/a^2)}{4u} \, ,\\&
w_2  = -\frac{3 c d}{2 u} s_c\, , \\&
\tilde{w}_2  = \frac{9 c^2}{4 \pi u} \, .
\end{aligned}
\end{equation}
(Note that the expressions of $\tilde{\mu}$ and $\tilde{w}_2$ are valid for $d=3$ only.) The form of the effective action is very similar to the one given in Sec.~\ref{sec:effective_theory-S1} for the periodic pattern. (The long-range link-link contribution cannot be determined from the slab geometry as it leads to subdominant contributions.) The functional dependence of the effective parameters in terms of the bare coupling constants is also remarkably similar: $w_2$ is exactly the same as in Sec.~\ref{sec:effective_theory-S1} and the effective external field $\tilde{\mu}$ only differs by a factor $-9 c^2/(4 u a^2)$.  There is a difference in the spatial decay of the pairwise antiferromagnetic effective interaction, which is found here to go as $1/r^{2d-1}$ while it goes as $1/r^{2d}$ for the periodic pattern. The interaction is however in both cases relatively short-ranged and the difference therefore does not seem to be significant.

In conclusion, our approximate procedure of determining the effective theory by matching a translationally invariant $1$-replica action with various choices of specific patterns of the overlap with the reference configuration appears to be quite robust with respect to this choice. Moreover, having performed this computation is important also for another reason, which goes as follows. Instead of Eq.~(\ref{eq:Heff-1}), another possibility to reproduce the functional dependence on $\ell$ and $\lambda$ of $\Delta {\cal S}_1$ in the case of the periodic cluster patterns, Eq.~(\ref{eq:DS1}) of Sec.~\ref{sec:periodic_cluster}, 
would be to use the following $1$-replica component of the effective Hamiltonian:
\begin{displaymath}
%\label{eq:Heff-1}
\Delta{\cal S}_{1,\rm eff}[\tau] \approx \mu \sum_i (1 - \tau^i) + w_2 \sum_{\langle i, j \rangle}
\tau^i (1 - \tau^j) %+ \frac{1}{2} \sum_{i \neq j} W(\vert  r_i-r_j\vert ) \tau^i \tau^j 
	+ \sum_{p=0}^\infty \frac{w_{4+p}}{L^{d+p}} \!\!\! \sum_{i_1 \neq \ldots \neq i_p \neq \langle i, i^\prime \rangle \neq \langle j j^\prime \rangle}
	\!\!\! \tau^{i_1} \cdots \tau^{i_p} [\tau^i (1 - \tau^{i^\prime}) \tau^j (1 - \tau^{j^\prime})]_{\rm sym}
\, ,
\end{displaymath}
with the same coupling constants as in Eq.~(\ref{eq:S1eff_couplings}), but with $\tilde{w}_2 = 0$ and $w_{4+p} = (-1)^p 3 c^2 / (4 u)$. This particular structure of the $1$-replica component of the effective Hamiltonian, without the scale-free pairwise interaction but with extra fully-connected $(p+4)$-body interactions between $p$ sites and two links, would indeed allow us to reproduce exactly Eq.~(\ref{eq:DS1}) at all orders. Nevertheless, it fails completely in the case of  the slab geometry, for which the presence of the pairwise interaction appears to be crucial to reproduce the functional form of $\Delta {\cal S}_1$ correctly.

\section{The fully connected $2^M$-KREM: Low-temperature variational approximations}
\label{app:REMFC}

This appendix is devoted to the analysis of the version of the REM with $2^M$ states (the $2^M$-KREM) on a fully connected lattice. We apply the low-temperature approximations developed for finite-dimensional liquids in the main text to this exactly solvable case already studied in paper~I by different methods.\cite{paperI} The goal is to assess the validity of these approximations. 

In order to construct the effective theory, we follow the procedure described in the main text (Sec.~\ref{sec:REMFC-check}). We consider $n+1$ replicas of the system and fix the overlap $\{ p_a^i \}$ of the replicas $a=1,\ldots,n$ with a given reference configuration. The starting point is the expression for the replicated action for the overlap $\{ p_a^i \}$ of the replica $a$, $a=1,\ldots,n$, with the reference replica $0$:
\begin{equation} 
\label{eq:Srep_KREM}
\begin{split}
e^{-\mathcal{S}_{\rm rep} [\{p_a\}]} &= \frac{1}{\overline{Z}} \overline{\sum_{\{ \mathcal{C}_i^\alpha \}} 
e^{ - \frac{\beta}{2\sqrt{N}} \sum_{i\neq j}\sum_{\alpha=0}^n 
E_{\langle i, j \rangle} (\mathcal{C}_i^\alpha, \mathcal{C}_j^\alpha)} \prod_{a,i} 
\delta_{p_a^i,\delta_{\mathcal{C}_i^0,\mathcal{C}_i^a}}}
= \frac{1}{\overline{Z}}  \sum_{\{ \mathcal{C}_i^\alpha \}} e^{\frac{M \beta^2}{8 N} \sum_{i\neq j} \sum_{\alpha,\beta=0}^n
\delta_{\mathcal{C}_i^\alpha,\mathcal{C}_i^\beta} \delta_{\mathcal{C}_j^\alpha,\mathcal{C}_j^\beta}}
\prod_{a,i} 
\delta_{p_a^i,\delta_{\mathcal{C}_i^0,\mathcal{C}_i^a}} \, .
\end{split}
\end{equation}
This expression corresponds to the so-called annealed approximation for handling the random energies (note however that the averages over the quenched disorder represented by the reference configuration are exactly handled). In principle the annealed approximation is only valid above the thermodynamic glass-transition (RFOT) temperature $T_K$ and at $T_K$. Yet, as discussed in paper~I (see in particular  Fig.~3 of paper~I), it gives reasonably good results also below but near $T_K$ for the fully-connected $2^M$-KREM, at least as far as the value of $\langle p \rangle$ is concerned. In any case, the situation in the glass phase below $T_K$ is not of key interest here.

%which is exact only above the thermodynamic glass transition but is expected to provide a good approximation even below. (This point will be further discussed.)

\subsection{``Zeroth-order'' approximation}
\label{app:REMFC-ann}

The ``zeroth-order'' approximation is obtained by setting $q_{ab}^i \equiv \delta_{\mathcal{C}_i^a,\mathcal{C}_i^b} = p_a^i p_b^i$ on all sites ({\it i.e.}, $\eta_{ab}^i = 0$). This leads to
\begin{equation} 
\label{eq:tau_annealed_REMFC}
\sum_{\alpha,\beta} \delta_{\mathcal{C}_i^\alpha,\mathcal{C}_i^\beta} \delta_{\mathcal{C}_j^\alpha,\mathcal{C}_j^\beta}
= 1 + n + 2 \sum_a p_a^i p_a^j + \sum_{a\neq b} q_{ab}^i q_{ab}^j 
= 1 + n + \sum_a p_a^i p_a^j + \sum_{a,b} p_a^i p_a^j p_b^i p_b^j \, ,
\end{equation}
The trace over the configurations $\{ \mathcal{C}_i^\alpha \}$ of Eq.~(\ref{eq:Srep_KREM}) can now be easily performed, as explained below. On each site $i$ the number of constrained replicas $a$ for which $p_a^i=0$ is given by $n - \sum_a p_a^i$. All these replicas must be in a configuration that is different than the reference one and each pair of replicas must be in different configurations. Therefore on site $i$ the number of configurations that are compatible with the ``zeroth-order'' approximation is given by
\begin{displaymath}
{{2^M-1} \choose {n - \sum_a p_a^i}} = 
\frac{\Gamma (2^M)}{\Gamma(2^M  - n + \sum_a p_a^i) \, \Gamma (1 + n - \sum_a p_a^i)} \, .
\end{displaymath}
Expanding the $\Gamma$-functions up to the second order in $n- \sum_a p_a^i$ (we are interested in the $1$ and $2$-replica components of the replicated action only) yields when $n\to 0$
\begin{equation}
\label{eq:ann_RMFC}
{{2^M-1} \choose {n - \sum_a p_a^i}}
 \approx \exp \Bigg \{ - \bigg( \frac{\Gamma^\prime (2^M )}{\Gamma (2^M)} + \gamma \bigg) \sum_a p_a^i 
 - \frac{1}{2} \bigg[\frac{\Gamma^{\prime \prime} (2^M )}{\Gamma (2^M )}  
 - \bigg( \frac{\Gamma^\prime (2^M )}{\Gamma (2^M)} \bigg)^2
 + \Gamma^{\prime \prime} (1) - \gamma^2 \bigg] \sum_{a,b} p_a^i p_b^i
\Bigg \} \, ,
\end{equation}
where $\gamma = - \Gamma^\prime (1)$ is the Euler constant. The expressions above can be rewritten in terms of the polygamma functions, defined as the logarithmic derivatives of the $\Gamma$-function: $\psi^{(m)} (z) = \textrm{d}^{m+1} \ln \Gamma (z) / \textrm{d} z^{m+1}$. Inserting Eqs.~(\ref{eq:tau_annealed_REMFC}) and~(\ref{eq:ann_RMFC}) into Eq.~(\ref{eq:Srep_KREM}) and using that $\psi^{(1)} (1) = \pi^2/6$ we obtain the effective action at the level of the ``zeroth-order'' approximation that is given in Eq.~(\ref{eq:S1S2-REM}) of the main text.

Eq.~(\ref{eq:S1S2-REM}) for the replicated action corresponds to the expansion in number of free replica sums associated with the cumulant expansion built from the following disordered Hamiltonian:
\begin{equation} 
\label{eq:disorder0-tau}
\beta {\cal H}_{\rm eff}^{(0)}[p] = -\sum_i (\tilde \mu^{(0)} + \d \tilde \mu_i) p^i 
- \frac 12\sum_{i \neq j} \Big( \frac{w_2^{(0)}}{N} + \frac{\delta \! w_{2,ij}}{\sqrt{N}} \Big)p^i p^j \, ,
\end{equation}
where
\begin{equation}
\begin{split}
\tilde \mu^{(0)} & = -\big [\psi^{(0)} (2^M) + \gamma \big]\, ,\\
w_2^{(0)} &= \frac{M \beta^2}{4} \, ,\\
\overline{\d \tilde \mu_i \d \tilde \mu_j}^{(0)} & = - \Big[ \psi^{(1)} (2^M) + \frac{\pi^2}{6} \Big ] \delta_{ij} \, , \\
\overline{\delta \! w_{2,ij} \delta \! w_{2,kl} }^{(0)} & = \frac{M \beta^2}{2} (\delta_{ik} \delta_{jl} + \delta_{il} \delta_{jk})\, ,\\
\overline{\d \tilde \mu_i \delta \! w_{2,jk} }^{(0)} & = 0\,,
\end{split}
\end{equation}
with $\overline{\d \tilde \mu_i}^{(0)} = \overline{w_{2,ij}}^{(0)}= 0$. Note the unphysical feature that the variance of the random chemical potential $\delta \tilde \mu_i$ is negative. This is a shortcoming of the approximation, which is cured when the $2$-replica component of the replicated action is computed (quasi) exactly, as done in paper~I. The value of the variance is however very small as it is exponentially suppressed in $M$ (it goes as $2^{-M}$ and is small already for $M=3$).

After introducing Ising spins variables in Eq.~(\ref{eq:disorder0-tau}) via the relation $p_a^i = (\sigma_a^i + 1)/2$,  we obtain the following disordered Hamilonian:
\begin{equation} 
\label{eq:Heff-REMFC-ann}
\beta {\cal H}_{\rm eff}^{(0)} [\sigma]=\mathcal S_0 -\sum_i (H + \d h_i) \sigma^i 
	- \frac{1}{2} \sum_{i \neq j} \Big( \frac{J_2}{N} + \frac{\delta \! J_{ij}}{\sqrt{N}} \Big)\sigma^i \sigma^j \, ,
\end{equation}
where $\mathcal S_0$ is a random term that does not depend on the Ising variables, the random bonds and random fields have a zero mean, $\overline{\d h_i}^{(0)} = \overline{\d J_{2,ij}}^{(0)} = 0$, and 
\begin{equation}
\label{eq_RFIMzero}
\begin{split}
H^{(0)} & = \frac{M \beta^2}{16}- \frac{\psi^{(0)} (2^M) + \gamma}{2}  \, ,\\
J_2^{(0)} &= \frac{M \beta^2}{16} \, ,\\
	\overline{\d h_i \d h_j}^{(0)} & =  \bigg (\frac{M \beta^2}{32} - \frac{1}{4} \big[ \psi^{(1)} (2^M) + \pi^2/6 \big ] \bigg ) \delta_{ij} + \frac{M \beta^2}{32 N} \, , \\
\overline{\delta \! J_{2,ij} \delta \! J_{2,kl} }^{(0)} & = \frac{M \beta^2}{8} (\delta_{ik} \delta_{jl} + \delta_{il} \delta_{jk})\, ,\\
\overline{\d h_i \delta \! J_{2,jk} }^{(0)} & = \frac{M \beta^2}{32}\frac {(\delta_{ij} + \delta_{ik})}{\sqrt N}\, .
\end{split}
\end{equation}
This disordered Ising model has a transition for an external field $H_c \approx 0$, which is the counterpart of the RFOT in the KREM. If what neglects the (small) effect of the cross-correlation between random fields and random couplings, the transition is exactly at $H_c^{(0)}= 0$, which corresponds to a RFOT  at  $\beta_K^{(0)} = \sqrt{8 ( \psi^{(0)} (2^M) + \gamma)/M}$. In the limit $M \to \infty$ one recovers $\beta_K^{(0)} \to \sqrt{8 \ln 2}$, which is the exact value in the standard mean-field limit.

%The disordered Ising model for generic parameters has a line of first-order transition terminating in a critical point at an external field $H_c^{(0)}\approx 0$ and a coupling $J_{2,c}^{(0)}$. The value of the external field at coexistence varies slightly with the external field due to the cross-correlation between the random bonds and the random fields. When the trajectory of the KREM in the phase diagram of the disordered Ising model [trajectory in which the parameters are related by Eq.~(\ref{eq_RFIMzero})] crosses the first-order transition (coexistence) line, this corresponds to the RFOT in the KREM at this zeroth order. It turns out that it exactly occurs at $H_c^{(0)}= 0$ in this case. The RFOT  temperature is thus $\beta_K^{(0)} = \sqrt{8 ( \psi^{(0)} (2^M) + \gamma)/M}$. In the limit $M \to \infty$ one recovers $\beta_K^{(0)} \to \sqrt{8 \ln 2}$, which is the exact value in the standard mean-field limit.

\begin{figure}
\includegraphics[scale=0.36]{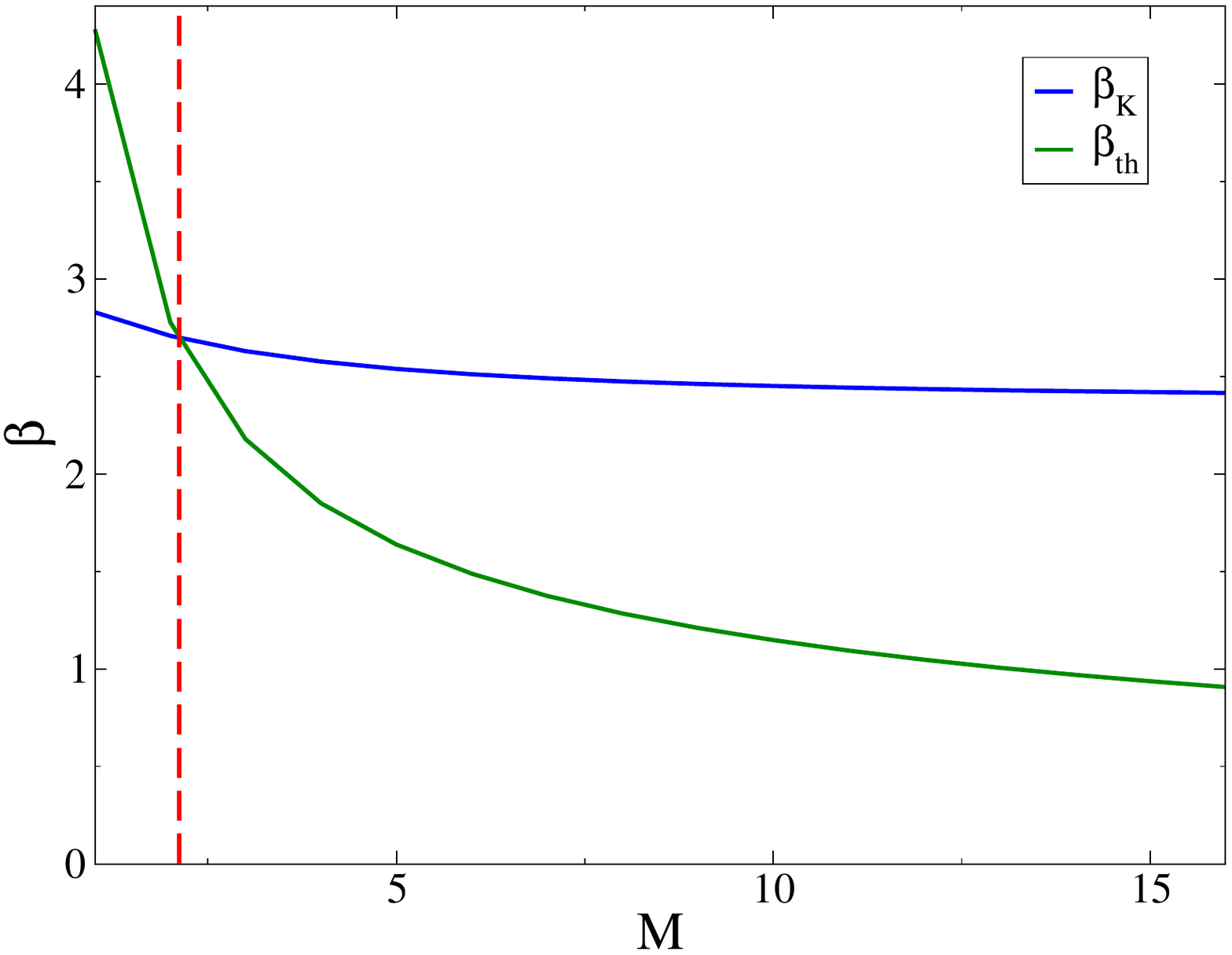}
\caption{Effective theory for the fully connected $2^M$-KREM: Transition line in the $\beta=1/T$-$M$ diagram for the ``zeroth-order'' approximation (when the covariance of the random field and random coupling is moreover neglected). The blue line marks the RFOT, $\beta_K^{(0)} = \sqrt{8( \psi^{(0)} (2^M) + \gamma)/M}$. The green line indicates the limit above which the approximate expression of the variance of the random field becomes negative, $\beta_{th}^{(0)} = \sqrt{8 (\psi^{(1)} (2^M) + \pi^2/6)/M}$. Below $M \approx 2.1$ where the two lines cross, the approximate description becomes meaningless. The asymptotic behavior for $M \to \infty$ is exact ($\beta_K^{(0)} \to \sqrt{8 \ln 2}$ and $\beta_{th}^{(0)} \to 0$).}
\label{fig_REM_SM}
\end{figure}

Since the variance of the random-field must be positive, {\it i.e}, $\delta h_i^2>0$, there is a threshold temperature, $T_{th}^{(0)}=1/\beta_{th}^{(0)}$, above which the approximate mapping to the effective disordered model is no longer justified: $\beta_{th}^{(0)} = \sqrt{8 (\psi^{(1)} (2^M) + \pi^2/6)/M}$. This is a consequence of the above noticed fact of a negative variance of the random chemical potential in the present approximation. The curves $\beta_K^{(0)}(M)$ and $\beta_{th}^{(0)}(M)$ are plotted in Fig.~\ref{fig_REM_SM}. For $M \lesssim 2$ one finds that $\beta_{th}^{(0)} > \beta_K^{(0)}$, {\it i.e.}, the approximate mapping is no longer valid. (In this region anyhow, the nature of the transition in the exact solution changes character and is no longer a RFOT.)  

\subsection{Variational approach beyond the ``zeroth-order'' approximation: The $1$-replica action}
\label{app:REMFC-S1}

We consider a random ``pinning'' configuration of the $p^i$'s as described in the main text. Due to the absence of geometry, this corresponds to setting the overlap with the reference configuration (for all constrained replicas) to be $1$ on the first $cN$ sites and $0$ on the other $(1-c)N$ sites. We introduce the matrices $q_{ab}^i \equiv \delta_{\mathcal{C}_i^\alpha,\mathcal{C}_i^\beta}$, %via the parametrization~(\ref{eq:parametrization}), 
and on all the sites $i$ where $p^i = 0$ [see Eq.~(\ref{eq:parametrization})]  we consider a 1-RSB ansatz: We divide the replicas $a=1, \ldots, n$ in $n/m$ blocks of $m$ replicas and set $q_{ab}^i = 1$ if $a$ and $b$  belong to the same block and zero otherwise (note that $m=1$ gives back the ``zeroth-order'' approximation).

After inserting this ansatz into Eq.~(\ref{eq:Srep_KREM}), and keeping only the terms of order $n$, the $1$-replica action can be expressed as 
\begin{equation} \label{eq:S1cm}
 \frac{\mathcal{S}_1 (c,m)}{N} = - \, \frac{M \beta^2}{8} \, \left[ m (1 - c^2) + 2 c^2 \right] 
- \big [\psi^{(0)}(2^M) + \gamma\big] \frac{1-c}{m} \, .
\end{equation}
Minimizing  with respect to $m$ then yields 
\begin{equation} \label{eq:mstar}
\begin{aligned}
& \textrm{if~~} \beta<\beta_K^{(0)} \qquad m_\star = \left \{
\begin{array}{ll}
1 & \textrm{if~} c<c_{\rm pin} (\beta) \, , \\
\frac{\beta_K^{(0)}}{\beta \sqrt{1+c}} & \textrm{if~} c \ge c_{\rm pin} (\beta) \, ,
\end{array}
\right .
\qquad 
\textrm{with~~} c_{\rm pin} (\beta) = \bigg( \frac{\beta_K^{(0)}}{\beta} \bigg)^{\!2} - 1 \, , \\
& \textrm{if~~} \beta \ge \beta_K^{(0)} \qquad m_\star = \frac{\beta_K^{(0)}}{\beta \sqrt{1+c}} \, ,
\end{aligned}
\end{equation}
where $\beta_K^{(0)}$ is the inverse of the ``zeroth-order'' critical temperature and the ``pinning'' concentration $c_{\rm pin} (\beta)$ (or rather its inverse) plays the role of the (point-to-set) pinning length $\ell_{\rm pin}$ found for finite-dimensional systems. After inserting the value of $m_\star$ into Eq.~(\ref{eq:S1cm}), one can easily compute the corrections to the zeroth-order description of the $1$-replica action due to the fluctuations of the overlaps $q_{ab}^i$ as $\Delta \mathcal{S}_1 (c) = \mathcal{S}_1 (c, m_\star) - \mathcal{S}_1 (c,1)$, with
\begin{displaymath} 
\begin{aligned}
& \textrm{if~~} \beta<\beta_K^{(0)} \qquad \frac{\Delta \mathcal{S}_1 (c)}{N} = \left \{
\begin{array}{ll}
0 & \textrm{~~if~} c<c_{\rm pin} (\beta) \, ,\\
\frac{M (1 - c)}{8} \left( \beta \sqrt{1+c} - \beta_K^{(0)} \right)^2 & \textrm{~~if~} 
c \ge c_{\rm pin} (\beta) \, ,
\end{array}
\right . \\
& \textrm{if~~} \beta \ge \beta_K^{(0)} \qquad
\frac{\Delta \mathcal{S}_1 (c)}{N} =  \frac{M (1 - c)}{8} \left( \beta \sqrt{1+c} - \beta_K^{(0)} \right)^2 \, .
\end{aligned}
\end{displaymath}
The second line of the above expression coincides with Eq.~(\ref{eq:DS1_REMFC}), which gives $\Delta \mathcal{S}_1 (c)$ for $\beta \ge \beta_K^{(0)}$.

\begin{figure}
\includegraphics[scale=0.36]{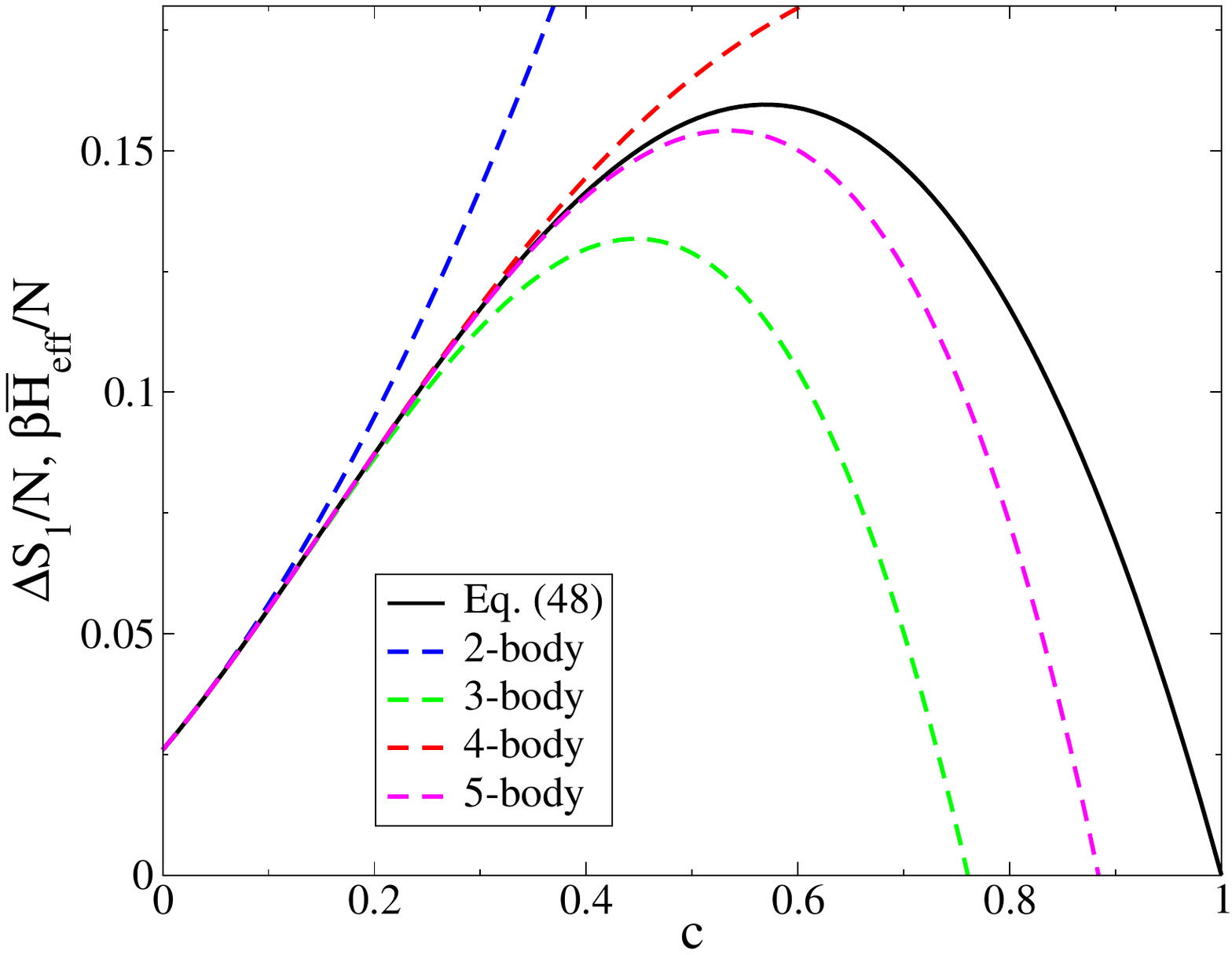}
\caption{Correction $\Delta \mathcal S_1$  to the first cumulant due to the fluctuations of the $q_{ab}^i$'s within the variational ``low-$T$'' approximate scheme as a function of the concentration (global overlap) $c$, for $\beta=1.1 \beta_K^{(0)}$  and $M=3$. The full black line corresponds to Eq.~(\ref{eq:DS1_REMFC}), while the dashed ones correspond to the results obtained after truncating the effective Hamiltonian in Eq.~(\ref{eq:Heff_REMFC-expansion}) to $2$- (blue), $3$- (green), $4$- (red), and $5$-body (magenta) interaction terms.}
\label{fig:cfr_REMFC}
\end{figure}

As explained in the main text, we now seek for an approximate effective Hamiltonian of the form of a linear combination of generic $q$-body interactions terms, Eq.~(\ref{eq:Heff_REMFC-expansion}), which is able to reproduce the functional dependence of $\Delta \mathcal{S}_1 (c)$. This amounts to expanding in powers of $c$ around $c=0$ and truncating the expansion. Focusing first on the regime $\beta \ge \beta_K^{(0)}$ and expanding Eqs.~(\ref{eq:DS1_REMFC}) and~(\ref{eq:Heff_REMFC-expansion}) up to the $5$-th order in $c$, we derive that
\begin{displaymath}
\begin{split}
\frac{\Delta {\cal S}_1 (c)}{N} & = \frac{M}{8} \bigg[ (\beta - \beta_K^{(0)})^2
+ \beta_K^{(0)} (\beta - \beta_K^{(0)}) c 
+ \frac{\beta (5 \beta_K^{(0)} - 4 \beta)}{4} c^2 - \frac{3 \beta \beta_K^{(0)}}{8} c^3
	+ \frac{13 \beta \beta_K^{(0)}}{64} c^4 - \frac{17  \beta \beta_K^{(0)}}{128} c^5 
	%+ \frac{49 M \beta \beta_K^{(0)}}{512} c^6 
	+ \ldots \bigg] \\
& = {\rm cst} - \mu c - (w_2/2) c^2 - (w_3/3!) c^3 - (w_4/4!) c^4 - (w_5/5!) c^5 %- (w_6/6!) c^6 
	+ \ldots \, ,
\end{split}
\end{displaymath}
from which one immediately extract the effective parameters $\mu$, $w_2$, $w_3$, $w_4$, etc, which are reproduces in Eq.~(\ref{eq:S1_fin_REMFC}) of the main text. The outcome of this procedure, with a truncation of $\Delta {\cal S}_1 (c)$ at several orders in $c$ up $c^5$ is shown in Fig.~\ref{fig:cfr_REMFC}. One can see that the description at the $5$th order is good over the whole range of $c$ but nonetheless deteriorates for $c\gtrsim 0.6$, and we have also considered an expansion up to the $10$th order (not shown here).

The above strategy does not work for temperatures higher than the ``zeroth-order'' RFOT temperature ($\beta<\beta_K^{(0)}$), since $\Delta \mathcal{S}_1 (c)$ has a nonanalyticity in $c_{\rm pin} (\beta)$ that cannot be reproduced by expanding Eq.~(\ref{eq:DS1_REMFC}) around zero. In this case, since $\Delta \mathcal{S}_1 (c)$ is identically zero for $c< c_{\rm pin} (\beta)$, all the first $N c_{\rm pin} (\beta)$-body couplings $w_p$ must vanish. The $1$-replica action can thus, at least formally, be written as
\begin{displaymath}
\Delta {\cal S}_1 (c) = -\sum_{q=N c_{\rm pin}(\beta)}^N \frac {w_q}{q!}
\sum_{\{\pi\}} p^{\pi(1)} \cdots p^{\pi(q)} (1 - p^{\pi(q+1)})
\cdots (1 - p^{\pi(N)}) \, ,
\end{displaymath}
where $\{ \pi \}$ is the set of all possible permutations of $\{ 1, \ldots, N \}$. In consequence, in order to reproduce the shape of $\Delta \mathcal{S}_1 (c)$ one needs {\it all} possible $q$-body interactions involving an extensive number of variables, from $q=N c_{\rm pin} (\beta)$ to $q = \infty$. This situation is analytically intractable. However, the variational approximation scheme proposed in Sec.~\ref{sec:approximations} is ``low-$T$'' in spirit and is therefore expected to be better justified for $T<T_K^{(0)}$. We can thus consider the regime $\beta<\beta_K^{(0)}$  as a pre-asymptotic regime. In the following, for simplicity, we will compute the coupling constants of the effective Hamiltonian only in the region $\beta>\beta_K^{(0)}$, and analytically continue them also in the high-temperature phase (but not very deep in this phase).

\subsection{Variational approximation for the $2$-replica action}
\label{app:REMFC-S2}

As explained in the main text, in order to compute the second cumulant of the effective action within our variational approach we have divided the $n$ constrained replicas in two groups of $n_1$ and $n_2$ replicas respectively, and considered random configurations of the $p_a^i$'s ($a=1,2$) with $c_1 N$ sites where $p^i$ is one for the first group of replicas and zero for the others, $c_2 N$ sites where $p^i$ is zero for the first group of replicas and one
for the others, $c_{12} N$ sites where $p^i$ is one for both group of replicas, and $c_0 N$ sites where $p^i$ is zero for both groups of replicas (with $c_0 = 1-c_1-c_2-c_{12}$). To compute the $2$-replica action we keep only the terms of order $n_1 n_2$ in the expression of the effective action, and take the limit $n_1 , n_2 \to 0$ in the end (see also paper~I).

The sum appearing in the exponential of Eq.~(\ref{eq:Srep_KREM}) can be expressed as
\begin{displaymath}
\begin{split}
 \sum_{i \neq j } \sum_{\alpha,\beta}
\delta_{\mathcal{C}_i^\alpha,\mathcal{C}_i^\beta} \delta_{\mathcal{C}_j^\alpha,\mathcal{C}_j^\beta}
=& N^2 \Big \{ \big( 1 + n_1 m_1^\star + n_2 m_2^\star \big ) 
( c_0^2  +  2 c_0 c_1 + 2 c_0 c_2 +
2 c_0 c_{12} + 2 c_1 c_2) \\
& + \big( 1 + n_1^2 + n_2 m_2^\star \big ) ( c_1^2 + 2 c_1 c_{12} )
+ \big( 1 + n_2^2 + n_1 m_1^\star \big ) ( c_2^2 + 2 c_2 c_{12} )
+ \big[1 + (n_1 + n_2)^2 \big] c_{12}^2 \Big \} \, .
\end{split}
\end{displaymath}
The only term of order $n_1 n_2$ in the above expression is $2 n_1 n_2 c_{12}^2 = 2 n_1 n_2 [(1/N)\sum_i p_1^i p_2^i ]^2$.

The trace over all possible configurations $\{\mathcal{C}_i^\alpha \}$ that are compatible with our variational ansatz gives the following combinatorial factor:
\begin{displaymath}
\sum_{\{ \mathcal{C}_i^\alpha \}} \prod_{a,i} 
\delta_{p_a^i,\delta_{\mathcal{C}_i^0,\mathcal{C}_i^a}} = 
\left[ { 2^M - 1 \choose n_1/m_1^\star } \right]^{c_2 N}
\left[ { 2^M - 1 \choose n_2/m_2^\star } \right]^{c_1 N}
\left[ { 2^M - 1 \choose n_1/m_1^\star + n_2/m_2^\star } \right]^{c_0 N} \, ,
\end{displaymath}
yielding a term in $\exp( -N [\psi^{(1)} (2^M) + \pi^2/6 ] (n_1 n_2)/(m_1^\star m_2^\star) c_0]$ in Eq.~(\ref{eq:Srep_KREM}). 

After collecting all the terms of order $n_1 n_2$ and using the fact that $c_0 = (1/N) \sum_i (1 - p_1^i) (1 - p_2^i)$, we then obtain
\begin{displaymath}
\mathcal{S}_2 [p_1^i, p_2^i] = \frac{M \beta^2}{4 N} \sum_{i,j} p_1^i  p_1^j p_2^i p_2^j
- \frac{1}{m_1^\star m_2^\star} \big [ \psi^{(1)} (2^M) + \frac{\pi^2}{6} \big] \sum_i p_1^i p_2^i \, ,
\end{displaymath}
where $m_1^\star$ and $m_2^\star$ are the values of $m$ that minimize the $1$-replica part of the replicated action ({\it i.e.}, the terms of order $n_1$ and $n_2$ separately), which are given by Eq.~(\ref{eq:mstar}). For $\beta \ge \beta_K^{(0)}$, one finds
\begin{displaymath} 
m_1^\star = \frac{\beta_K^{(0)}}{\beta \sqrt{1 + c_1 + c_{12}}} \, , 
\qquad
\textrm{and}
\qquad
m_2^\star = \frac{\beta_K^{(0)}}{\beta \sqrt{1 + c_2 + c_{12}}} \, .
\end{displaymath}
Bu using again the fact that $c_1 + c_{12} = (1/N) \sum_i p_1^i$ and $c_2 + c_{12} = (1/N) \sum_i p_2^i$ and expanding the expressions above for small $c_1$, $c_2$ and $c_{12}$, we obtain
\begin{displaymath}
\begin{split}
\frac{1}{m_1^\star m_2^\star} \sum_i p_1^i p_2^i &\approx
\left ( \frac{\beta}{\beta_K^{(0)}} \right)^{\!2} \bigg[ \sum_{i} p_1^i p_2^i
+ \frac{1}{2N} \sum_{i,j} p_1^i p_2^i ( p_1^j + p_2^j) + \frac{1}{4N^2} \sum_{i, j, k} p_1^i p_1^j p_2^i p_2^k
-  \frac{1}{8N^2} \sum_{i ,j, k} p_1^i p_2^i ( p_1^j p_1^k + p_2^j p_2^k)   
\ldots \bigg] \, .
\end{split}
\end{displaymath}
This term generates higher-order correlations between random couplings and random chemical potentials which, however, are not proportional to $M$ and stay of order $1$ for large $M$.

Collecting all the terms we finally obtain the expression of the second cumulant of the effective action given in Eq.~(\ref{eq:S1S2_REMFC}) of the main text.

\section{Construction of the approximate effective theory for the finite-dimensional lattice version of the $2^M$-KREM } 
\label{app:REM-Kac}

The starting point is similar to that of the fully connected version in Eq.~(\ref{eq:Srep_KREM}),
\begin{equation} 
\label{eq:SrepKac}
\begin{split}
e^{-\mathcal{S}_{\rm rep} [\{p_a^i\}]} &= \frac{1}{\overline{Z}} \overline{\sum_{\{ \mathcal{C}_i^\alpha \}} 
e^{ - \frac{\beta}{\sqrt{d}} \sum_{\langle i,j\rangle}\sum{\alpha=0}^n 
E_{\langle i, j \rangle} (\mathcal{C}_i^\alpha, \mathcal{C}_j^\alpha)} \prod_{a,i} 
\delta_{p_a^i,\delta_{\mathcal{C}_i^0,\mathcal{C}_i^a}}}
= \frac{1}{\overline{Z}}  \sum_{\{ \mathcal{C}_i^\alpha \}} e^{\frac{M \beta^2}{2 d} \sum_{\langle i,j\rangle} \sum_{\alpha,\beta=0}^n
\delta_{\mathcal{C}_i^\alpha,\mathcal{C}_i^\beta} \delta_{\mathcal{C}_j^\alpha,\mathcal{C}_j^\beta}}
\prod_{a,i} 
\delta_{p_a^i,\delta_{\mathcal{C}_i^0,\mathcal{C}_i^a}} \, ,
\end{split}
\end{equation}
and, as for the fully connected case, the first steps of the approximation scheme (Secs.~\ref{sec:2-state_variables} and \ref{sec:constraints}) are exactly satisfied by the model.

\subsection{The ``zeroth-order'' approximation}

The ``zeroth-order'' approximation can be obtained following the same procedure as described in Appendix~\ref{app:REMFC-ann}: We set $q_{ab}^i \equiv \delta_{\mathcal{C}_i^a,\mathcal{C}_i^b} = p_a^i p_b^i$ on all sites ({\it i.e.}, $\eta_{ab}^i = 0$),  insert this ansatz into Eq.~(\ref{eq:SrepKac}),  and perform the trace over the configurations ${\cal C}_i^\alpha$. This leads to the following replicated action,
\begin{displaymath} 
\begin{aligned}
	{\mathcal S}_{\rm rep}^{{(0)}} [ \{ p_a^i \}]  & = \sum_a\Big ( - \frac{M \beta^2}{2 d} \sum_{ \langle i , j \rangle } p_a^i p_a^j + 
\big[ \psi^{(0)} (2^M) + \gamma \big]  \sum_i p_a^i  \Big ) \\
	& \qquad \qquad -\frac 12 \sum_{ab}\Big (
 \frac{M \beta^2}{d} \sum_{\langle i , j \rangle } p_a^i p_a^j p_b^i p_b^j - \big[  \psi^{(1)} (2^M) 
+ \frac{\pi^2}{6} \big] \sum_{i} p_a^i p_b^i\Big ) \, ,
 \end{aligned}
\end{displaymath}
which we have truncated at the $2$-replica level. This has exactly the same form as the replicated action for the fully connected case in Eq.~(\ref{eq:S1S2-REM}).
%corresponds to the expansion in cumulants (trucated at the second one) of a disordered Hamiltonian as in  with $\tilde \mu=$, $w_2=$, $\overline{\d \mu_i^2}= M \beta^2/8 - (\psi^{(1)} (2^M) + \pi^2/6)/4$ and $\overline{\delta \! w_{ij}^2} = M \beta^2/(8 d)$.
%${\cal H}_{\rm eff}^{(0)} = \sum_i (H + \d h_i) \sigma^i - \sum_{\langle i,j \rangle} (J + \delta\! J_{ij}) \sigma^i \sigma^j$, with
%$H = M \beta (\beta_K^{(0)} - \beta)/2$, $J = M \beta^2/8 d$, $\overline{\d h_i^2}= M \beta^2/8 - (\psi^{(1)} (2^M) + \pi^2/6)/4$ and $\overline{\delta \! J_{ij}^2} = M \beta^2/(8 d)$.

\subsection{Variational approach}

To implement the low-$T$ variational approach beyond the zeroth-order result, we choose the same ``periodic-cluster pattern'' for the overlap with the reference configuration as the one described in Sec.~\ref{sec:periodic_cluster} (see also the left panel Fig.~\ref{fig:sketch}), {\it i.e.}, a periodic arrangement of the $p_a^i$'s in which cubes of side $\lambda$ with $p_a=1$ on all their sites are regularly placed on the lattice with a distance $\ell$ between the centers of two neighboring cubes, with $\ell \gg \lambda$; everywhere outside the cubes, $p_a=0$.

On a $d$-dimensional hyper-cube of linear size $L$, the total number of sites is ${\cal N} = L^d$ and the total number of cubes is ${\cal N}_c = (L/\ell)^d$. We start with the $1$-replica action and evaluate the different terms of  Eq.~(\ref{eq:SrepKac}) for the pattern described above and for $p_a^i = p^i$~$\forall a$. On the sites where $p^i = 0$ we consider a variational 1-RSB ansatz, {\it i.e.}, we divide the replicas in $n/m$ blocks of size $m \times m$ such that if $a$ and $b$ are in the same block both replicas are in the same configuration as the reference one ({\it i.e.}, ${\cal C}_i^a = {\cal C}_i^b = {\cal C}_i^0$), whereas if $a$ and $b$ do not belong to the same block the two replicas are in different configurations, which are also different from the reference one ({\it i.e.}, ${\cal C}_i^a \neq {\cal C}_i^0$, ${\cal C}_i^b \neq {\cal C}_i^0$, ${\cal C}_i^a \neq {\cal C}_i^b$). Thus, on a link between two sites $i$ and $j$ where $p^i = p^j = 1$, we have that $\sum_{\alpha,\beta} \delta_{\mathcal{C}_i^\alpha,\mathcal{C}_i^\beta} \delta_{\mathcal{C}_j^\alpha,\mathcal{C}_j^\beta} = (1 + n)^2$. Conversely, on all the other links $\sum_{\alpha,\beta} \delta_{\mathcal{C}_i^\alpha,\mathcal{C}_i^\beta} \delta_{\mathcal{C}_j^\alpha,\mathcal{C}_j^\beta} = 1 + n m$.

The number of links between two sites with $p = 1$ is ${\cal N}_L^{(1,1)} = {\cal N}_c ( d \lambda^d - d \lambda^{d-1})$. The number of all the other links is then $d L^d - {\cal N}_L^{(1,1)} = d L^d [1 - \lambda^{d-1} (\lambda - 1)/\ell^d]$. In consequence,
\begin{displaymath}
\frac{M \beta^2}{2 d} \sum_{\langle i,j\rangle} \sum_{\alpha,\beta}
\delta_{\mathcal{C}_i^\alpha,\mathcal{C}_i^\beta} \delta_{\mathcal{C}_j^\alpha,\mathcal{C}_j^\beta} = 
\frac{M \beta^2 L^d}{2} \Big[ (1+n)^2 \frac{\lambda^{d-1}(\lambda - 1)}{\ell^d} + n m \Big( 1 - 
\frac{\lambda^{d-1}(\lambda - 1)}{\ell^d}\Big)\Big] \, .
\end{displaymath}
In order to trace over all possible configurations $\{\mathcal{C}_i^\alpha\}$ compatible with our variational ansatz, we simply need to compute how many ways there are to choose $n/m$ configurations among the $2^M - 1$ configurations that are different from the reference one on all the sites $i$ where
$p^i = 0$. The number of such sites is ${\cal N}_0 = L^d - \lambda^d {\cal N}_c = L^d [1 - (\lambda/\ell)^d]$. We thus obtain the following combinatorial factor, which we expand up to second order in $n$:
\begin{displaymath}
\left[{{2^M-1}\choose{n/m}}\right]^{{\cal N}_0} 
= \left[\frac{\Gamma(2^M)}{\Gamma\left(1+\frac{n}{m}\right) \Gamma\left(2^M - \frac{n}{m}\right)} \right]^{{\cal N}_0}
\simeq \exp \left\{ 
{\cal N}_0 \left[ \left ( \psi^{(0)} (2^M) + \gamma \right) \frac{n}{m} 
- \frac{1}{2} \left( \psi^{(1)} (2^M) + \frac{\pi^2}{6} \right) \left( \frac{n}{m} \right)^2 
\right] \right \}\, .
\end{displaymath}
After putting all the terms together and keeping only the terms of order $n$, the $1$-replica action can be expressed as
\begin{equation} \label{eq:DS1-Kac}
\mathcal{S}_1(\lambda,\ell,m) \simeq - L^d \left \{  
\frac{M d \beta^2}{2 d} \left[ \frac{2 \lambda^{d-1}(\lambda - 1)}{\ell^d} + m \left( 1 - 
\frac{\lambda^{d-1}(\lambda - 1)}{\ell^d}\right) \right] + 
\left[ 1 - \left( \frac{\lambda}{\ell} \right)^{\! d} \right] \frac{\psi^{(0)} (2^M) + \gamma}{m} \right \} \, .
\end{equation}
(We recall that for $m=1$ we get back the result of the ``zeroth-order'' approximation.) Taking the derivative of Eq.~(\ref{eq:DS1-Kac}) with respect to $m$,  one gets (as expected, the result depends on whether the temperature is above or below the ``zeroth-order'' RFOT temperature $T_K^{(0)}=1/\beta_K^{(0)}= [2 ( \psi^{(0)} (2^M) + \gamma )/M]^{-1/2}$)
\begin{displaymath} 
\begin{aligned}
& \textrm{if~} \beta<\beta_K^{(0)} \textrm{~~~~} m_\star = \left \{
\begin{array}{ll}
1 & \textrm{~~if~} \ell>\ell_{\rm pin} (\beta) \, , \\
\frac{\beta_K^{(0)}}{\beta} \sqrt{\frac{\ell^d - \lambda^d}{\ell^d - \lambda^{d-1}(\lambda-1)}}
 & \textrm{~~if~} \ell \le \ell_{\rm pin} (\beta) \, ,
\end{array}
\right .
\textrm{~~~with~~} \ell_{\rm pin} (\beta) = 
\left( \frac{(\beta_K^{(0)})^2 \lambda^d - \beta^2 \lambda^{d-1} (\lambda - 1)}{(\beta_K^{(0)})^2 - \beta^2} 
\right)^{\!\frac{1}{d}} \, , \\
& \textrm{if~} \beta \ge \beta_K^{(0)}  \textrm{~~~~}  m_\star = 
\frac{\beta_K^{(0)}}{\beta} \sqrt{\frac{\ell^d - \lambda^d}{\ell^d - \lambda^{d-1}(\lambda-1)}} \, ,
\end{aligned}
\end{displaymath}
where $\ell_{\rm pin}$ is akin the ``pinning'' correlation length, which is a specific instance of a point-to-set correlation length. Inserting these expressions into Eq.~(\ref{eq:DS1-Kac}), we obtain after some simple algebra $\Delta \mathcal{S}_1$ (per site) as a function of $\ell$, $\lambda$, and of the temperature, as
\begin{displaymath} 
\begin{aligned}
& \textrm{if~~} \beta<\beta_K^{(0)} \qquad \frac{\Delta \mathcal{S}_1}{L^d} = \left \{
\begin{array}{ll}
0 & \textrm{~~if~} \ell>\ell_{\rm pin} (\beta) \, ,\\
\frac{M}{2 \ell^d} \left( \beta_K^{(0)} \sqrt{\ell^d - \lambda^d} - \beta
\sqrt{\ell^d - \lambda^{d-1} (\lambda - 1) } \right)^2 & \textrm{~~if~} 
\ell \le \ell_{\rm pin} (\beta) \, ,
\end{array}
\right . \\
& \textrm{if~~} \beta \ge \beta_K^{(0)} \qquad
\frac{\Delta \mathcal{S}_1}{L^d} =  
\frac{M}{2 \ell^d} \left( \beta_K^{(0)} \sqrt{\ell^d - \lambda^d} - \beta
\sqrt{\ell^d - \lambda^{d-1} (\lambda - 1) } \right)^2 \, .
\end{aligned}
\end{displaymath}
Since we are interested in studying the system significantly below the mean-field RFOT at $T_K^{(0)}$, we will mostly focus in the following on the case $\beta > \beta_K^{(0)}$ which corresponds to $\ell_{\rm pin} = \infty$. After expanding the non mean-field part of the $1$-replica action in powers of $\lambda/\ell$, we  find
\begin{displaymath}
\frac{\Delta \mathcal{S}_1(\lambda,\ell)}{L^d} \approx \frac{M}{2} \left [ 
(\beta - \beta_K^{(0)})^2 + \beta (\beta - \beta_K^{(0)}) \frac{\lambda^{d-1}}{\ell^d} - (\beta - \beta_K^{(0)})^2 
\frac{\lambda^d}{\ell^d} + \frac{\beta \beta_K^{(0)}}{4} \, \frac{\lambda^{2d-2}}{\ell^{2d}} + \frac{\beta \beta_K^{(0)}}{4} \,
\frac{\lambda^{3d-2}-\lambda^{3d-3}/2}{\ell^{3d}}  \right] \, .
\end{displaymath}
As explained in Sec.~\ref{sec:approximations}, our strategy will now consist in finding an approximate ansatz for the effective action that allows us to reproduce this result by means of an effective external source and effective $2$- and multi-body interactions in a translationally invariant theory. More specifically, we choose the effective $1$-replica action of the form of Eq.~(\ref{eq:Heff-1}):
\begin{displaymath} 
\Delta \mathcal S_{1,\rm eff}[p] = \mu \sum_i (1 - p^i) + w_2 \sum_{\langle i, j \rangle}
p^i (1 - p^j) + \frac{\tilde{w}_2}{2} \sum_{|i- j|>a} \frac{p^i 
p^j}{|i-j|^{2d}} + \frac{w_4}{L^d} \!\!\! \sum_{\langle i, i^\prime \rangle \neq \langle j j^\prime \rangle}
\!\!\! [p^i (1 - p^{i^\prime}) p^j (1 - p^{j^\prime}) ]_{\rm symm} 
\, ,
\end{displaymath}
which for the specific pattern of the $p^i$'s chosen here gives (in the continuum limit)
\begin{displaymath}
\frac{\Delta \mathcal S_{1,\rm eff}}{L^d} = \mu \left[ 1 - \frac{\lambda^d}{\ell^d} \right] + d w_2 \frac{\lambda^{d-1}}{\ell^d}
+ \frac{\tilde{w}_2 \Omega_d}{2 d} \, \frac{\lambda^{2d}}{\ell^{3d}} + w_4 d^2 \frac{\lambda^{2d-2}}{\ell^{2d}} \, ,
\end{displaymath}
Thus, the choice of the effective parameters such that $\Delta \mathcal S_{1,\rm eff}$ best reproduces the functional dependence of $\Delta \mathcal{S}_1$ on $\ell$ and $\lambda$ is
\begin{displaymath}
\begin{split}
\mu & = \frac{M}{2} (\beta - \beta_K^{(0)})^2 \, ,\\
w_2 & = \frac{M \beta (\beta - \beta_K^{(0)})}{2 d} \, ,\\
\tilde{w}_2 & \approx \frac{M d \beta \beta_K^{(0)}}{8 \Omega_d} \, ,\\
w_4 &= \frac{M \beta \beta_K^{(0)}}{8d^2} \, .
\end{split}
\end{displaymath} 
Note that the functional dependence of the term of order $\ell^{-3d}$ in $\Delta \mathcal{S}_1$ is not exactly reproduced by the effective description, since the former is proportional to $\lambda^{3d-2}-\lambda^{3d-3}/2$ and the latter to $\lambda^{2d}$. Here we have chosen the value of $\tilde{w}_2$ such that the two terms are equal for $\lambda=1$.

The behavior of $\Delta {\cal S}_1$ {\it above} $T_K^{(0)}$ can also be reproduced by the effective $1$-replica action with the same coefficients $w_2$, $\tilde{w}_2$, and $w_4$ as given above, by introducing a finite interaction range $\ell_{\rm pin}$ for the $2$- and $4$-body interaction terms.

Following the strategy described in Appendix~\ref{app:REMFC-S2} for the fully connected model, we can also compute the $2$-replica effective action within the variational approximation. Neglecting all higher-order disorder correlations, we finally obtain
\begin{displaymath} 
\begin{split}
& {\mathcal S}_1 [ p^i] = \frac{M \beta \beta_K^{(0)}}{2} \sum_i p^i - \frac{M \beta (2 \beta - \beta_K^{(0)})}{2d}
\sum_{\langle i, j \rangle} p^i p^j
+ \frac{\tilde{w}_2}{2} \sum_{\vert  r_i-r_j\vert >a} \frac{p^i 
p^j}{|i-j|^{2d}} + \frac{w_4}{L^d} \!\!\! \sum_{\langle i, i^\prime \rangle \neq \langle j j^\prime \rangle}
\!\!\! [p^i (1 - p^{i^\prime}) p^j (1 - p^{j^\prime}) ]_{\rm symm} \, , \\
& {\mathcal S}_2 [ p_1^i, p_2^i] =
\frac{M \beta^2}{d} \sum_{\langle i , j \rangle } p_1^i p_1^j p_2^i p_2^j 
	- \left( \frac{\beta}{\beta_K^{(0)}} \right)^{\!2} \Big[ \psi^{(1)} (2^M) 
	+ \frac{\pi^2}{6} \Big]
 \sum_{i} p_1^i p_2^i + \ldots \, .
 \end{split}
\end{displaymath}
As found before for the overlap field theory of glass-forming liquids and for the fully connected KREM, the above expressions correspond to the cumulants of an effective disordered Hamiltonian, $\b {\cal H}_{\rm eff} [p]$. By going from the overlap variables $p^i = 0,1$ to the Ising spins, $\sigma^i = \pm 1$ via the relation $p^i = (1 + \sigma_i)/2$, one finally obtains the following disordered Ising Hamiltonian:
\begin{equation} \label{eq:Heff_final_REMKac}
	\begin{aligned}
	\b {\cal H}_{\rm eff}[\sigma] &= {\cal S}_0 -\sum_i (H + \d h_i) \sigma^i 
- \sum_{\langle i , j \rangle } ( J_2 + \delta \! J_{ij} )\sigma^i \sigma^j 
+ \frac{\tilde{J}_2}{2} \sum_{\vert  r_i-r_j\vert  >a} \frac{\sigma^i 
\sigma^j}{|r_i-r_j|^{2d}} \\
	& \qquad \qquad + \frac{J_4}{L^d} \!\!\! \sum_{\langle i, i^\prime \rangle \neq \langle j j^\prime \rangle}
\!\!\! [(1 + \sigma^i) (1 - \sigma^{i^\prime}) (1 + \sigma^j) (1 - \sigma^{j^\prime}) ]_{\rm symm} \, ,
\end{aligned}
\end{equation}
where the applied uniform source $H$, the effective couplings $J_2$, $\tilde{J}_2$, $J_4$, and the second cumulants of the random variables are given by
\begin{equation} \label{eq:coefficients_REMKac}
\begin{split}
H & = \frac{M \beta}{2} \Big ( \beta_K^{(0)} - \beta + \frac{\beta_K^{(0)}}{16 a^d} \Big ) \, ,\\
J_2 & = \frac{M \beta (2 \beta - \beta_K^{(0)})}{8 d} \, , \\
\tilde J_2 & \approx \frac{M d \beta \beta_K^{(0)}}{32 \Omega_d} \, ,\\
J_4 &= \frac{M \beta \beta_K^{(0)}}{128d^2} \, \\
\overline{\d h_i \d h_j} & = \bigg [ \frac{M \beta^2}{8} -  \frac{1}{4} \bigg( \frac{\beta}{\beta_K^{(0)}} \bigg)^{\!2} 
\Big( \psi^{(1)} (2^M) + \frac{\pi^2}{6} \Big) \bigg ]\delta_{ij} + \frac{M \beta^2}{16} \mathbb{C}_{ij} \, , \\
\overline{\delta \! J_{ij} \delta \! J_{kl}} & = \frac{M \beta^2}{32 d} 
( \delta_{ik} \delta_{jl} + \delta_{il} \delta_{jk}) \, , 
\end{split}
\end{equation}
where $\mathbb{C}_{ij}$ is the connectivity matrix of the lattice.  (${\cal S}_0$ is a random term that does not depend
on the Ising variables.) The requirement of a positive variance for the random field imposes that $M\gtrsim2.1$.

In conclusion, the effective Hamiltonian has exactly the same random-field + random-bond Ising form as that found for describing glass-forming liquids. As in the latter case, the configurational entropy is renormalized by a positive factor, implying that the thermodynamic glass transition temperature is lowered with respect to its mean-field value. For this specific model the fluctuations of the ferromagnetic coupling always stay smaller than the average value, which excludes the possibility of having an effective theory in the class of an Ising spin-glass in a field as advocated in Ref.~[\onlinecite{Moore}].

\subsection{Estimate of the transition}
\label{app:REM-Kac-estimate}

Following the steps described in Sec.~\ref{sec:effective_theory-estimation}, one can use the approximate effective theory to estimate whether the thermodynamic glass transition exists as a function of the number of states $M$ and of the dimensions $d$. In order to do this, we ``project'' the effective Hamiltonian~(\ref{eq:Heff_final_REMKac}) onto a standard short-range RFIM. As explained in Sec.~\ref{sec:4body}, performing a Hubbard-Stratanovich transformation and a saddle-point calculation allows one to decouple the $4$-body link-link interaction, yielding a ``renormalized'' value of $J_2$ [see Eq.~(\ref{eq:Heff_psi})]. Taking $\psi = 0$ (which provides a lower bound for the effective ferromagnetic coupling) leads to $J_2 \to J_2 + 2 d J_4$. On the other hand, as discussed in Sec.~\ref{sec:effective_theory-estimation}, the pairwise antiferromagnetic interaction disfavors magnetically ordered phases, and its effect can be taken into account as an effective decrease of the the short-range ferromagnetic coupling,
\begin{displaymath}
J_2^{\rm eff} \approx J_2 + 2 d J_4 - \frac{\tilde J_2 \Omega_d }{2d} 
\int_a^\infty r^{-d-1} \, {\rm d} r = \frac{M \beta}{8 d} \Big[ 2 \beta - \frac{\beta_K^{(0)}}{8}
\Big( 7 + \frac{1}{8 a^d} \Big ) \Big] 
\, .
\end{displaymath}
As a first approximation, we neglect the random-bond disorder and all higher-order disorder correlations.

A transition can then only take place in zero external field, $H=0$, {\it i.e.}, when
\begin{displaymath}
\beta_c = \beta_K^{(0)} \Big( 1 + \frac{1}{16 a^d} \Big ) \, .
\end{displaymath}
As discussed in Sec.~\ref{sec:effective_theory-estimation}, a transition  exists  in the $3$-dimensional (standard) short-range RFIM with a Gaussian distributed random field provided that $\sqrt{\Delta_h}/J_2^{\rm eff} \lesssim 1.2$.\cite{Nattermann,middleton-fisher,martinmayor-fytas} We postulate a rough generalization of this criterion in higher dimensions by assuming that  in $d$-dimension a transition exists for $\sqrt{\Delta_h}/(d J_2^{\rm eff}) \lesssim 0.4$. (The upper bound for the ratio is rather 0.5 in $d=4$\cite{fytas_4,middleton_4} and in $d=5$,\cite{fytas_5} but this makes no qualitative difference.) By inserting the expression of $\Delta_h$ given in Eq.~(\ref{eq:coefficients_REMKac}) into this condition we obtain the minimum value of $M$ for which a thermodynamic glass transition can occur as a function of the dimensionality of the lattice.

The results are listed in the table below and plotted in Fig.~\ref{fig:estimate_rem}. They can also be compared to the recent predictions of a Migdal-Kadanoff real-space renormalization group study of the same model\cite{Angelini}, as discussed in the main text (see also Fig.~\ref{fig:estimate_rem}).
\begin{equation} 
\label{tab:estimate_rem}
\begin{array}{c|c}
d & M \\
\hline
3&  47.2571\\
4&  35.2346\\
5&  28.0212\\
6&  23.2122\\
7&  19.7772\\
8&  17.2009\\
9&  15.1972\\
10&  13.5943\\
11&  12.2828\\
12&  11.19\\
14&  9.47323\\
16&  8.18658\\
18&  7.18718\\
20&  6.38935\\
25&  4.96131\\
30&  4.02112\\
40&  2.87458
\end{array}
\end{equation}


\begin{thebibliography}{99}
\bibitem{footnote_2-state}
Note that the argument we have used to motivate the description of the overlaps in terms of $2$-state variables is based on average quantities, in particular on the shape of the potential $U(q)$. However, the potential, which is akin to the Franz-Parisi potential already discussed, fluctuates from region to region in a picture of finite-dimensional glass-formers sketched in Fig. 2 of paper~I.\cite{paperI} One should work at temperatures such that the typical local potentials in most regions present two well-formed minima.
\bibitem{footnote_continuum}
The continuum limit is justified when the configurations of the $\tau_a^i$'s that dominate the thermodynamics have a small to moderate concentration of sites where $\tau_a^i=1$. This should be valid for temperatures above the thermodynamic glass transition of the exact model, and therefore still below the mean-field RFOT at $T_K^{MF}$. As we are mainly interested by the existence and the properties of the thermodynamic glass transition and not by those of the putative ideal glass phase itself, considering the continuum limit is therefore sensible.
\bibitem{S0} 
The role of the $\sigma$-independent random energy term $\mathcal S_0$ is to ensure that the variance of the effective hamiltonian is zero in the low-overlap state, $\sigma^i=-1\; \forall i$.
\bibitem{footnote_incompatible}
Although the absence of a terminal critical point in the ($T$,$\epsilon$) diagram precludes the presence of a nonzero thermodynamic transition in $\epsilon=0$, the reverse may not be true. For instance, one knows from spin plaquette models that a $T_K=0$ is compatible with a terminal critical point at a nonzero temperature.\cite{jack_plaquette,rulquin_plaquette}
\bibitem{footnote_bare-parameters}
One of the reason for the uncertainty in the estimation is that Ref.~[\onlinecite{Dzero}] uses the action (\ref{eq_replica_action_wolynes}) at the saddle-point level to obtain predictions and, hence, an estimate of the parameters, i.e., as if (\ref{eq_replica_action_wolynes}) were a free energy. Instead, in our case, (\ref{eq_replica_action_wolynes}) is the action governing the fluctuations of the overlap field, thus its parameters are "bare" coupling constants that cannot be directly connected to correlation functions and thermodynamic behavior.
\bibitem{footnote_2D}
A recent simulation study\cite{berthier2d} gives evidence that the overlap fluctuations in a $2$-$d$ glass-forming liquid model behave quite differently than in a $3$-$d$ model.
\end{thebibliography}
\end{document}